\documentclass[twocolumn]{aastex701}

\usepackage{CJKutf8}
\usepackage{amsmath}

\usepackage{natbib}

\newcommand{\oiiitext}{[O~{\sc iii}]}
\newcommand{\oiiihb}{[O~{\sc iii}]/H$\beta$}

\newcommand{\Lbol}{$L_{\rm bol}$}
\newcommand{\edd}{$\lambda_{\rm Edd}$}
\newcommand{\logLbol}{log($L_{\rm bol}$/erg s$^{-1}$)}

\newcommand{\Mbh}{$M_{\rm BH}$}
\newcommand{\logMbh}{log($M_{\rm BH}$/$M_\odot$)}

\newcommand{\fwhmhb}{FWHM$_{\rm H\beta}$}
\newcommand{\fwhmha}{FWHM$_{\rm H\alpha}$}

\newcommand{\mum}{\ifmmode{\rm \mu m}\else{$\mu$m}\fi}
\newcommand{\vdisp}{$\vdisp$}

\newcommand{\wjiu}{W$_{90}$}

\newcommand{\Lwu}{{$\lambda L_{\lambda}(5100)$}}

\newcommand{\vjiuba}{{$v_{98}$}}

\newcommand{\ewo}{ EW$_{\rm [O{\scriptsize III}]}$ }
\newcommand{\ewfe}{ EW$_{\rm Fe{\scriptsize II}}$ }
\newcommand{\ewhb}{ EW$_{\rm H{\beta}, broad}$ }
\newcommand{\RFe}{ R$_{\rm Fe}$ }

\newcommand{\oii}{{[O$\,${\scriptsize II}] $\lambda$$\lambda$3726,3729}}
\newcommand{\oiii}{{[O$\,${\scriptsize III}] $\lambda$5007}}
\newcommand{\oiiiab}{{[O$\,${\scriptsize III}] $\lambda$$\lambda$4959,5007}}

\newcommand{\civ}{\hbox{C$\,${\scriptsize IV} $\lambda$1549}}
\newcommand{\civtext}{\hbox{C$\,${\scriptsize IV}}}

\newcommand{\ciiitext}{\hbox{C$\,${\scriptsize III}]}}
\newcommand{\siivtext}{\hbox{Si$\,${\scriptsize IV}]}}

\newcommand{\niitext}{\hbox{[N$\,${\scriptsize II}]}}
\newcommand{\siitext}{\hbox{[S$\,${\scriptsize II}]}}
\newcommand{\nii}{\hbox{[N$\,${\scriptsize II}] $\lambda$$\lambda$6548,6583}}
\newcommand{\sii}{\hbox{[S$\,${\scriptsize II}] $\lambda$$\lambda$6716,6731}}

\newcommand{\ha}{\hbox{H$\alpha$}}
\newcommand{\hb}{\hbox{H$\beta$}}
\newcommand{\hg}{\hbox{H$\gamma$}}

\newcommand{\mgii}{\hbox{Mg$\,${\scriptsize II}} $\lambda$2800}
\newcommand{\mgiitext}{\hbox{Mg$\,${\scriptsize II}}}
\newcommand{\feii}{\hbox{Fe$\,${\scriptsize II}}}

\newcommand{\kms}{km s$^{-1}$} 
\newcommand{\msun}{M$_{\odot}$}

\newcommand{\ergs}{erg\,s$^{-1}$}

\begin{document}

\title{A JWST/NIRSpec Integral Field Unit Survey of Luminous Quasars at $z \sim$ 5--6 (Q-IFU): \\
Rest-frame Optical Nuclear Properties and Extended Nebulae}

\author[0000-0003-3762-7344]{Weizhe Liu \begin{CJK}{UTF8}{gbsn}(刘伟哲)\end{CJK}}
\affiliation{Steward Observatory, University of Arizona, 933 N. Cherry Ave., Tucson, AZ 85721, USA}
\email{wzliu@arizona.edu}

\correspondingauthor{Weizhe Liu}
\email{wzliu@arizona.edu}

\author[0000-0003-3310-0131]{Xiaohui Fan}
\affiliation{Steward Observatory, University of Arizona, 933 N. Cherry Ave., Tucson, AZ 85721, USA}
\email{}

\author[0000-0003-1245-5232]{Richard Green}
\affiliation{Steward Observatory, University of Arizona, 933 N. Cherry Ave., Tucson, AZ 85721, USA}
\email[]{}

\author[0000-0002-6184-9097]{Jaclyn B. Champagne}
\affiliation{Steward Observatory, University of Arizona, 933 N. Cherry Ave., Tucson, AZ 85721, USA}
\email[]{}

\author[0000-0002-5768-738X]{Xiangyu Jin}
\affiliation{Steward Observatory, University of Arizona, 933 N. Cherry Ave., Tucson, AZ 85721, USA}
\email[]{}

\author[0000-0002-6221-1829]{Jianwei Lyu}
\affiliation{Steward Observatory, University of Arizona, 933 N. Cherry Ave., Tucson, AZ 85721, USA}
\email[]{}

\author[0000-0003-4924-5941]{Maria Pudoka}
\affiliation{Steward Observatory, University of Arizona, 933 N. Cherry Ave., Tucson, AZ 85721, USA}
\email[]{}

\author[0000-0003-0747-1780]{Wei Leong Tee}
\affiliation{Steward Observatory, University of Arizona, 933 N. Cherry Ave., Tucson, AZ 85721, USA}
\affiliation{Department of Astronomy and Astrophysics, The Pennsylvania State University, 525 Davey Lab, University Park, PA 16802, USA}
\email{}

\author[0000-0002-7633-431X]{Feige Wang}
\affiliation{Department of Astronomy, University of Michigan, 1085 S. University Ave., Ann Arbor, MI 48109, USA}
\email[]{}

\author[0000-0001-5287-4242]{Jinyi Yang}
\affiliation{Department of Astronomy, University of Michigan, 1085 S. University Ave., Ann Arbor, MI 48109, USA}
\email[]{}

\author[0000-0003-3307-7525]{Yongda Zhu}
\affiliation{Steward Observatory, University of Arizona, 933 N. Cherry Ave., Tucson, AZ 85721, USA}
\email[]{}

\author[0009-0008-9125-9208]{Nayera Abdessalam}
\affiliation{Steward Observatory, University of Arizona, 933 N. Cherry Ave., Tucson, AZ 85721, USA}
\email[]{}

\author[0000-0002-4321-3538]{Haowen Zhang}
\affiliation{Canadian Institute for Theoretical Astrophysics, 60 St George St, Toronto, ON M5S 3H8}
\email[]{}

% \author[]{(Add your author info to 00\_Authorship.tex)}
% \affiliation{}
% \email{}

\begin{abstract}
It remains debatable how billion-solar-mass supermassive black holes (SMBHs) form and evolve within the first billion years. We report results from a James Webb Space Telescope (JWST)/NIRSpec integral field unit (IFU) survey of 27 luminous quasars at $z \sim 5$--$6$, enabling a systematic investigation of their key physical properties and the associated, extended line emission. Our sample hosts SMBHs with masses $\log(M_{\mathrm{BH}}/{M_\odot}) \sim 8.6$--$9.7$ and Eddington ratios of $\sim 0.1$--$2.6$ based on \hbox{H$\beta$}, and the \hbox{H$\beta$}-based and \hbox{H$\alpha$}-based \Mbh\ are broadly consistent with each other. Our sample may have a slightly smaller median BH mass and larger median Eddington ratio than lower-redshift quasars within the same luminosity range, although the difference could still be explained by the {systematic uncertainties}. They generally follow the empirical correlations between {the} [O\,\textsc{iii}] $\lambda$5007 equivalent widths and bolometric luminosities or Eddington ratios formed by lower-redshift quasars. The majority of them fall within the Eigenvector~1 planes formed by lower-redshift quasars. Nevertheless, a subset of the sample shows enhanced, blueshifted [O\,\textsc{iii}] emission associated with fast outflows. Spatially extended [O\,\textsc{iii}] line emission is detected in 6 objects and shows morphologies and kinematics consistent with merging activities and/or turbulent and clumpy interstellar media (ISM). {We find a possible kpc-scale quasar-AGN pair in one system, although the data could also be tentatively explained as quasar radiative feedback shaping the ISM of a merging companion galaxy.} Our results provide crucial insight into the rapid growth of SMBHs and the gaseous environments they reside in at $z \sim$ 5--6.
\end{abstract}

\section{Introduction} \label{sec:intro}

Luminous quasars powered by $\sim$$10^8$--$10^{10}$ $M_\odot$ supermassive black holes ({SMBHs}) are already in place within $\sim$1 billion years after the Big Bang. It remains unclear how they grow so rapidly \citep[e.g.,][]{Fan2006,Volonteri2012,Wu2015,Banados2018,Matsuoka2019b, Onoue2019,Shen2019,Yang2021, Schindler2020,Wang2021,Farina2022,Fan2023}. 
To deepen our understanding of the early growth of such giants, it is crucial to obtain reliable measurements of their BH {masses}. 
For high-$z$ Type 1 quasars, BH masses are usually obtained using the single-epoch virial method, which assumes virial motions for the line-emitting gas within the broad line region (BLR). The BLR sizes are derived from the quasar continuum luminosity following empirical correlations between the two (i.e., the R--L relation), and the conversion from observables to BH mass is further calibrated with active galactic nucleus (AGN) with reverberation mapping (RM) BH mass measurements.
Rest-frame UV and optical broad emission lines, \civ, \mgii, \ha, and \hb, are the workhorses for virial BH mass measurements \citep[e.g.,][]{McLureJarvis2002,McLureDunlop2004,Vestergaard2002,VestergaardPeterson2006,GreeneHo2005,Shen11a,Barth2015,Dietrich2002,Dietrich2009,DietrichHamann2004,Netzer2004,Sulentic2004, Sulentic2006, Greene2010, Assef2011, Ho2012, Runnoe2013a, Brotherton2015a, Plotkin2015, ShemmerLieber2015, Zuo2015}. Among them, \hb\ is deemed the most reliable tracer based on comparisons with both RM-based BH mass and other single-epoch virial BH mass estimators \citep[e.g.,][]{ShenLiu2012,Shen2013review}.
At $z \gtrsim5$, BH masses are mainly derived from \mgii\ or \civ\ in the rest-frame UV \citep[e.g][]{Yang2021,Schindler2020} due to the
challenge of observing faint objects beyond the K-band.
With the advent of JWST, high S/N rest-frame optical spectra can be obtained efficiently \citep[e.g.,][]{Marshall2023,Yang2023b,Loiacono2024,Decarli2024,Liu2024b,Lyu2025}, which starts to confirm the large BH masses of those quasars with the more reliable tracer, like \hb.

{Rest-frame optical spectroscopy also allows for close examinations of well-known empirical correlations of quasars, like Eigenvector~1 \citep[EV1; e.g.,][]{BorosonGreen1992} and ``Baldwin effect'' \citep[e.g.,][]{Baldwin1977}, which reflect fundamental properties of BH accretion. 
{Specifically, the ``Baldwin effect'' originally refers to the empirical anti-correlation between the equivalent widths (EWs) of broad emission lines and the continuum luminosities of quasars. The effect was first discovered for \civ, and one popular explanation is the softer UV continuum shapes and perhaps increasing gas metallicity in more luminous AGN \citep[e.g.,][]{Binette1989,Zheng1993,Korista1998,Dietrich2002}, which lowers the ion populations having high ionization potentials. It has also been suggested that the ``Baldwin effect'' may arise from the combination of increasing luminosities and decreasing EWs of emission lines along with Eddington ratios \citep[e.g.,][]{BaskinLaor2004,Richards2011}. Such ``Baldwin effect'' has also been observed in narrow lines like \oiii\ \citep[e.g.,][]{Steiner1981,Dietrich2002,Netzer2004,Sulentic2004,Zhang2011,SternLaor2012a,SternLaor2012b}, which may result from the same physical mechanisms described above. Nevertheless, other explanations have also been proposed. For example, the narrow line regions (NLRs) of luminous quasars may become gravitationally unbound and lost due to strong outflows or other mechanisms \citep[e.g.,][]{Netzer2004}. } 
Finally, rest-frame optical spectroscopy also unveils the size and kinematics of NLR and more spatially extended ionized gas through emission lines like \oiii\ \citep[e.g.,][and references therein]{Netzer2004,Nesvadba2008,Carniani2015,Liu2013b,Perna2015,Shen2016,Liu2020,Liu2024b,Liu2026a,Vayner2023b,Yang2023b}.}

{At cosmic noon, such spectral studies of the most luminous quasars, such as the WISE/SDSS selected hyperluminous (WISSH) quasars sample \citep[e.g.,][]{Bischetti2019} and extremely red quasars \citep[e.g.,][]{zaka16b,Perrotta2019}, have provided benchmarks for some of the most massive rapidly accreting supermassive black holes and the most extreme quasar-driven outflows.
It is thus critical to understand whether the physical properties of these luminous quasars stay the same or not during the early epoch of cosmic history. 
Previous studies based on the rest-frame UV spectra suggest no significant redshift evolution for the majority of quasar spectral properties up to $z\sim$ 6--7. Recent JWST studies probing the rest-frame optical spectral properties of these $z>5$ quasars again suggest no significant redshift evolution, despite the limited sample size \citep[e.g.,][]{Yang2023b}. Nonetheless, $z>5$ quasars likely show an elevated incidence of broad absorption line (BAL) features indicative of powerful winds\citep[e.g.,][]{Bischetti2022}. In addition, several studies also report an increase of \civtext\ blueshifts \cite{Meyer2019,Schindler2020}. Finally, one latest study suggests that galaxy-scale outflows traced by blueshifted and broad \oiiitext\ emission lines are more frequent in $z>5$ quasars \citep[][]{Liu2026a}. 
}

%based on small samples also suggest no significant evolution 

%provide key diagnostics of SMBH accretion physics and feedback \citep{}. 

{It is therefore imperative to statistically investigate the rest-frame optical spectral properties of $z \gtrsim$ 5 quasars and compare them with those of the lower-redshift objects.
Moreover, integral field unit (IFU) observations are essential to further improve our understanding of $z>5$ quasars by not only capturing the spectra from the quasars themselves but also enabling spatially-resolved spectroscopic diagnostics of various physical properties of their host galaxies and line-emitting nebulae \citep[e.g.,][]{Marshall2023,Marshall2025a,Marshall2025b,Loiacono2024,Decarli2024,Liu2024b,Liu2026a,Wolf2026}. }

In \citet[][]{Liu2026a}, we introduced a JWST cycle 2 NIRSpec/integral field unit (IFU) survey program \#3428 (PI W. Liu; Dubbed: Q-IFU), which observed 27 luminous quasars at $z \sim$ 5--6. In that paper,
we presented novel results on the detection of more frequent extreme galaxy-scale outflows in the $z \sim$ 5--6 quasars from the survey program than in the lower-z quasars.
In this work, we report other key rest-frame optical spectral properties of these $z \sim$ 5--6 quasars and their host galaxies/companion galaxies from the same program.  This paper is organized as follows. In Section \ref{sec:data}, we describe the sample selection, observations and data reduction. In Section \ref{sec:fit}, the spectral fitting and analysis are presented. The discussions on the quasar nuclear spectral properties and extended line emission are presented in \ref{sec:quasar} and \ref{sec:extended}, respectively. Finally, our main conclusions are summarized in Section \ref{sec:conclusion}. 
Throughout the paper, we assume a $\Lambda$CDM cosmology with $H_0 =$ 70 km s$^{-1}$ Mpc$^{-1}$, $\Omega_{\rm m}$ = 0.3, and $\Omega_{\rm \Lambda} = 0.7$. An arcsecond corresponds to 5.71--6.45 kpc at the redshifts of our objects.

\section{Sample Selection, Observation and Data Reduction} \label{sec:data}

The sample selection, observations and data reduction are described in detail in \citet{Liu2026a}.  
In the following, we briefly summarize their key aspects.

\begin{figure}
    \centering
    \plotone{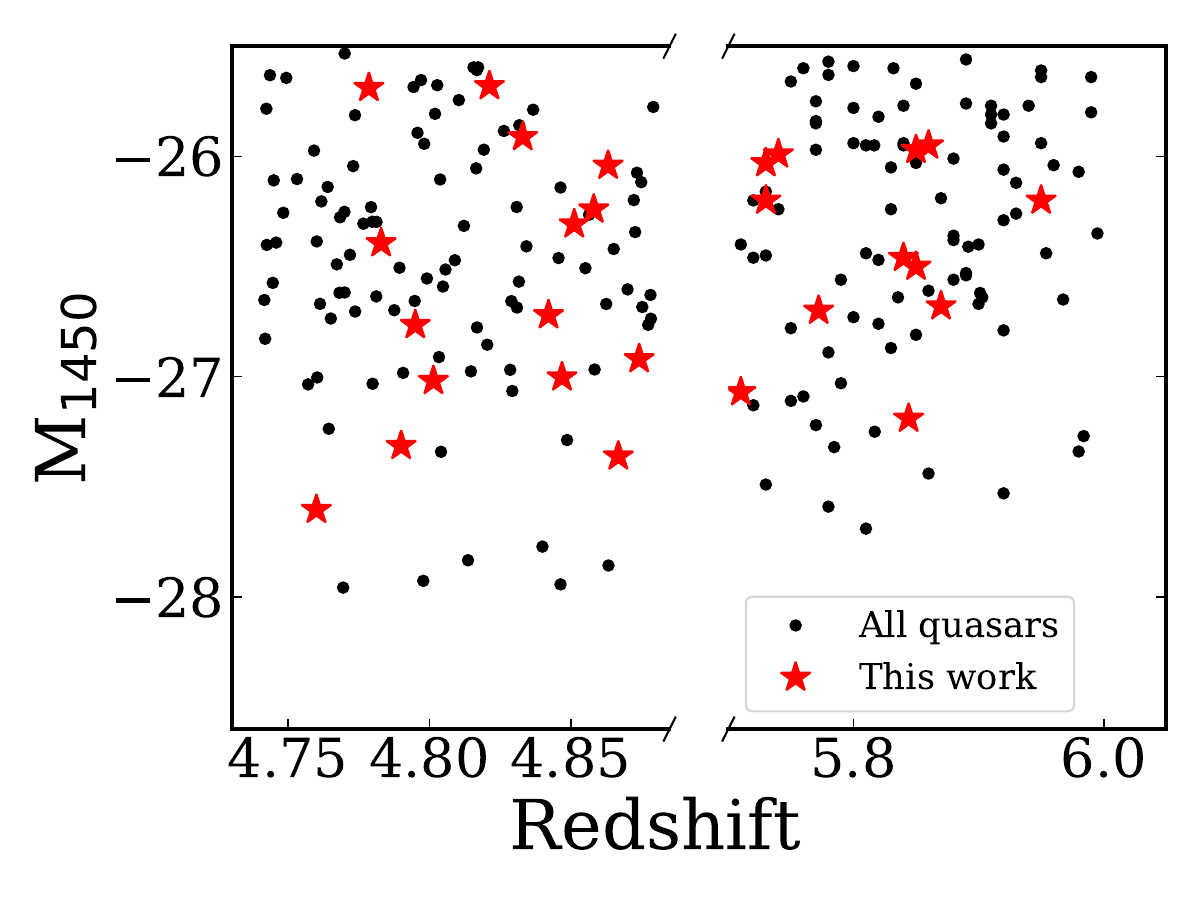}
    \caption{M$_{1450}$ (absolute magnitude at 1450 \AA) versus redshift for our objects (red) and all other spectrally-confirmed luminous quasars within the same redshift ranges from the literature (black) in Table \ref{tab:redshift}.} 
    \label{fig:z_1450}
\end{figure}

Our objects are observed with JWST NIRSpec/IFU mode \citep{Bok2022, Jak2022} through the JWST cycle 2 survey program \#3428. The parent sample consists of 100 spectroscopically-confirmed luminous quasars at two narrow redshift ranges $z \sim$ 4.74 -- 4.88 and $z \sim$ 5.66 -- 6 with absolute magnitude at rest-frame 1450 \AA\ M$_{1450}$ $\lesssim$ $-$25.5, which are randomly chosen from the sources published in \citet{sdssdr12q,Wang2016,Fan2023} to maximize the sky coverage. The redshift windows are chosen to place key spectral features (\hb, \oiiitext, Fe, etc) in the sweet spots of the NIRSpec gratings. In the end, we obtained data for 27 objects, which are randomly chosen from the parent sample based on JWST scheduling. Their locations on the redshift vs M$_{1450}$ plane are shown in Fig. \ref{fig:z_1450}.
The sources at $z\sim$ 4.74--4.88 and $z\sim$ 5.66--6 are observed in configurations G235H/F170LP and G395H/F290LP, with corresponding wavelength coverage of 1.66--3.05 and 2.87–5.14~\mum, respectively. The gratings have a nominal resolving power $\lambda / \Delta\lambda \simeq 2700$, corresponding to a velocity resolution $\sim 110$ \kms.
We use a 2-point small cycling dither pattern and a NRSIRS2RAPID readout pattern. The total on-source exposure time is 466.8s per target, which gives a 3-$\sigma$ limiting surface brightness of $\sim$1$\times$10$^{-17}$ erg s$^{-1}$ cm$^{-2}$  \AA$^{-1}$ arcsec$^{-2}$ at rest-frame 5008 \AA. The field-of-view (FoV) of the IFU observations is $\sim 3\arcsec \times 3 \arcsec$. The data are available on the Mikulski Archive for Space Telescopes (MAST) at the Space Telescope Science Institute, which can be accessed via \dataset[10.17909/s4b6-ng62]{http://dx.doi.org/10.17909/s4b6-ng62}.

The IFU data were reduced following the general 3 stages of JWST\ Science Calibration Pipeline (version ``1.14.0'' and the context file ``jwst\_1293.pmap''), combined with customized software and scripts to replace or improve certain steps in the public pipeline.
Specifically, after stage 1, we subtract the correlated detector noise (1/$f$ noise) in the count rate images using NSCLean \citep{NSclean}. After stage 2, we apply further sigma-clipping in each calibrated 2D spectral image to remove outliers \citep{Vayner2023b,VeilleuxLiu2023,Liu2024b,Liu2026a}. The final IFU data cube is reconstructed with the ``drizzle'' method adopting a spatial pixel (spaxel) size of 0\arcsec.1, and a master background is built from the spaxels covering the blank sky and then subtracted.
The final quasar spectrum is extracted from each IFU data cube with r$=$0\arcsec.1--0\arcsec.5 apertures to maximize the S/N of their spectra and remove artificial spectral oscillations. Aperture correction is applied to each spectrum based on the curve of growth analysis results from the NIRSpec/IFU data of the standard star TYC 4433-1800-1 (proposal ID: 1128) over the same wavelength range.

\section{Spectral Analysis}
\label{sec:fit}

\subsection{Spectral Fitting}

\subsubsection{\hb, \hg, and {\rm\oiiitext} region}
\label{sec:312}

For each object, we adopt the public software, PyQSOFit \citep{pyqsofit,Shen2019} with additional customized scripts to model the spectrum. Specifically, the quasar pseudo-continuum is fit with a power-law and empirical \feii\ templates from \citet[][]{BorosonGreen1992,Vestergaard2001}. {In the fitting, the amplitudes of the power-law and \feii\ templates were allowed to be any positive values. The spectral index of the power-law, and the velocity and line width (FWHM) of the \feii\ template were allowed to vary freely in the ranges of [$-$5, 3], [$-$3000\kms, 3000\kms], and [1200\kms, 18000\kms], respectively.
Modeling the complex profiles of \feii\ emission can cause non-trivial uncertainties for the best-fit \oiiitext\ line profiles \citep[e.g.,][]{Kovacevic2010,Park2022}.
As an example, \citet{Schmidt2018} have fit both the observed and \feii\ subtracted spectra of {a sample of narrow-line Seyfert 1 galaxies
(NLSy1s)} with strong \feii\ emission, and reported that the asymmetric component of \oiiitext\ emission line was affected the most, with uncertainties of 5--10\% and 10--20\% for the FWHM and flux of the asymmetric component of \oiiitext\ line, respectively.
To evaluate such uncertainties in our fits, we fit our spectra with {two other} popular iron templates from \citet{VC04} and \citet{Kovacevic2010}, in addition to the default template mentioned above. The rest-frame \oiiitext\ equivalent width (\ewo) and \oiiitext\ kinematics measured with the two \feii\ templates are $\sim$0.7--2.8$\times$ (median 1.0$\times$) and $\sim$0.7--2.0$\times$ (median 1.0$\times$) of those derived with the default template, respectively. Overall, we conclude that our \oiiitext\ measurements are in general not substantially affected by the choice of \feii\ templates.}

For emission lines, \hb\ and \hg\ are fit with up to 3 broad Gaussian components and 1 narrow Gaussian component. The kinematics (i.e., the velocity and velocity dispersion) of each corresponding broad Gaussian component in the two lines are tied together. The \oiiitext\ doublet is fit with up to 2 Gaussian components where the kinematics of the narrow Gaussian component is tied to those of the narrow Gaussian components of \hb\ and \hg. The only exception is for J0859$+$2520, where the narrow \oiii\ component is highly blueshifted with respect to \hb\ and the kinematics of the two are untied. 
The number of Gaussian components adopted in the final best-fit model is determined based on the Bayesian information criterion (BIC), where the model with the lowest BIC statistics is adopted.
All spectra with their best-fit models can be found in Fig. \ref{fig:specfit} of the Appendix \ref{sec:appendix}, {except for those already presented in \citet{Liu2026a}}. The derived measurements are listed in Table \ref{tab:quasar}.
{For the two major comparison quasar samples adopted in this study, the cosmic noon quasars from \cite{Shen2016} and the SDSS z $<$ 1 quasars, their spectra are fit with the same software PyQSOFit and \feii\ templates as our sample. Therefore, the choice of \feii\ templates does not affect the results based on the comparisons with these samples.}

\subsubsection{\ha\ region}

For the 12 targets at $z \sim$ 6, \ha\ emission lines are also covered by our spectra. We fit the quasar pseudo-continuum with the same approach as described in Sec. \ref{sec:312}, and the \ha\ with up to 3 broad Gaussian components and
1 narrow Gaussian component. For J1141$+$7114 and J1327$+$5732, narrow \nii\ and \sii\ emission lines are also detected and fit with 1 narrow Gaussian component with its kinematics tied to that of the narrow \ha. The ratio of the \niitext\ doublet is set to the theoretical value,  1:3. For J1327$+$5732, such fit leads to an unphysical \siitext\ doublet ratio;  we set the line widths of the narrow \ha, \niitext, and \siitext\ the same as that of the narrow \hb. The best-fits for all of these 12 objects are shown in Fig. \ref{fig:haexample}.

\subsection{Systemic Redshift}
\label{sec:32}

\begin{deluxetable}{ccccc c}
\tablecaption{Target Information\label{tab:redshift}}
\tablecolumns{15}
\tabletypesize{\scriptsize}
\tablewidth{\textwidth}
\tablehead{
\colhead{Object} & \colhead{RA} & \colhead{DEC} & \colhead{$z_{\rm ref}$} & \colhead{ref.} & \colhead{$z_{\rm sys}$} \\
(1) & (2) & (3) & (4) & (5) & (6) }
%\colnumber
\startdata
J0732+3256 & 07:32:31.28 & +32:56:18.33 & 4.760 & a & 4.771 \\
J0756+0218 & 07:56:22.37 & +02:18:20.17 & 5.730 & b & 5.762 \\
J0759+1800 & 07:59:07.57 & +18:00:54.7 & 4.778 & c & 4.796 \\
J0807+1328 & 08:07:15.11 & +13:28:05.1 & 4.870 & c & 4.880 \\
J0829+0303 & 08:29:07.62 & +03:03:56.56 & 5.850 & b & 5.855 \\
J0831+4046 & 08:31:22.57 & +40:46:23.3 & 4.820 & c & 4.900 \\
J0840+5624 & 08:40:35.09 & +56:24:19.90 & 5.844 & b & 5.837 \\
J0850+3246 & 08:50:48.25 & +32:46:47.90 & 5.870 & b & 5.830 \\
J0859+2520 & 08:59:31.29 & +25:20:19.5 & 4.780 & c & 4.779 \\
J0927+2001 & 09:27:21.82 & +20:01:23.70 & 5.772 & b & 5.768 \\
J0941+5947 & 09:41:08.35 & +59:47:25.7 & 4.861 & c & 4.860 \\
J0953+6910 & 09:53:55.90 & +69:10:52.62 & 5.840 & b & 5.918 \\
J1050+4627 & 10:50:05.11 & +46:27:35.5 & 4.844 & c & 4.837 \\
J1100+5800 & 11:00:41.94 & +58:00:01.3 & 4.776 & c & 4.759 \\
J1102+6635 & 11:02:47.29 & +66:35:19.6 & 4.803 & c & 4.787 \\
J1116+5853 & 11:16:33.75 & +58:53:22.04 & 5.730 & b & 5.721 \\
J1134+3928 & 11:34:15.21 & +39:28:26.0 & 4.795 & c & 4.826 \\
J1141+7119 & 11:41:43.06 & +71:19:25.07 & 5.860 & b & 5.851 \\
J1245+4348 & 12:45:10.13 & +43:48:37.9 & 4.820 & c & 4.890 \\
J1257+6349 & 12:57:57.47 & +63:49:37.20 & 5.950 & b & 6.012 \\
J1327+5732 & 13:27:41.33 & +57:32:38.43 & 5.740 & b & 5.751 \\
J1328+4445 & 13:28:25.16 & +44:45:00.2 & 4.810 & c & 4.820 \\
J1342+5838 & 13:42:43.46 & +58:38:50.0 & 4.855 & c & 4.852 \\
J1436+5007 & 14:36:11.74 & +50:07:06.90 & 5.850 & b & 5.840 \\
J1458+3327 & 14:58:05.99 & +33:27:23.0 & 4.830 & c & 4.851 \\
J1620+5202 & 16:20:45.64 & +52:02:46.65 & 4.790 & a & 4.791 \\
J1621+5155 & 16:21:00.92 & +51:55:48.79 & 5.710 & b & 5.614 \\
\enddata
\tablecomments{(4): Redshifts of our objects in the literature. (5): references for column (4): a. \citet{Wang2016} b. \citet{Fan2023}, c. \citet{sdssdr12q}. (6): redshifts determined from our data. See \ref{sec:32} for details. }
\end{deluxetable}

\begin{figure}
    \centering
    \plotone{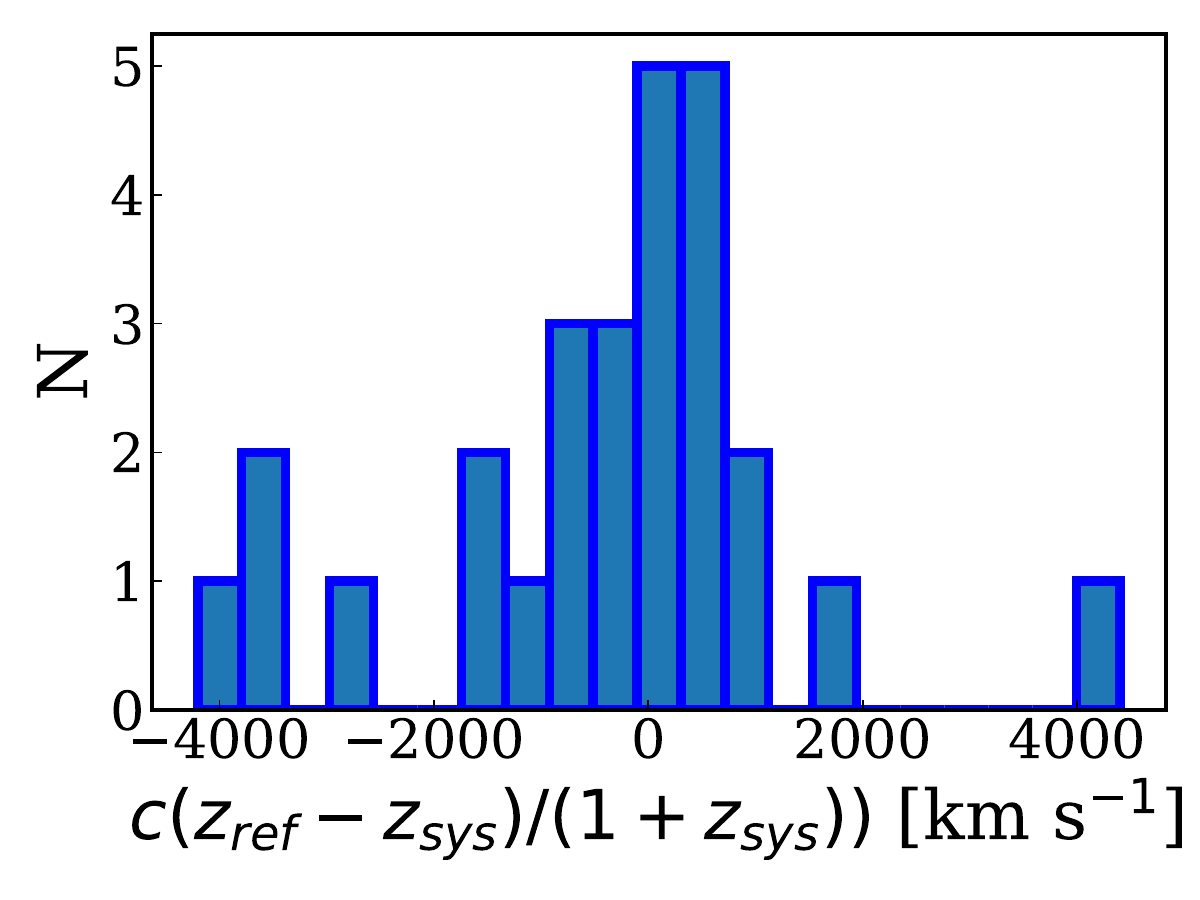}
    \caption{Difference of the systemic redshifts between those determined from our data ($z_{sys}$) and those from the literature which are based on rest-frame UV emission lines including \civtext, \ciiitext\ and/or \siivtext\  ($z_{ref}$).}
    \label{fig:redshift}
\end{figure}

The determination of the systemic velocities of our objects is described in detail in \citet{Liu2026a}, which is briefly summarized here: whenever \oiiitext\ is detected, the systemic redshift is determined from the peak of the relatively narrower \oiiitext\ component, if it exists and is not significantly blueshifted with respect to the \hb\ emission line. Otherwise, the systemic redshift is determined from the peak of the overall \hb\ emission line profile. The final results are listed in Table \ref{tab:redshift}.
In Fig. \ref{fig:redshift}, we compare these systemic redshifts with those reported in the literature \citep{sdssdr12q,Wang2016,Fan2023}, which are mainly based on rest-frame UV emission lines including \civtext, \ciiitext\ and/or \siivtext. We refer the reader to the corresponding references for more detailed descriptions of their redshift determination. Our mean (median) systemic velocities are $\sim$400 ($\sim$40) \kms\ larger than those UV-based ones, and the largest offsets between the two reach $\sim$$\pm$4000 \kms. Such redshift differences are seen in previous studies of lower redshift quasars \citep[e.g.,][and references therein]{Hewett2010,Matthews2023}, where the average blueshift of rest-frame UV emission lines is usually attributed to the presence of strong nuclear quasar winds. A more detailed analysis of the differences in various redshift determination approaches is deferred to future work.

\subsection{Nuclear Properties}

{ 
The bolometric luminosities are derived adopting 5100\AA\ continuum luminosity (\Lwu) with 
 a bolometric correction factor of 9.26 \citep{Richards2006}. This factor is widely adopted for high-$z$ quasar studies \citep[e.g.,][]{Marshall2023,Loiacono2024,Yang2023b,Liu2024b,Liu2026a} and thus allows for straightforward comparisons. The same correction factor is applied to all other quasar samples discussed in this work. Nevertheless, many studies suggest smaller correction factors \citep[e.g.,][]{Marconi2004,Runnoe2012,Duras2020}, claiming that a larger factor results from double counting the UV photons that are re-emitted in the infrared \citep[e.g.,][]{Marconi2004,NemmenBrotherton2010,Runnoe2012}. A full range of 5.53--12.45 for bolometric correction factors are reported previously as compiled by \citet{Runnoe2012}. Therefore, the bolometric luminosities and thus Eddington ratios of our objects may be up to $\sim$60\% lower or $\sim$34\% higher. Furthermore, it is widely suggested that the correction factor decreases with increasing quasar optical luminosity \citep[e.g.,][]{Runnoe2012,Krawczyk2013,Netzer2019}, which again translates to lower bolometric luminosities and thus Eddington ratios for our sample.} 
 
The \hb-based BH masses are derived from the scaling relation
in \citet{Vestergaard2006}:

{\begin{equation}
\begin{split}
{\rm log}(M_{\rm BH}/M_\odot) = {\rm log}\left\{\left[\frac{{\rm FWHM}({\rm H\beta})}{1000\ {\rm km\ s}^{-1}}\right]^2\right. \\ 
\left.\left[\frac{\lambda L_\lambda(5100 )}{10^{44}\ {\rm erg\ s}^{-1}}\right]^{0.5}
\right\}
+ (6.91 \pm{0.02})
\end{split}
\label{eq:BH}
\end{equation}}

Here FWHM(\hb) are calculated for the entire line profile of the broad \hb\ emission lines. The Eddington ratios (\edd) are then derived following $L_{\rm Edd} = 1.26\times10^{38} (M_{\rm BH}/M_\odot)\ erg s^{-1} $. Overall, our objects have black hole masses of log($M_{\rm BH}/M_\odot)\sim$8.6--9.7 and \edd\ $\sim$\ 0.1--2.6.
These \hb-based measurements are the fiducial values for the BH mass related quantities presented in this paper and are listed in Table \ref{tab:quasar}.

Similarly, whenever possible, we also derive the \ha-based BH mass following \citep{GreeneHo2005}:

{\begin{equation}
\begin{split}
{\rm log}(M_{\rm BH, H\alpha}/M_\odot) = {\rm log}\bigg\{(2.0^{+0.4}_{-0.3}\times10^6) \\
\left[\frac{{\rm FWHM}({\rm H\alpha})}{1000\ {\rm km\ s}^{-1}}\right]^{2.06\pm{0.06}}
\left[\frac{L({\rm H\alpha})}{10^{42}\ {\rm erg\ s}^{-1}}\right]^{0.55\pm{0.02}}\bigg\}
\end{split}
\label{eq:BHHa}
%\end{eqnarray}
\end{equation}}
Here, FWHM(\ha) are calculated for the entire line profile of the broad \ha\ emission line. $L({\rm H\alpha})$ is the luminosity of the broad \ha\ emission line. These results are also summarized in Table \ref{tab:quasar}. {For all other quasar samples presented in this work, their corresponding properties are calculated in the same way as {described} above}. %unless stated otherwise.}

\begin{figure*}[!ht]
\begin{minipage}[t]{\textwidth}
\centering
\includegraphics[width=\textwidth]{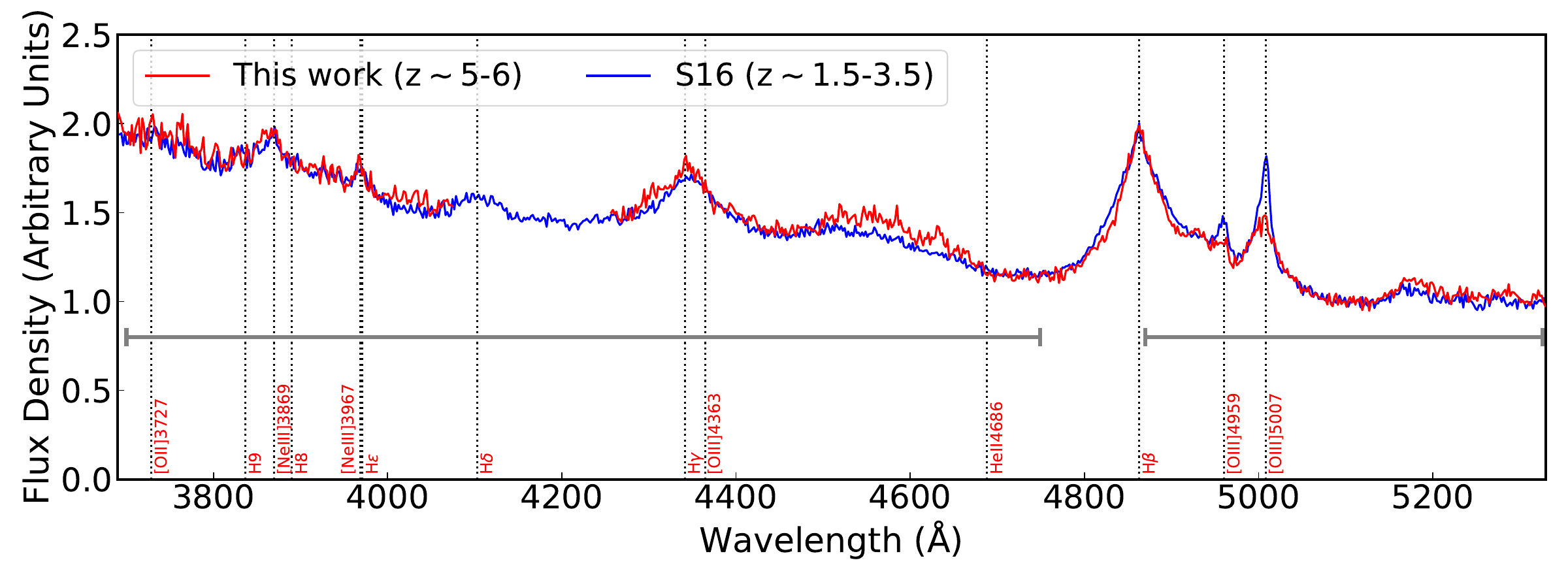}
\end{minipage}

\begin{minipage}[t]{\textwidth}
\centering
\includegraphics[width=\textwidth]{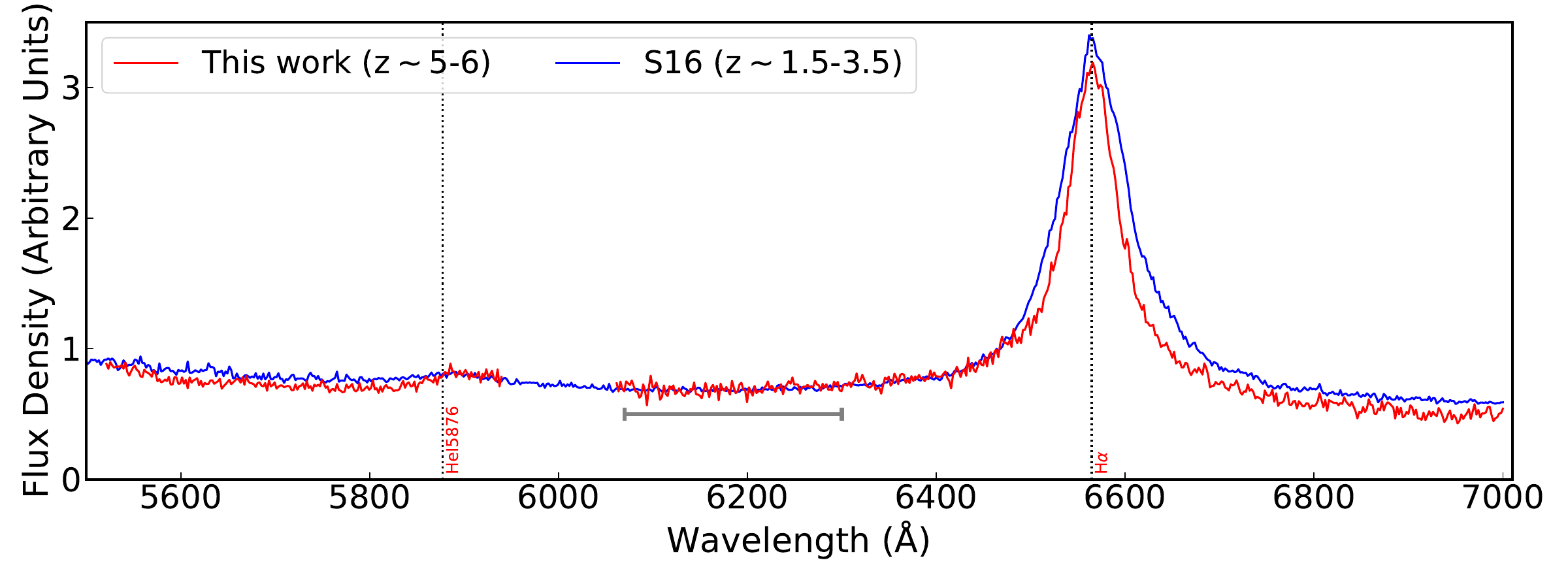}
\end{minipage}
    \caption{Composite spectra of our $z \sim$ 5--6 sample (red) and the $z \sim$ 1.5--3.5 {\textit{S16 sample}} matched in the same luminosity range. The two spectra are further normalized at 5100 \AA. The locations of the major quasar emission lines are indicated by the dotted lines. The horizontal gray bars indicate the locations of the major blended Fe emission.}
    \label{fig:composite}
\end{figure*}

\subsection{Composite Spectrum}
\label{34}
To compare the spectral properties of our high-redshift quasars with lower-redshift ones, {we construct a comparison sample at cosmic noon from \citet{Shen2016} where all quasars falling within the same bolometric luminosity range as our sample (\logLbol$\sim$46.8--47.7) are chosen.} In short, this comparison sample, referred to as {\textit{S16 sample}} hereafter, represents typical Type 1 quasars at $z \sim$ 1.5--3.5 with rest-frame optical spectra of similar S/N to those of our sample. {Matching the quasar luminosity range is essential,} as the luminosity not only correlates with accretion power but also influences various emission line properties, including the rest-frame equivalent widths \citep[EW; e.g., the ``Baldwin effect'';][]{Baldwin1977} and line shapes \citep[e.g., ][]{Richards2002, Shen2008b, Richards2011}. 
The median composite spectra of our objects and the comparison sample are constructed following a similar approach adopted by \citet{Shen2016} and \citet{VandenBerk2001}. Specifically, each spectrum is first shifted to the rest-frame and normalized by the mean flux density between 4400--5100 \AA. The median spectrum of each sample is then built from the normalized spectra. This approach better preserves the relative strength of the emission lines.

As shown in Fig. \ref{fig:composite}, the composite spectrum of our objects closely resembles that of {\textit{S16 sample}}. Our high-$z$ sample shows spectral characteristics fully consistent with luminous, Type 1 quasars, including blue power-law continuum with little dust obscuration, blended broad Fe emission and strong line emission (Balmer lines, \oiiiab\ doublet, etc). 
A faint \oii\ emission line is tentatively detected (S/N $\sim$ 2) in the composite spectrum, while \niitext\ and \siitext\ lines remain undetected. %the broad \ha\ lines.  
Nevertheless, these faint emission lines are clearly detected in a few objects. For example, \oii\ is detected in 1 object, J0807$+$1328. \niitext\ and \siitext\ are detected in 2 objects, J1141$+$7114 and J1327$+$5732.

The most prominent difference between the composite spectra of the two samples lies in the \oiiitext\ emission lines. They are weaker (with respect to \hb) but more blueshifted and broader in our sample than those in the {\textit{S16 sample}}. As both two samples are matched to the same narrow bolometric luminosity range, the difference in \oiiitext\ is not primarily caused by the well-known ``Baldwin effect'' observed in low-$z$ quasars, a well-known trend of decreasing \ewo\ with increasing luminosity \citep[e.g.,][]{Baldwin1977,Stern2013,Zhang2013}. Instead, the larger blueshift and line width of \oiiitext\ in our sample mainly reflect the presence of more frequent fast outflows in our sample as investigated in detail in \citet{Liu2026a}. We will expand more on this in Sec. \ref{sec:43}.
In addition, more subtle differences between the two composite spectra also exist: For our sample, the \hb\ and \ha\ emission lines are narrower and the Fe emission is stronger.

\begin{figure}[!ht]
\centering
\includegraphics[width=0.5\textwidth]{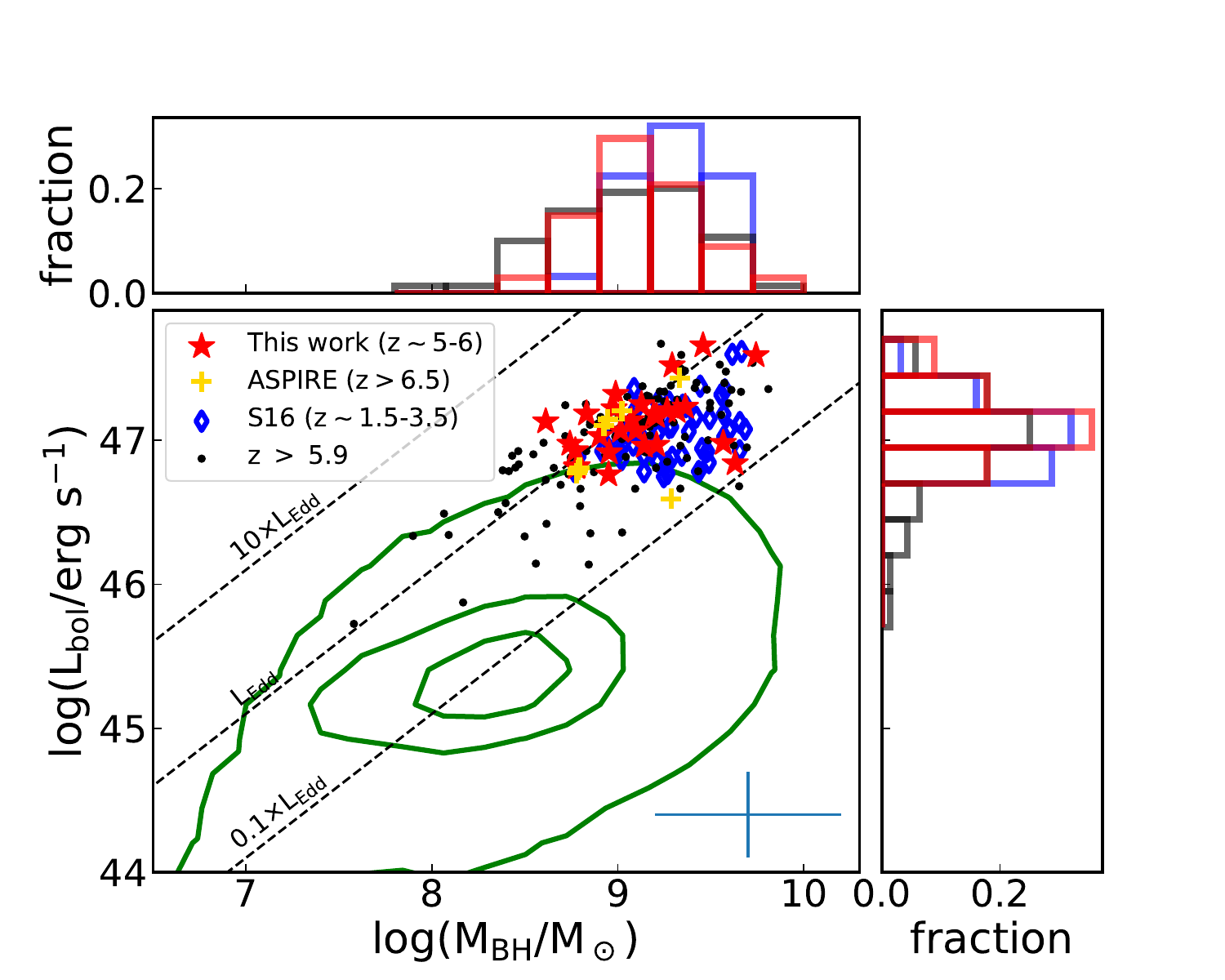}
\caption{Bolometric luminosity versus BH mass for our objects (red), {\textit{S16 sample}} (blue), ASPIRE quasars at $z \sim$ 6.5--6.8 (orange) and all z $>$ 5.9 quasars from \citet[][black]{Fan2023}. The green contours represent the distributions of 99.998\%, 90\% and 50\% of all z $<$ 1 SDSS quasars from \citet{wu_spectroscopic_2023}. The dashed lines show the locations of constant accretion rates at 0.1, 1 and 10 times the Eddington luminosity. All BH masses are based on \hb, except for those of quasars from \citet{Fan2023}  which are based on \mgiitext. The {systematic uncertainties} for the bolometric luminosity and \hb\ BH mass are indicated by the blue cross. The top and right panels show the fraction histograms for our objects, {\textit{S16 sample}} and all $z>5.9$ quasars.}
\label{fig:mbhlbol}
\end{figure}

\begin{figure*}[!ht]
\begin{minipage}[t]{0.33\textwidth}
\centering
\includegraphics[width=\textwidth]{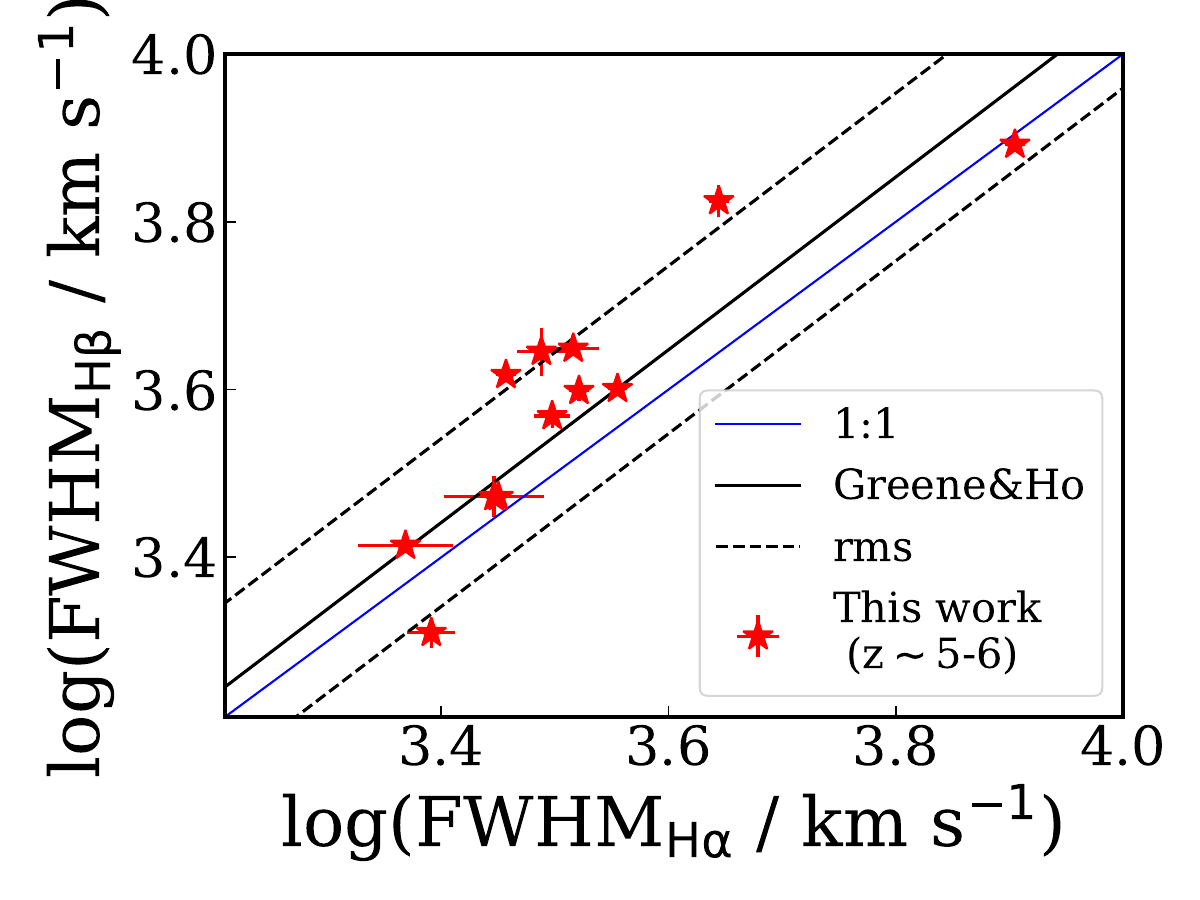}
\end{minipage}
\begin{minipage}[t]{0.33\textwidth}
\centering
\includegraphics[width=\textwidth]{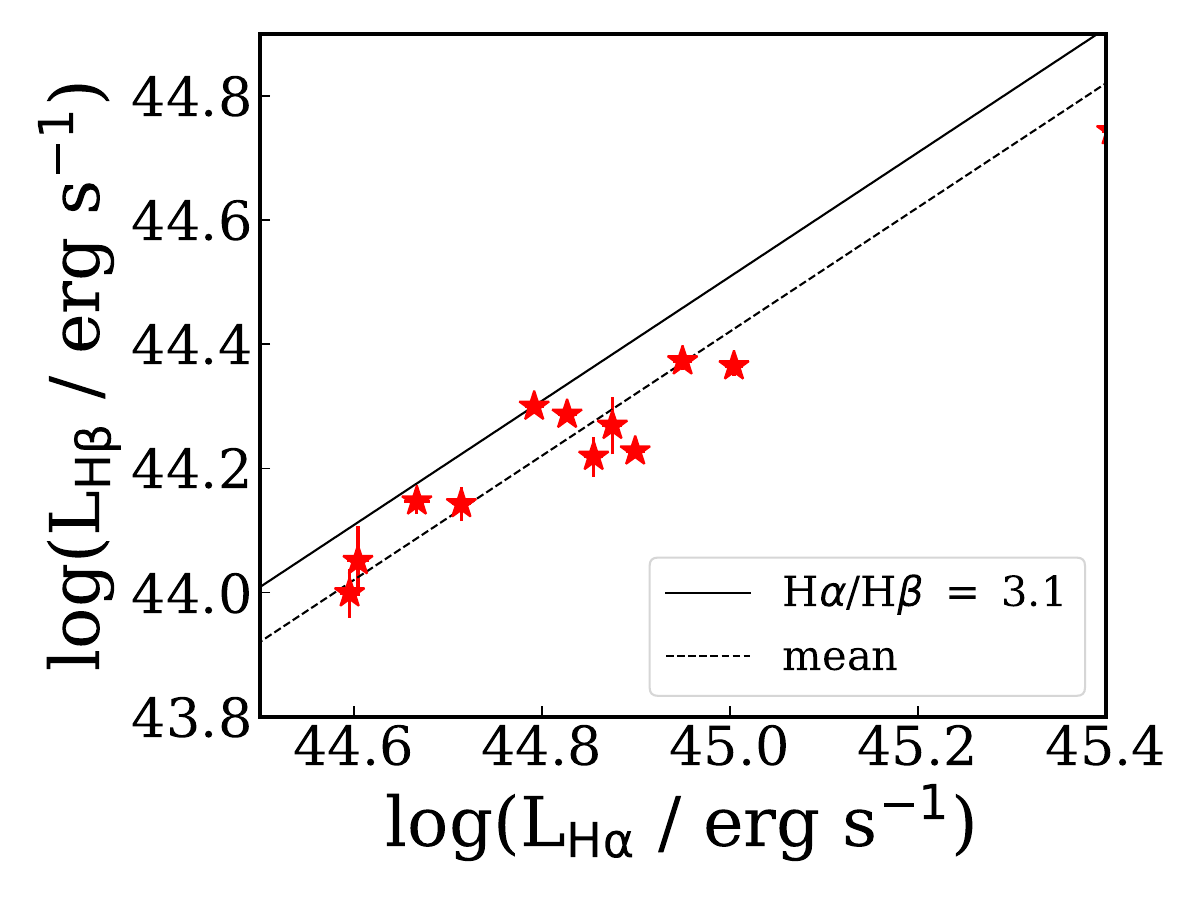}
\end{minipage}
\begin{minipage}[t]{0.33\textwidth}
\centering
\includegraphics[width=\textwidth]{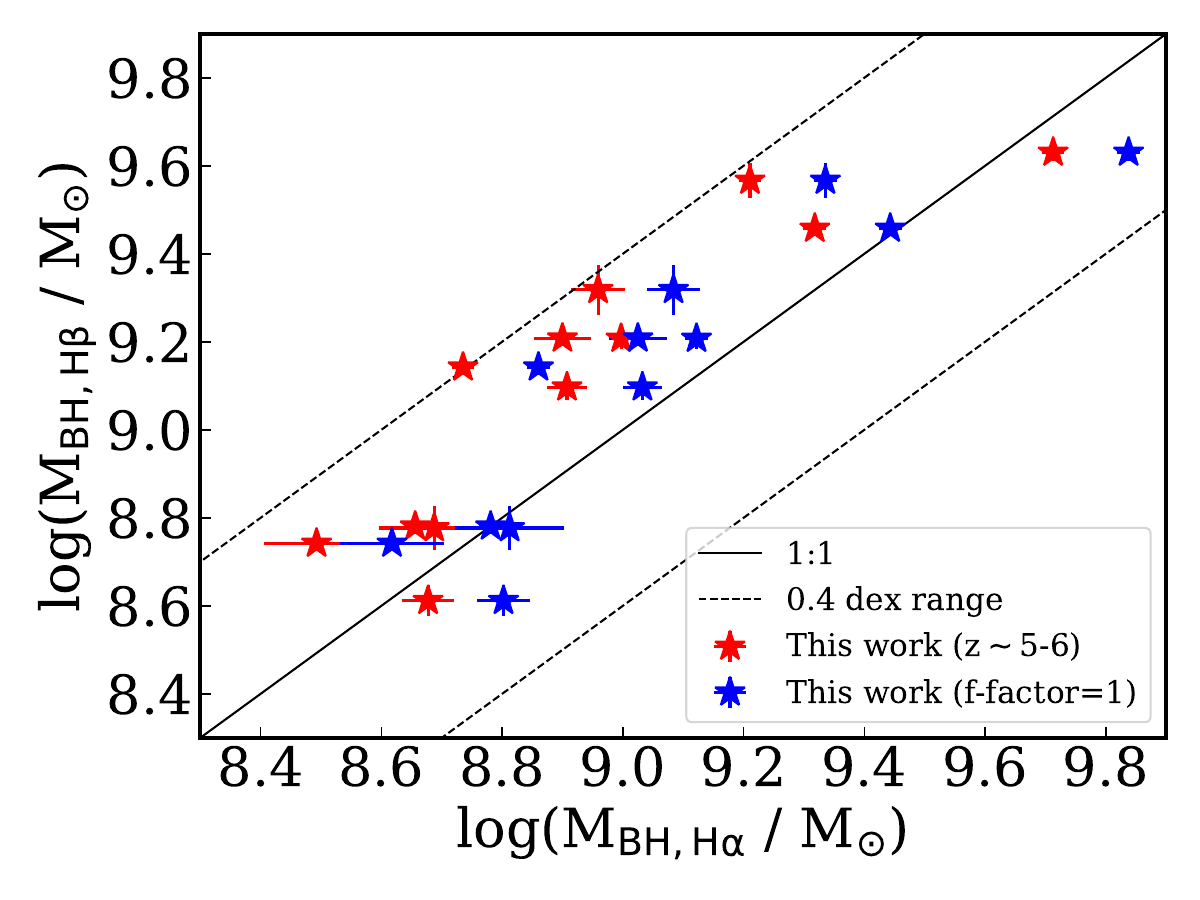}
\end{minipage}
\caption{\textbf{Left:} \ha\ FWHM versus \hb\ FWHM. The blue line indicates the 1-to-1 equality. The black solid and dotted lines indicate the best-fit relation of z $<$ 0.35 SDSS quasars and the associated 0.1 dex rms from \citet{GreeneHo2005}. \textbf{Middle:} \ha\ luminosity versus \hb\ luminosity. The solid line indicates the theoretical expectation of 3.1 and the dotted line indicates the mean value of our sample (3.8). \textbf{Right:} {BH masses based on \ha\ and \hb. The values adopting equations \ref{eq:BH} and \ref{eq:BHHa} are shown in red, and those adopting an ``f-factor'' of 1 for the \ha-based BH masses are shown in blue. For the latter, the \ha-based values increase by 0.125 dex (see Sec. \ref{42} for more details).} The error bars only reflect the measurement errors. The solid and dotted lines indicate the 1-to-1 equality and the $\pm{0.4}$ dex ranges, respectively.  Overall, the relative \ha\ and \hb\ properties of our sample are consistent with those observed in low-$z$ quasars.}
\label{fig:hahb}
\end{figure*}

\section{Quasar Spectral Properties}
\label{sec:quasar}
\subsection{\hb\ Black Hole Mass}

Our data provide reliable, \hb-based BH masses for our objects for the first time. Overall, our sample exhibits broad \hb\ lines with FWHM $\sim$ 2040 -- 7180 \kms\ and BH masses with log($M\rm _{BH}$/\msun) $\sim$ 8.6 -- 9.7. 
The locations of our objects in the bolometric luminosity -- BH mass plane are shown in Fig. \ref{fig:mbhlbol}. As expected, our sample occupies the region of high bolometric luminosity ({\logLbol} $\sim$ 46.8--47.7) and large BH mass with respect to the overall SDSS z $<$ 1 quasar population gathered from \citet{wu_spectroscopic_2023}, with high Eddington ratios ($\sim$ 0.1--2.6). They share almost the same BH mass and Eddington ratio ranges as $z>5.9$ quasars with reliable \mgiitext-based BH mass from \citet{Fan2023} when matched in bolometric luminosity and other high-$z$ quasars with \hb-based BH mass, like $z>6.5$ ASPIRE quasars \citep{Yang2023b}. Our results confirm that these luminous quasars are 10$^9$ \msun\ SMBHs with high accretion rates within the first $\sim$1 billion years after the Big Bang, which is in line with previous studies at slightly lower \citep[e.g., $z \sim$ 4.8; \mgiitext-based BH mass]{Trakhtenbrot2011} and higher redshifts \citep[e.g., $z>6.5$; \hb-based BH mass][]{Yang2023b} drawing the picture that the early emergence of these luminous quasars requires continuous/frequent active accretion \citep[see][and references therein]{Fan2023}.    

When compared to the {\textit{S16 sample}} at cosmic noon within the same bolometric luminosity range, our objects seem to show, on average, slightly smaller BH masses and higher Eddington ratios: Our sample has a median log($M\rm _{BH}$/\msun) of 9.1 and a median \edd\ of 0.9.
The {\textit{S16 sample}} has a median log($M\rm _{BH}$/\msun) of 9.3 and a median \edd\ of 0.4. 
However, the systematic uncertainty of \hb-based BH mass is on the order of 0.5 dex \citep[e.g.][]{McLureJarvis2002,Vestergaard2006,Shen2013review}.
{We therefore use a Monte Carlo approach to estimate the 90\% confidence levels of the median BH mass and \edd.
For each measurement, we assumed the uncertainty follows a Normal distribution with a standard deviation of 0.5 dex. We then generated 10$^4$ mock realizations of the sample by randomly perturbing each measurement according to its uncertainty distribution. The median value was computed for each realization, and the 16th and 84th percentiles of these median values were then used to derive the 90\% confidence level.} 
For BH mass log($M\rm _{BH}$/\msun), we obtain 8.9--9.3 for our sample and 9.2--9.5 for the {\textit{S16 sample}}. For \edd, we obtain 0.5--1.3 for our sample and 0.3--0.5 for the {\textit{S16 sample}}. Therefore, the differences between the median BH mass and \edd\ of the two samples are insignificant. 
Overall, our results cannot conclusively prove that the BH mass and \edd\ of the two samples are systematically offset from each other.

The rest-frame UV emission line \mgii\ is another widely used single-epoch BH mass estimator of high-$z$ quasars and is accessible from the ground.
Only two objects (J0759$+$1800, J0807$+$1328) from our sample have reliable \mgii-based BH mass (log(M$_{\rm BH}$/M$_\odot$ $=$ 9.0 and 9.2) from the literature \citep{Trakhtenbrot2011}. They are consistent with the \hb-based BH masses (log(M$_{\rm BH}$/M$_\odot$ $=$ 9.1 and 9.3) derived from our data given the {systematic uncertainty} (0.5 dex) of these BH mass measurements \citep[e.g.][]{McLureJarvis2002,Vestergaard2006,Shen2013review}. This agreement is consistent with the findings from previous studies \citep[e.g.,][]{Yang2023b,Marshall2023,Liu2024b,Loiacono2024} at similar redshifts where the difference between the two BH mass measurements is up to $\sim$0.5 dex.

\subsection{\ha\ vs \hb}
\label{42}

The relative profiles of \ha\ and \hb\ reflect the physical properties of BLR, including their kinematic and ionization structures. In this section, we compare the \ha\ and \hb\ properties for the 12 objects where both lines are covered. 

In the left panel of Fig. \ref{fig:hahb}, we compare the FWHMs of the broad \ha\ and \hb\ lines. All 12 objects fall within the rms scatter ($\sim$0.1 dex) of the best-fit relation for a sample of 162 $z\leq0.35$ SDSS quasars with order-of-magnitude lower luminosities \citep[43$\lesssim$log(\Lwu)$\lesssim$46;][]{GreeneHo2005}.
The average FWHM of broad \hb\ ($\sim$4140 \kms) of our sample is slightly larger than that of broad \ha\ ($\sim$3530 \kms). As \hb\ is emitted preferentially in regions with higher density and/or higher ionization parameter than \ha\ \citep{Osterbrock2006} and that line width generally decreases with radius, this suggests that the density or the ionization parameter of the BLR increases with decreasing radii, as is observed in low-$z$ quasars.
The mean value of FWHM$_{\rm \hb}$/FWHM$_{\rm \ha}$ ratio is $\sim$1.17, which is identical to that of the $z\leq0.35$ SDSS quasars \citep[][]{GreeneHo2005} and that found in nearby quasars by an even earlier study \citep{OsterbrockShuder1982}. It indicates that the FWHM$_{\rm \hb}$/FWHM$_{\rm \ha}$ ratio does not vary in quasars across a large dynamical range of luminosities and redshifts.

The relation between the \ha\ and \hb\ luminosities of our sample is shown in the middle panel of Fig. \ref{fig:hahb}. The mean \ha/\hb\ ratio is $\sim$3.8 (for the combined broad and narrow components; The ratio for the broad component alone is $\sim$3.9). This ratio is close to that ($\sim$3.5) of the aforementioned $z\leq0.35$ SDSS quasars and that of the value seen in the SDSS quasar composite spectrum from \citet{VandenBerk2001}, and only mildly larger than the theoretical limit of 3.1 \citep{Osterbrock2006}. This again suggests strong similarities of the BLR in these quasars, despite their significant differences in luminosities and cosmic epochs. This also suggests that there is little dust extinction in the nuclear region along the line-of-sight of our objects as expected.

The comparison of BH masses derived from \ha\ and \hb\ is shown in the right panel of Fig. \ref{fig:hahb}.
Overall, the two BH masses broadly agree with each other. On average, the \ha-based BH mass is slightly lower ($\sim$0.2 dex) than the \hb-based ones, and the maximum difference is $\sim$0.4 dex, which is consistent with the findings from previous studies on individual $z>6$ quasars \citep{Marshall2023,Loiacono2024}. Such a difference is smaller than the {systematic uncertainty} of \hb-based BH mass ($\sim$0.5 dex). 
Nevertheless, a significant portion of the difference between the two BH masses is caused by the virial factor or ``f-factor'' adopted in each recipe (i.e., Eqs. \ref{eq:BH} and \ref{eq:BHHa}). As emphasized in \citet{Restrepo2022}, for the \ha-based mass, \citet{GreeneHo2005} assumes an ``f-factor'' of 0.75, while for the \hb-based mass, \citet{VestergaardPeterson2006} implicitly assumes an "f-factor" of 1. If we adopt an ``f-factor'' of 1 for both equations, then the \ha-based masses will increase by 0.125 dex, making the average difference between the two BH masses decrease to $\sim$0.075 dex. {The true value of ``f-factor'' for these quasars are unknown. Therefore, we show \ha-based BH masses derived with both ``f-factors'' in Fig. \ref{fig:hahb} simultaneously.}

\begin{deluxetable*}{ccccc ccccc c}[!h]
\tablecaption{Sample Properties}
\tablecolumns{11}
\tablewidth{\textwidth}
\tablehead{
\colhead{Object} & \colhead{log($L_{\mathrm{bol}}$)} &                                             
   \colhead{\ewo}  &
  \colhead{\RFe} & \colhead{log($L_{\mathrm{H\beta}}$)} &
  \colhead{\fwhmhb}  &
  \colhead{log(M$_{\mathrm{BH}}$)} & \edd & \colhead{log($L_{\mathrm{H\alpha}}$)} &
   \colhead{\fwhmha}  &   \colhead{log(M$_{\mathrm{BH, H\alpha}}$)} \\
 &  [erg s$^{-1}]$ & [\AA] & & [erg s$^{-1}$] & [\kms] & [\msun] &  & [erg s$^{-1}$] & [\kms] & [\msun] \\
(1) & (2) & (3) & (4) & (5) & (6) & (7) & (8) & (9) & (10) & (11) }
%\colnumber
\startdata
J0732+3256 & 47.5 & $<$1.1 & 2.37$\pm{0.03}$ & 44.53$\pm{0.01}$ & 3570$\pm{82}$ & 9.29$\pm{0.02}$ & 1.34$\pm{0.07}$ & - & - & - \\
J0756+0218 & 46.8 & $<$1.2 & 2.29$\pm{0.06}$ & 44.00$\pm{0.04}$ & 2966$\pm{59}$ & 8.78$\pm{0.02}$ & 0.87$\pm{0.05}$ & 44.60$\pm{0.01}$ & 2820$\pm{56}$ & 8.66$\pm{0.02}$ \\
J0759+1800 & 47.2 & 77.1$\pm{25.7}$ & 1.26$\pm{0.02}$ & 44.87$\pm{0.03}$ & 3448$\pm{398}$ & 9.13$\pm{0.10}$ & 1.05$\pm{0.38}$ & - & - & - \\
J0807+1328 & 47.2 & 16.8$\pm{5.9}$ & 1.56$\pm{0.05}$ & 44.57$\pm{0.02}$ & 4139$\pm{83}$ & 9.26$\pm{0.02}$ & 0.70$\pm{0.04}$ & - & - & - \\
J0829+0303 & 47.1 & 25.6$\pm{9.1}$ & 0.84$\pm{0.03}$ & 44.27$\pm{0.05}$ & 3702$\pm{123}$ & 9.10$\pm{0.03}$ & 0.74$\pm{0.06}$ & 44.88$\pm{0.01}$ & 3145$\pm{113}$ & 8.91$\pm{0.03}$ \\
J0831+4046 & 47.2 & $<$1.7 & 1.35$\pm{0.07}$ & 44.24$\pm{0.05}$ & 2548$\pm{51}$ & 8.83$\pm{0.02}$ & 1.79$\pm{0.09}$ & - & - & - \\
J0840+5624 & 47.2 & 26.0$\pm{5.6}$ & 0.99$\pm{0.02}$ & 44.37$\pm{0.01}$ & 3968$\pm{110}$ & 9.21$\pm{0.02}$ & 0.72$\pm{0.05}$ & 44.95$\pm{0.01}$ & 3320$\pm{66}$ & 9.00$\pm{0.02}$ \\
J0850+3246 & 47.2 & $<$0.7 & 1.41$\pm{0.03}$ & 44.33$\pm{0.02}$ & 4418$\pm{290}$ & 9.32$\pm{0.06}$ & 0.61$\pm{0.10}$ & 45.00$\pm{0.01}$ & 3077$\pm{151}$ & 8.96$\pm{0.04}$ \\
J0859+2520 & 47.0 & 56.0$\pm{22.6}$ & 0.50$\pm{0.01}$ & 44.74$\pm{0.01}$ & 3024$\pm{169}$ & 8.90$\pm{0.05}$ & 1.06$\pm{0.15}$ & - & - & - \\
J0927+2001 & 47.0 & $<$1.2 & 2.22$\pm{0.04}$ & 44.14$\pm{0.03}$ & 4145$\pm{83}$ & 9.14$\pm{0.02}$ & 0.52$\pm{0.03}$ & 44.71$\pm{0.01}$ & 2864$\pm{57}$ & 8.74$\pm{0.02}$ \\
J0941+5947 & 47.2 & 13.6$\pm{5.6}$ & 0.67$\pm{0.03}$ & 44.73$\pm{0.02}$ & 2960$\pm{59}$ & 8.98$\pm{0.02}$ & 1.38$\pm{0.07}$ & - & - & - \\
J0953+6910 & 47.0 & $<$2.6 & 1.30$\pm{0.05}$ & 44.05$\pm{0.06}$ & 2594$\pm{52}$ & 8.74$\pm{0.02}$ & 1.36$\pm{0.07}$ & 44.60$\pm{0.01}$ & 2336$\pm{226}$ & 8.49$\pm{0.09}$ \\
J1050+4627 & 47.1 & 37.4$\pm{2.9}$ & $<$1.08 & 44.41$\pm{0.02}$ & 3455$\pm{69}$ & 9.07$\pm{0.02}$ & 0.92$\pm{0.04}$ & - & - & - \\
J1100+5800 & 47.2 & 42.9$\pm{7.0}$ & 0.41$\pm{0.01}$ & 44.70$\pm{0.01}$ & 3873$\pm{77}$ & 9.19$\pm{0.02}$ & 0.77$\pm{0.04}$ & - & - & - \\
J1102+6635 & 46.9 & $<$3.2 & 0.51$\pm{0.08}$ & 44.23$\pm{0.10}$ & 3424$\pm{632}$ & 8.95$\pm{0.16}$ & 0.73$\pm{0.61}$ & - & - & - \\
J1116+5853 & 46.8 & 8.6$\pm{4.2}$ & 0.92$\pm{0.04}$ & 44.15$\pm{0.02}$ & 2965$\pm{167}$ & 8.78$\pm{0.05}$ & 0.86$\pm{0.13}$ & 44.67$\pm{0.01}$ & 2795$\pm{282}$ & 8.69$\pm{0.09}$ \\
J1134+3928 & 46.8 & $<$7.3 & 1.74$\pm{0.08}$ & 44.18$\pm{0.05}$ & 3702$\pm{74}$ & 8.95$\pm{0.02}$ & 0.53$\pm{0.04}$ & - & - & - \\
J1141+7119 & 46.8 & 112.1$\pm{12.9}$ & 1.77$\pm{0.09}$ & 44.28$\pm{0.01}$ & 7800$\pm{156}$ & 9.63$\pm{0.02}$ & 0.13$\pm{0.01}$ & 44.81$\pm{0.01}$ & 8031$\pm{161}$ & 9.71$\pm{0.02}$ \\
J1245+4348 & 47.3 & $<$1.7 & 0.78$\pm{0.02}$ & 44.65$\pm{0.02}$ & 2820$\pm{79}$ & 8.99$\pm{0.02}$ & 1.71$\pm{0.12}$ & - & - & - \\
J1257+6349 & 47.1 & $<$0.9 & 1.36$\pm{0.03}$ & 44.22$\pm{0.03}$ & 2042$\pm{83}$ & 8.61$\pm{0.04}$ & 2.62$\pm{0.26}$ & 44.86$\pm{0.01}$ & 2462$\pm{119}$ & 8.68$\pm{0.04}$ \\
J1327+5732 & 47.0 & 55.9$\pm{7.0}$ & $<$0.69 & 44.20$\pm{0.01}$ & 6680$\pm{301}$ & 9.57$\pm{0.04}$ & 0.21$\pm{0.02}$ & 44.88$\pm{0.01}$ & 4408$\pm{88}$ & 9.21$\pm{0.02}$ \\
J1328+4445 & 47.2 & 44.8$\pm{6.4}$ & $<$0.55 & 44.68$\pm{0.02}$ & 4574$\pm{271}$ & 9.36$\pm{0.05}$ & 0.59$\pm{0.09}$ & - & - & - \\
J1342+5838 & 46.9 & 10.0$\pm{6.2}$ & 0.71$\pm{0.04}$ & 44.29$\pm{0.03}$ & 2754$\pm{536}$ & 8.77$\pm{0.17}$ & 1.14$\pm{1.06}$ & - & - & - \\
J1436+5007 & 47.0 & 9.9$\pm{8.4}$ & 0.99$\pm{0.03}$ & 44.30$\pm{0.01}$ & 4455$\pm{89}$ & 9.21$\pm{0.02}$ & 0.46$\pm{0.02}$ & 44.79$\pm{0.01}$ & 3282$\pm{174}$ & 8.90$\pm{0.05}$ \\
J1458+3327 & 47.1 & 8.1$\pm{6.0}$ & 1.04$\pm{0.04}$ & 44.37$\pm{0.03}$ & 3370$\pm{584}$ & 9.01$\pm{0.15}$ & 0.89$\pm{0.65}$ & - & - & - \\
J1620+5202 & 47.6 & 51.1$\pm{14.5}$ & 1.19$\pm{0.04}$ & 44.82$\pm{0.01}$ & 5763$\pm{115}$ & 9.74$\pm{0.02}$ & 0.56$\pm{0.03}$ & - & - & - \\
J1621+5155 & 47.7 & $<$0.6 & 0.35$\pm{0.01}$ & 44.74$\pm{0.01}$ & 3990$\pm{80}$ & 9.46$\pm{0.02}$ & 1.26$\pm{0.06}$ & 45.41$\pm{0.01}$ & 3590$\pm{72}$ & 9.32$\pm{0.02}$ \\
\enddata
\label{tab:quasar}
\tablecomments{(2): quasar bolometric luminosity determined from the 5100 \AA\ continuum luminosity. (3): {rest-frame equivalent width of \oiii}, where the 3-$\sigma$ upper limits assuming a line width FWHM of 500 \kms\ are listed for non-detections. (4): Fe strength, defined as the ratio of \ewfe\ over \ewhb. (5)--(8): \hb\ luminosity, FWHM, BH mass and Eddington ratio. (9)--(11): \ha\ luminosity, FWHM and BH mass. Only the measurement errors are reported.}
\end{deluxetable*}

\subsection{{\rm \oiiitext} Strength and Kinematics}
\label{sec:43}

\begin{figure*}[!ht]
\begin{minipage}[t]{0.5\textwidth}
\centering
\includegraphics[width=\textwidth]{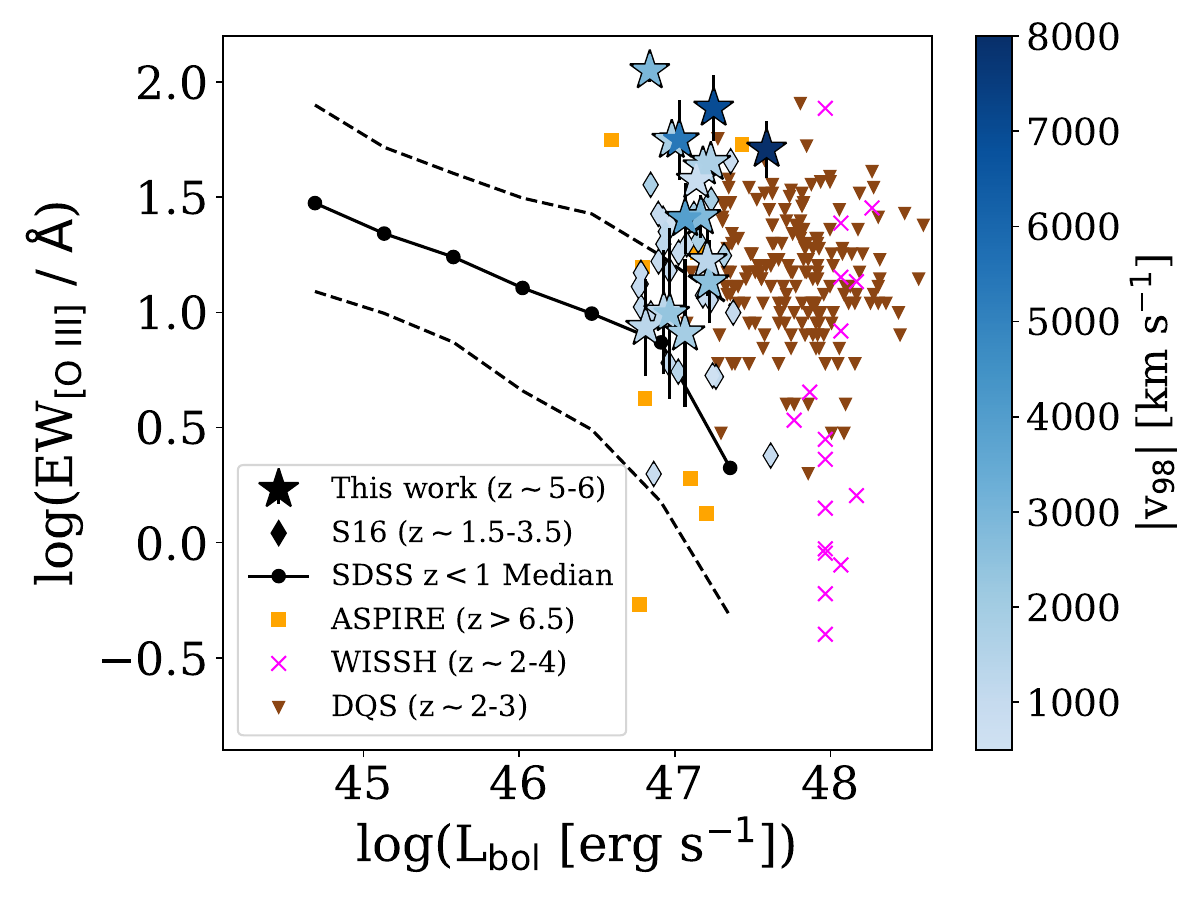}
\end{minipage}
\begin{minipage}[t]{0.5\textwidth}
\centering
\includegraphics[width=\textwidth]{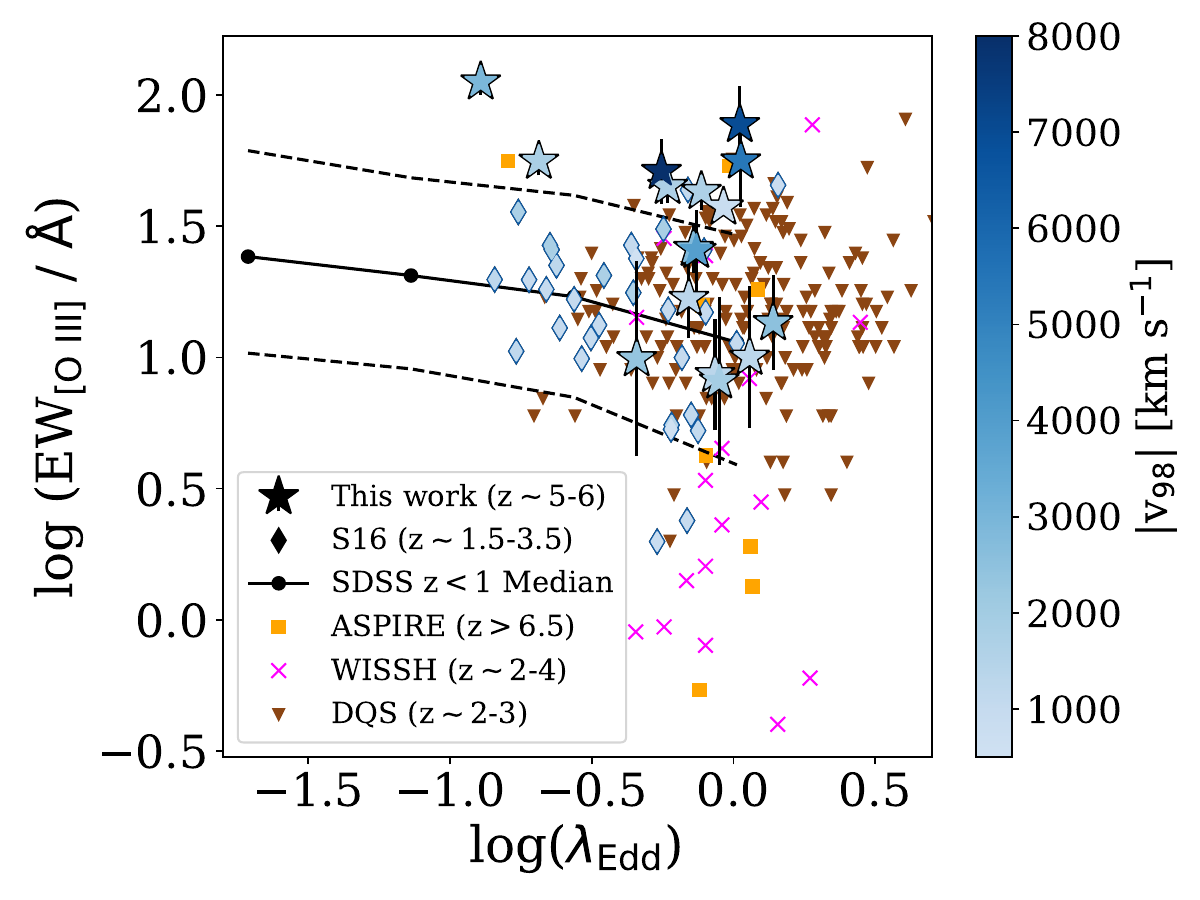}
\end{minipage}
\caption{Rest-frame \oiiitext\ equivalent widths as a function of bolometric luminosities (left), and Eddington ratios (right) for our objects (stars) and {\textit{S16 sample}} (diamonds). They are color-coded by the \vjiuba\ of their \oiiitext\ emission lines. The median and 16th and 84th percentiles of z $<$ 1 SDSS quasars are indicated by the black line with dots and dashed lines, respectively. {Also plotted are the ASPIRE sample (orange squares), the WISSH sample (magenta crosses) and DQS sample (brown downward triangles)}. {For our sample and DQS sample, the objects with no \oiiitext\ detections are omitted.} }
\label{fig:ewo3}
\end{figure*}

{In \citet{Liu2026a}, we reported \oiiitext\ emission line detections (S/N $>$ 3) in 16 out of the sample of 27 $z \sim$ 5--6 quasars as examined here. This sample exhibits significantly higher ($>$3.9$\times$) incidence of extremely fast \oiiitext\ outflows (with velocities greater than 2700 \kms) when compared to those in lower-redshift quasars. Readers are referred to that paper for more details. Here, we further explore the \oiiitext\ strength and kinematics of our objects with \oiiitext\ detections (Fig. \ref{fig:ewo3}). The \oiiitext\ kinematics adopted here were {reported in \citet[][]{Liu2026a}}.
To put our measurements into a better context, we have also included several well-studied quasar samples from previous studies as shown in Fig. \ref{fig:ewo3}. These include the \textit{S16 sample} as introduced in Sec. \ref{34}, the ASPIRE sample, the WISSH sample \citep[e.g.,][]{Bischetti2019,WISSH2018}, the distant quasar survey (DQS) sample \citep[e.g.,][]{Matthews2021,Matthews2023}, and the SDSS z $<$ 1 quasars compiled by \citet{wu_spectroscopic_2023}. Specifically, the ASPIRE quasars have similar bolometric luminosities to ours and are at slightly higher redshifts ($z>6.5$). The WISSH and DQS quasars exhibit higher \Lbol\ (\logLbol\ $\simeq$ 47 -- 49, median: 48.0), slightly higher \Mbh\ (\logMbh\ $\simeq$ 8.7--10.6, median: 9.6) and \edd\ ($\sim$0.2--9, median: 1.2), and are at lower redshifts ($z\sim$ 2--4). 
Blueshifted \oiiitext\ emission lines tracing fast outflows have also been reported in many of these objects \citep[e.g.,][]{wissh2017,Yang2023b}.}

{As introduced in Sec. \ref{sec:intro}, the rest-frame \oiiitext\ equivalent widths \ewo\ generally decrease with increasing bolometric luminosity \Lbol\ (i.e., ``Baldwin effect'') or Eddington ratio \edd\ \cite[e.g.,][]{ShenHo2014}. 
Similar to many other quasars with \logLbol\ / \ergs $\gtrsim$ 46.6 from the various samples shown in Fig. \ref{fig:ewo3} , many of our objects follow the expectation from the ``Baldwin effect'' and the trend of decreasing \ewo\ along with increasing \edd. Nevertheless, for these luminous quasars with \logLbol\ $\simeq$ 46.6 -- 49 and \edd\ $\simeq$ 0.1--3, a wide range of \ewo\ ($\simeq$ 0.3--110) is seen. % with \logLbol\ / \ergs $\gtrsim$ 46.5. 
Such a large range of \ewo\ in luminous quasars has also been reported in previous studies and explained with different hypotheses.
For example, \citet[][]{Netzer2004} suggests that some luminous quasars could lose their dynamically unbound NLR, whereas others could be experiencing violent star formation that produces a large quantity of high-density gas. \citet[][]{bask05o3} attributes it to varying covering factors, electron densities and ionization parameters in different objects.}

{Considering our sample alone, the \ewo\ still spans more than two orders of magnitude, while the \Lbol\ and \edd\ span merely one order of magnitude. 
There is an exceptional group of objects exhibiting excesses of \ewo\ (left panel of Fig. \ref{fig:ewo3}) compared to the ``Baldwin effect''.
Likewise, a few exceptions with elevated \ewo\ are also seen with respect to the general trend between \ewo\ and \edd\ (right panel of Fig. \ref{fig:ewo3}), although the \ewo\ boost in them are less prominent compared to the case for \Lbol. }
In both cases, the objects with large \ewo\ usually show highly blueshifted \oiiitext\ profiles (large $|$\vjiuba$|$; \vjiuba\ is the velocity where 98\% of the \oiiitext\ line flux are redshifted with respect to it). For our sample and the {\textit{S16 sample}} combined, there is a positive trend between \ewo\ and $|$\vjiuba$|$ and \wjiu\ (the line width enclosing 90\% of the total line flux) in logarithm, with Pearson correlation coefficients of 0.49 and 0.41 (where $\pm{1}$ indicate perfect linear correlations and 0 indicates no correlation), and $p$-values of 2$\times$10$^{-4}$ and 3$\times$10$^{-3}$ (with a null-hypothesis that the two variables are uncorrelated), respectively. On the other hand, for our sample and the {\textit{S16 sample}} combined, the dependence of \ewo\ on the bolometric luminosities, BH masses and Eddington ratios, if any, are quite weak (with Pearson correlation coefficients of 0.007, 0.12, and $-$0.11). In summary, the boosted \ewo\ in our objects is primarily related to the emission from the fast outflows.
{This scenario is consistent with many previous studies of lower-$z$ quasars/AGN. For example, as proposed in  \citet{Netzer2004},  luminous quasars cannot maintain a NLR as large as one would expect from a simple extrapolation from lower luminosity AGN. Instead, the NLR may be easily disrupted by the outflows, and/or become gravitationally unbound due to other reasons. This is in line with our hypothesis that the boosted \oiiitext\ emission in certain objects within our sample is mainly associated with fast outflows.}

\begin{figure*}[!htb]
\begin{minipage}[t]{0.5\textwidth}
 \centering
\includegraphics[width=\textwidth]{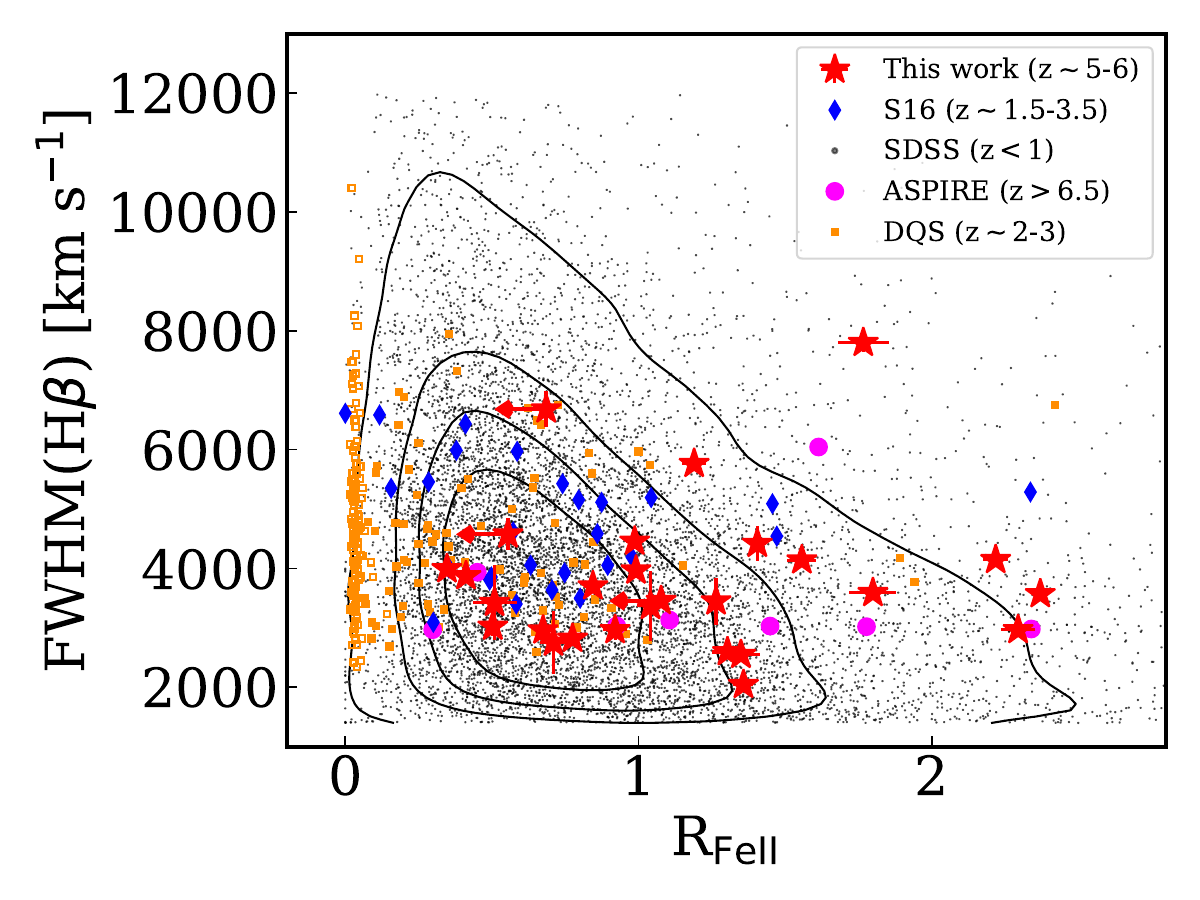}
\end{minipage}
 \begin{minipage}[t]{0.5\textwidth}
 \centering
\includegraphics[width=\textwidth]{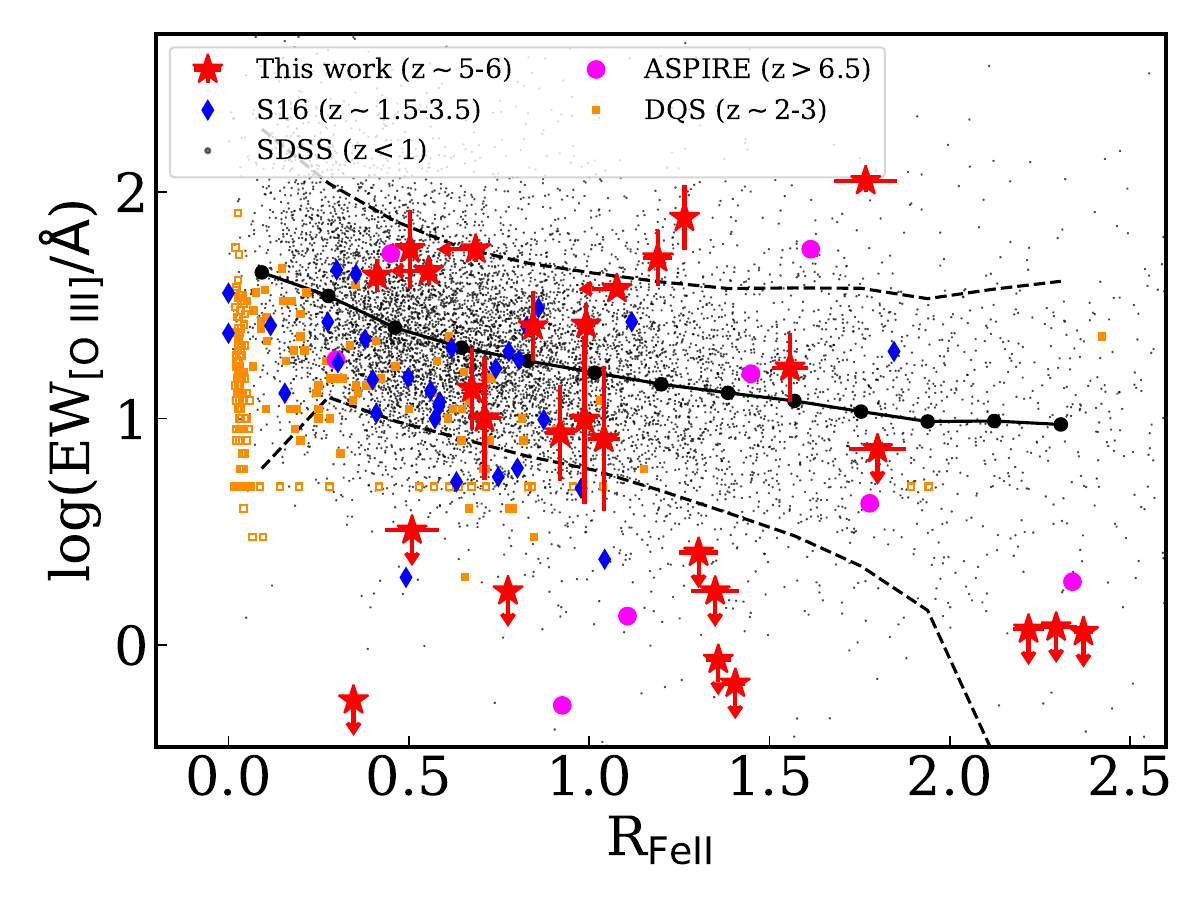}
\end{minipage}
\caption{EV1 planes defined by broad \hb\ FWHM versus \RFe\ (left) and \ewo\ versus \RFe\ (right) for our sample (red stars), {\textit{S16 sample} (blue diamonds), ASPIRE sample (magenta circles), DQS sample (orange squares) and SDSS z $<$ 1 quasars (gray dots). For the DQS sample, the hollow symbols indicate objects with upper limits on either \RFe\ or \ewo.} In the left panel, the contours correspond to the 99\%, 90\%, 75\% and 50\% percentiles of the the distribution of the SDSS sample. In the right panel, the solid and dashed lines indicate the median, 10th and 90th percentiles of the distribution of the SDSS sample.}
\label{fig:EV1}
\end{figure*}

\subsection{Eigenvector 1 Relations}
Previous studies suggest that lower-z quasars ($z\lesssim3$) follow well-defined correlations despite a diverse range of spectroscopic properties. The most prominent one of these correlations, known as Eigenvector 1 (EV1), links many quasar properties to the strength of the optical \feii\ and \oiiitext\ emission. It is widely believed that the main driver of EV1 is the Eddington ratio, and more recently, orientation has also been suggested to play a role in the diversity of quasar phenomenology \citep{BorosonGreen1992,Boroson2002,Marziani2001,Marziani2003,ShenHo2014}.
In Fig. \ref{fig:EV1}, we examine the locations of our high-$z$ quasar sample in the EV1 planes, namely the correlations of \ewo\ and FWHM$_{\rm H\beta}$ with \RFe. {They are again compared with the same quasar samples examined in Sec. \ref{sec:43} and Fig. \ref{fig:ewo3}, except for the WISSH sample whose exact \RFe\ measurements, to our knowledge, have not been reported publicly so far.}
Here \RFe\ is the ratio of \ewfe\ over \ewhb, where \ewfe\ is measured between rest-frame 4434 \AA\ and 4684 \AA.

Within their small \Lbol\ range ($\sim$1 dex), our objects show a broad range of Fe strength, with \RFe\ range from undetected to $\sim$2.5. 
In the two-dimensional EV1 plane of \RFe\ versus \fwhmhb, all but one of our objects fall within/close to the 99\% contours of z $<$ 1 SDSS quasars. The only exception in our sample shows larger \RFe\ given its \fwhmhb\, which is rare but not unseen in z $<$ 1 SDSS quasars and cosmic noon {\textit{S16 sample}}. The median \fwhmhb\ of our sample ($\sim$3600 \kms) is smaller than that of {\textit{S16 sample}} ($\sim$ 5000 \kms), consistent with our findings from the composite spectrum (Fig. \ref{fig:composite}).
Meanwhile, it should be noted that at a fixed \RFe, the dispersion in \fwhmhb\ could be dominated by the orientation effect \citep[e.g.,][]{ShenHo2014}. Therefore, the lower \fwhmhb\ in our sample may be partly caused by a combination of the orientation effect and small sample size.

In the EV1 plane of \ewo\ versus \RFe, {for our $z \sim$ 5--6 objects with \oiiitext\ detections,} all but three show \ewo\ falling within the 10\%--90\% range of SDSS z $<$ 1 quasars. The three objects beyond the 10\%--90\% SDSS range show significantly larger \ewo\ than the {\textit{S16 sample}}. At the same bolometric luminosity range, while the overall detection rates of \oiiitext\ emission line in our sample and the {\textit{S16 sample}} are similar ($\sim$60\% with S/N $>$ 3), the median \ewo\ of these detections in our sample ($\sim$32 \AA) is larger than those of {\textit{S16 sample}} ($\sim$16 \AA). {In the right panel of Fig. \ref{fig:EV1}, we have also included the 3-$\sigma$ upper limits of \ewo\ for objects with no \oiiitext\ detections in our sample. While all but four of them show weaker \ewo\ than 90\% of SDSS z $<$ 1 quasars at a given \RFe, two out of the eight ASPIRE quasars also exhibit comparably weak \oiiitext\ emission. Moreover, the \ewo\ upper limits in our sample depend on the assumed \oiiitext\ line width (500 \kms) and are thus highly uncertain. Therefore, no object in our sample robustly exhibits significantly lower \ewo\ than those reported in previous studies, especially for those targeting $z>5$ quasars.}

The 3 ``outliers'' with large \ewo\ in our sample, J0759$+$1800, J1141$+$7119, and J1620$+$5202, have prominent, blueshifted \oiiitext\ emission line tracing fast outflows that are among the fastest ones in our sample. {Similarly, the two objects with the highest \ewo\ in the ASPIRE sample, J0224$-$4711 and J2232$+$2930, also show clear evidence of fast \oiiitext\ outflows \citep[See Fig. 1 in][]{Yang2023b}.} This is again in line with the findings shown in Fig. \ref{fig:ewo3} and Sec. \ref{sec:43}, where there is a positive trend between \ewo\ and \oiiitext\ outflow velocity in our sample. This reinforces the picture that the boost of \oiiitext\ strength some objects within our sample is caused by the presence of fast \oiiitext\ outflows \citep{Liu2026a}. On the other hand, the systematically lower \ewo\ observed in the {\textit{S16 sample}} than in the $z < 1$ SDSS quasars, can be explained well by the ``Baldwin effect'' as the mean luminosity of {\textit{S16 sample}} is higher than the whole population of $z < 1$ SDSS quasars.

Additionally, the \RFe\ itself may also be taken as a crude indicator of the metallicity of the BLR. Our sample exhibits a \RFe\ range similar to that of $z < 1$ SDSS quasars, suggesting that the BLR of our $z \sim$ 5--6 quasars already have metallicity comparable to those of nearby quasars. 
In our sample, 3 objects show \RFe\ larger than 99\% of low-$z$ SDSS quasars at a given \fwhmhb.
Additionally, 20 out of our 27 objects show \RFe $>$ 0.6, the value corresponds to solar metallicity as found in a previous study \citep{Netzer2007}.
These may imply super-solar metallicity in the BLR of a significant portion of our high-$z$ quasars and a very short time scale of metal enrichment in them.

{Furthermore, our sample exhibits distributions similar to that of the ASPIRE sample at $z>6.5$ \citep{Yang2023b} in both EV1 planes, which is expected due to the similarity of the two samples  (e.g., in their \Lbol, \Mbh, \edd, redshifts, etc.). Nonetheless, while the maximum values of \RFe\ are comparable between our sample and the DQS sample, the former shows larger median \RFe (1.0) than the latter ($<$0.04), despite their more similar ranges of \fwhmhb\ and \ewo. This could be caused by the different \Lbol, \Mbh, \edd, redshifts, and other properties of the two samples. }

Overall, most of the objects in our $z \sim$ 5--6 sample fall within the EV1 planes formed by the low-$z$ quasars. 
These high-$z$ luminous quasars likely possess BLR/NLR as mature as their low-$z$ siblings, despite the much more limited evolution time available for the former. In addition, our high-$z$ sample contains a population of sources with significantly stronger, more blueshifted and broader \oiiitext\ line emission than those of low-$z$ quasars, which is due to the presence of fast galaxy-scale outflows in those systems.

\section{Spatially Extended Emission} 
\label{sec:extended}

\begin{figure*}[!ht]
\begin{minipage}[t]{\textwidth}
 \centering
\includegraphics[width=\textwidth]{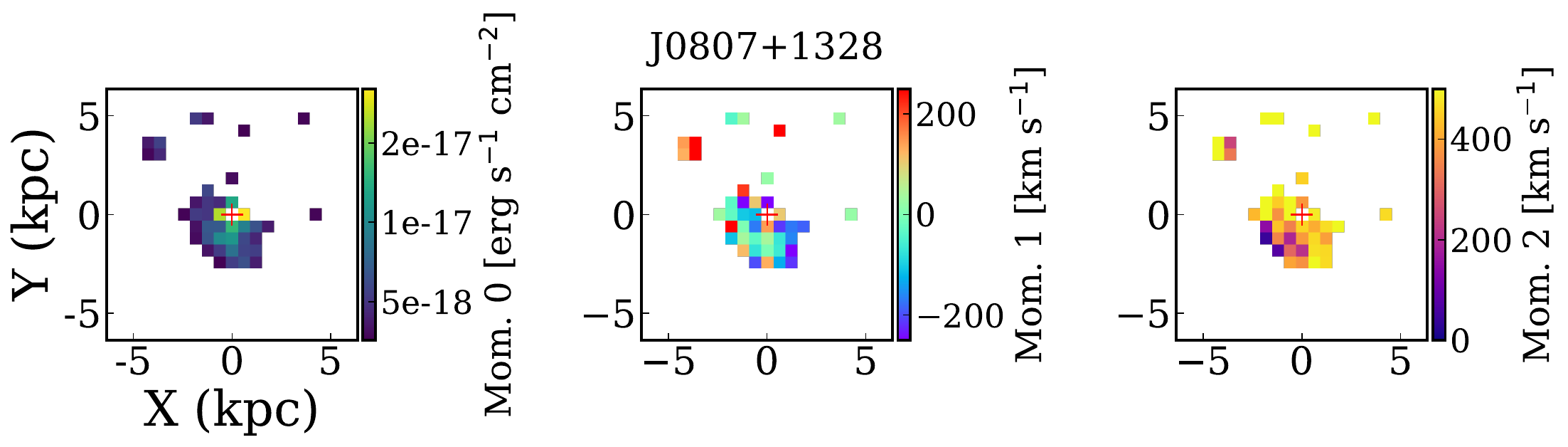}
\end{minipage}
\begin{minipage}[t]{\textwidth}
 \centering
\includegraphics[width=\textwidth]{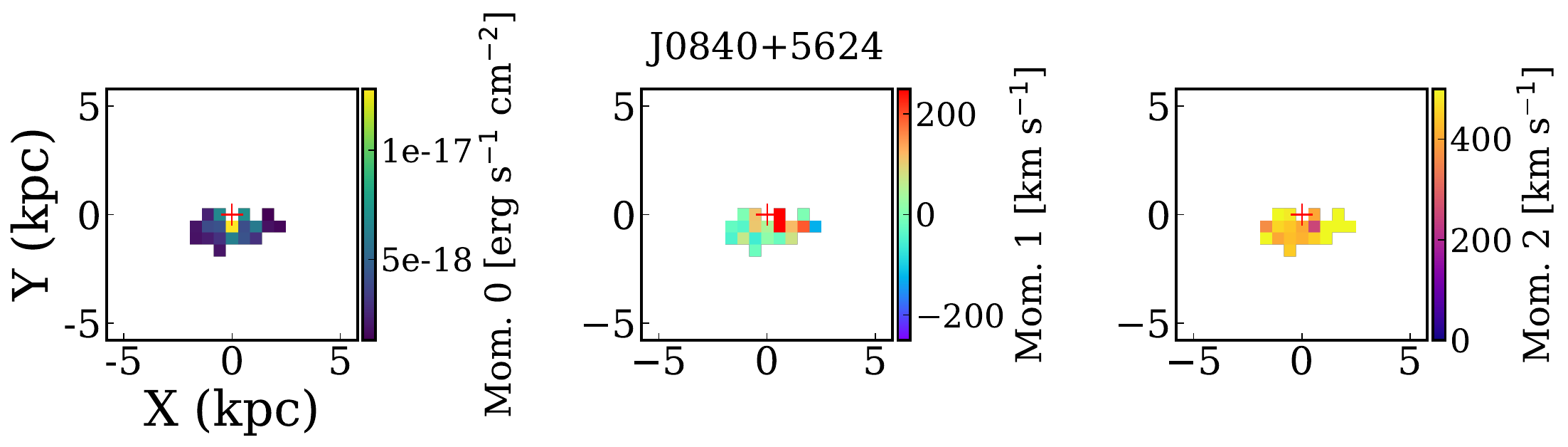}
\end{minipage}
\begin{minipage}[t]{\textwidth}
 \centering
\includegraphics[width=\textwidth]{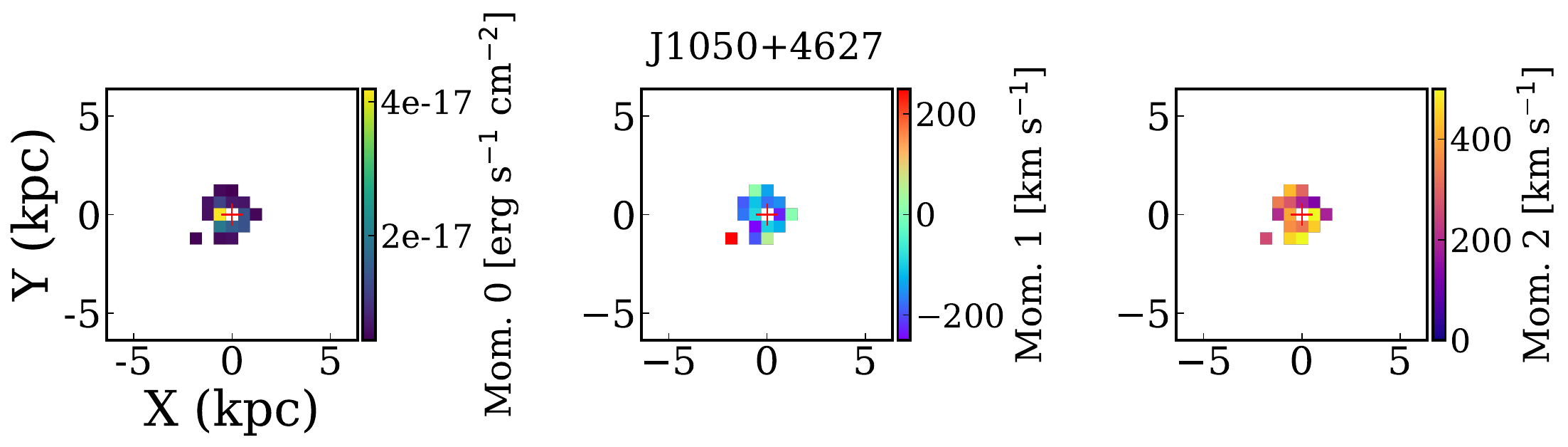}
\end{minipage}
\caption{PSF-subtracted IFU maps for the first 3 of the 6 objects with extended \oiiitext\ line emission. For each object, moment 0 (flux) , moment 1 (velocity) and moment 2 (velocity dispersion) are shown from left to right. Spaxels with S/N $<$ 3 are omitted from the maps.}
\label{fig:o3map}
\end{figure*}

\begin{figure*}[!ht]
\begin{minipage}[t]{\textwidth}
 \centering
\includegraphics[width=\textwidth]{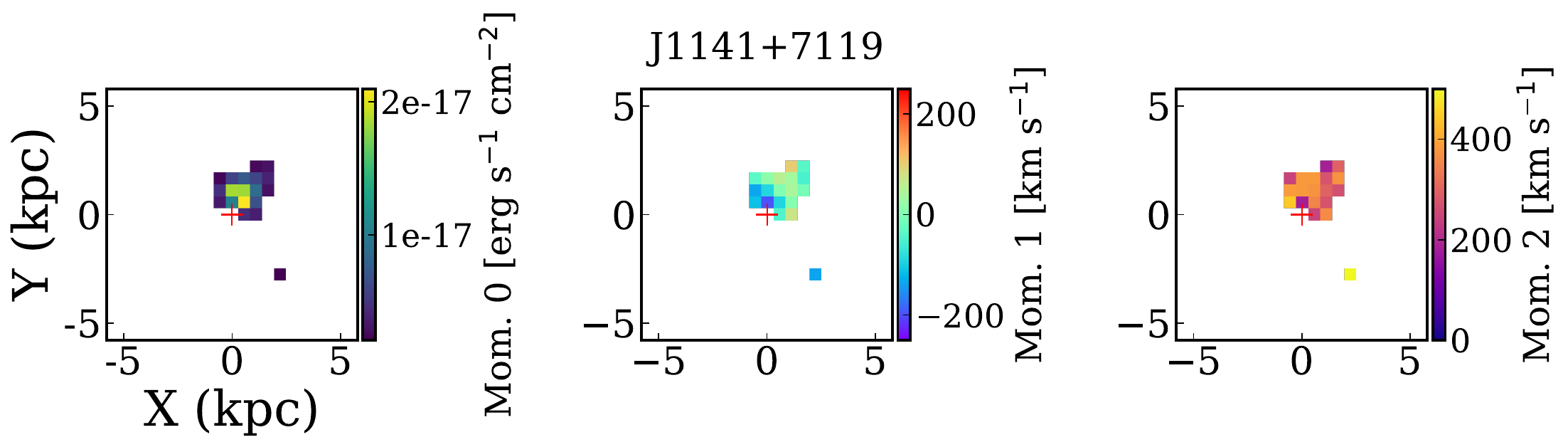}
\end{minipage}
\begin{minipage}[t]{\textwidth}
 \centering
\includegraphics[width=\textwidth]{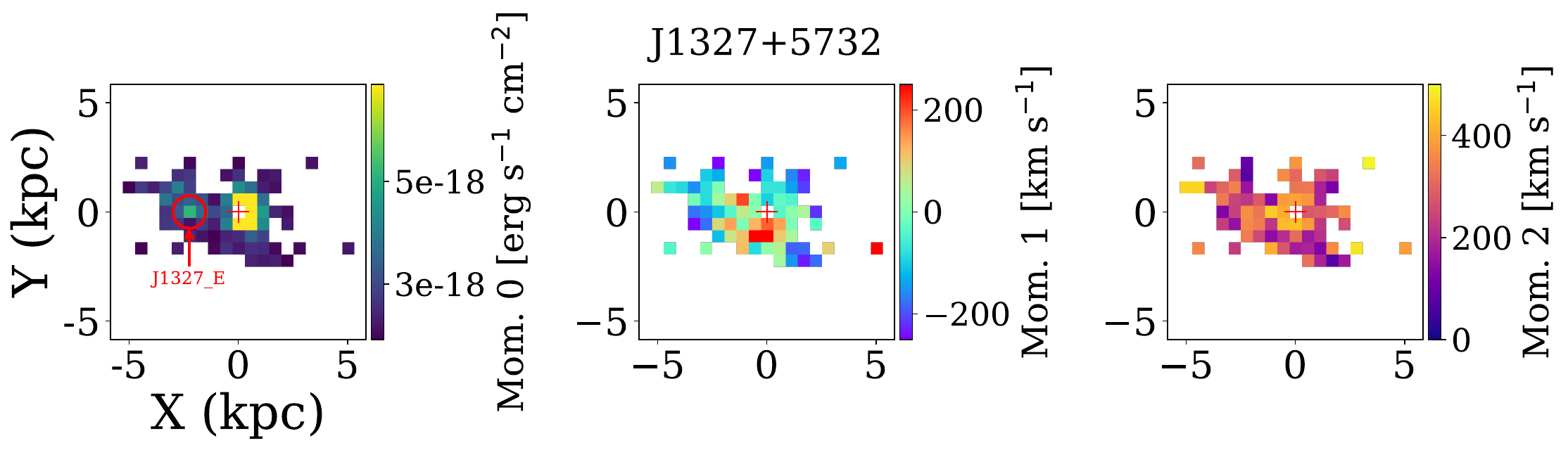}
\end{minipage}
\begin{minipage}[t]{\textwidth}
 \centering
\includegraphics[width=\textwidth]{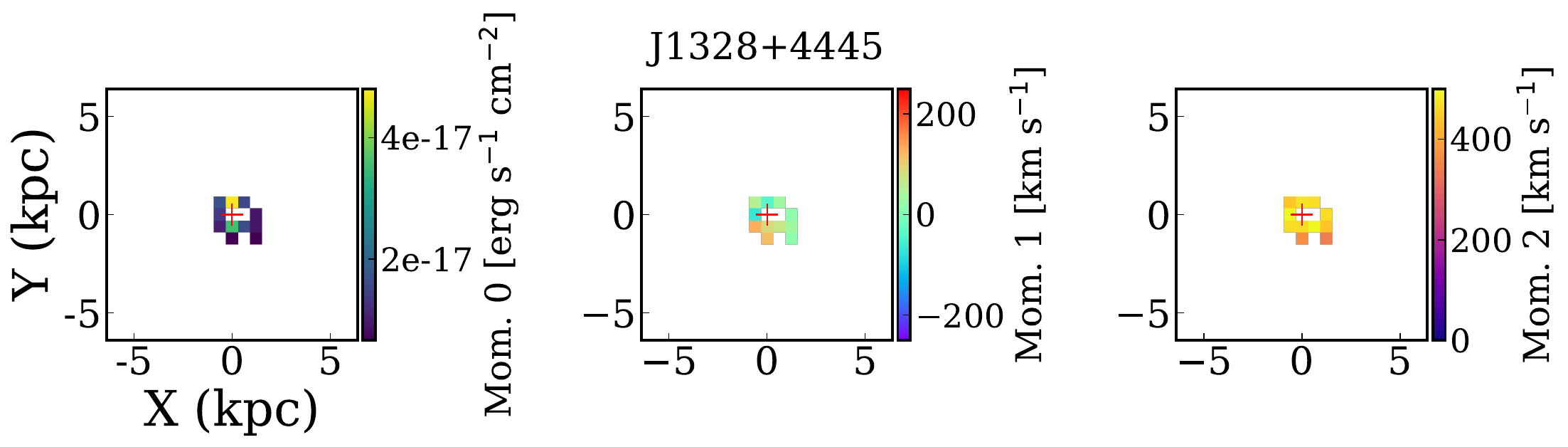}
\end{minipage}
\caption{Same as Fig. \ref{fig:o3map} but for the remaining 3 objects. For J1327$+$5732, the red circle indicates the aperture to extract the spectrum for the merging companion J1327\_E. See Sec. \ref{sec:51} for details.}
\label{fig:o3map2}
\end{figure*}

{We first search for dual/binary quasars in our sample by identifying two distinct point sources with broad (FWHM $>$ 1200 \kms) \ha\ and/or \hb\ emission lines.
Among the 27 objects, we see no evidence of such dual/binary quasars.} This is consistent with the low detection rates of them at $z \gtrsim 5$ where only a few objects have been reported in the literature so far \citep[e.g.,][]{Yue2023,Matsuoka2024,Yue2026}.

Our IFU data also provide a glance at the extended line emission from the quasar host galaxy and/or companion galaxy within our sample, which unveils the gaseous environments they live in \citep[e.g.,][]{VeilleuxLiu2023,Liu2024b,Liu2026a}.
We apply a simple PSF subtraction by first constructing a 2D PSF model using the flux map of the broad \hb\ emission line wing within the velocity range of [$-$1500,$-$500] and [$+$500, $+$1500] \kms, which relies on the fact that the broad \hb\ emission comes from the spatially unresolved BLR of the quasar. Next, we extract a ``pure'' quasar spectrum from the spaxel with the maximum wavelength-integrated total flux (i.e., the spaxel where the quasar is located). Finally, we scale-down the flux level of the ``pure'' quasar spectrum adopting the value of the PSF model at each spaxel and subtract it from the spectrum of the same spaxel, which gives the final PSF-subtracted spectrum for that spaxel. 

We see clear evidence of extended \oiiitext\ line emission from the host galaxy and/or companion galaxy in 6 objects, J0807$+$1328, J0840$+$5624, J1050$+$4627, J1141$+$7119 J1327$+$5732, and J1328$+$4445, which gives a detection rate of 6/27 or $\sim$22\% in our sample. This incidence rate should be treated as a lower limit since i) more diffuse, fainter emission will be missed by our shallow observation (only $\sim$8 min exposures, $\sim$1--5\% of other NIRSpec/IFU programs targeting individual high-$z$ quasars) and ii) our PSF subtraction assumes that the brightest spaxel is pure quasar emission, which may still be contaminated by host galaxy emission in reality and could thus result in an oversubtraction of the quasar PSF. 
The PSF-subtracted, moment 0 (flux), moment 1 (velocity), and moment 2 (velocity dispersion) maps of \oiii\ of the 6 objects are shown in Fig. \ref{fig:o3map} and \ref{fig:o3map2}. They are calculated over the [$-$1000, $+$1000] \kms\ range that is wide enough for our objects after the PSF subtraction. The extended line emission shows asymmetric morphologies in all 6 objects, although the evidence in J1050$+$4627 and J1328$+$4456 is weak. The velocities (moment 1) of these extended \oiiitext\ emission are in general modest, mostly within $\pm{300}$ \kms\ from the systemic velocities. The velocity dispersions (moment 2) of them range from $\sim$200 to $\sim$500 \kms. These characteristics may be consistent with late-stage merging activities within these systems, implying frequent merger events in these $z \sim$ 5--6 quasars. Nevertheless, the \oiiitext\ line emission is tracing ionized gas rather than stellar components within these systems. Therefore, these emission may alternatively trace the turbulent and/or clumpy interstellar medium of these quasar host galaxies. {Moreover, as stated in \citet{Liu2026a}, the radial profiles of the blueshifted \oiiitext\ line emission of the 6 fastest outflows in our sample are clearly extended on kpc scale, although our shallow observations could not robustly spatially resolve the outflowing gas spaxel by spaxel. Readers are referred to that paper for details.}
Similar extended line emission properties are also reported from deeper IFU observations of other individual high-$z$ quasars, tracing merging activity, companion galaxies and/or outflows \citep[e.g.][]{Decarli2024,Liu2024b,Marshall2023,Marshall2025a,Marshall2025b}. 
 In all, these high-$z$ quasars in our sample likely live in busy close environments with frequent mergers. Future deeper JWST NIRSpec/IFU data combined with more detailed quasar PSF subtraction are needed to better decipher the extended line emission in these objects and reveal more potential fainter extended emission in the rest of the sample.

\subsection{J1327$+$5732}
\label{sec:51}
\begin{figure*}[!ht]
\begin{minipage}[t]{0.33\textwidth}
 \centering
\includegraphics[width=\textwidth]{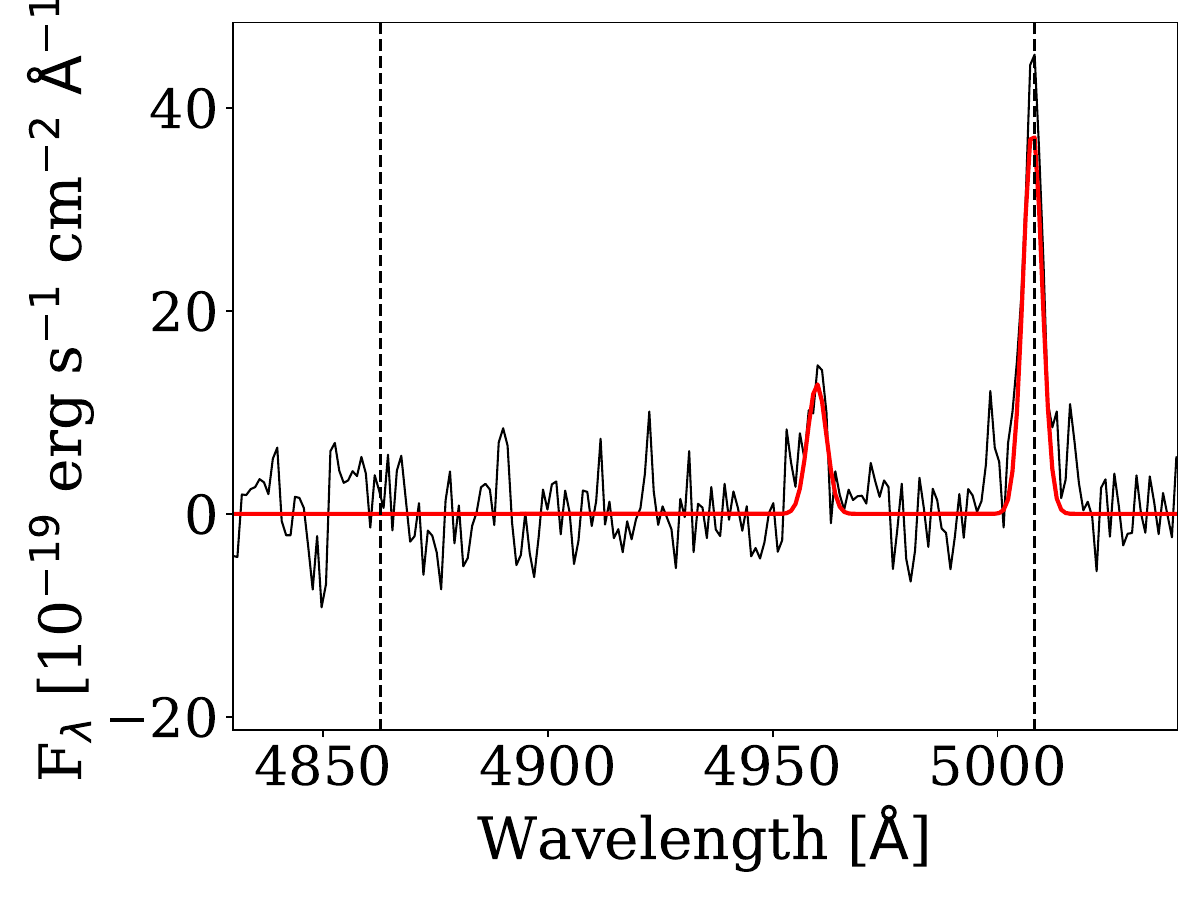}
\end{minipage}
\begin{minipage}[t]{0.33\textwidth}
 \centering
\includegraphics[width=\textwidth]{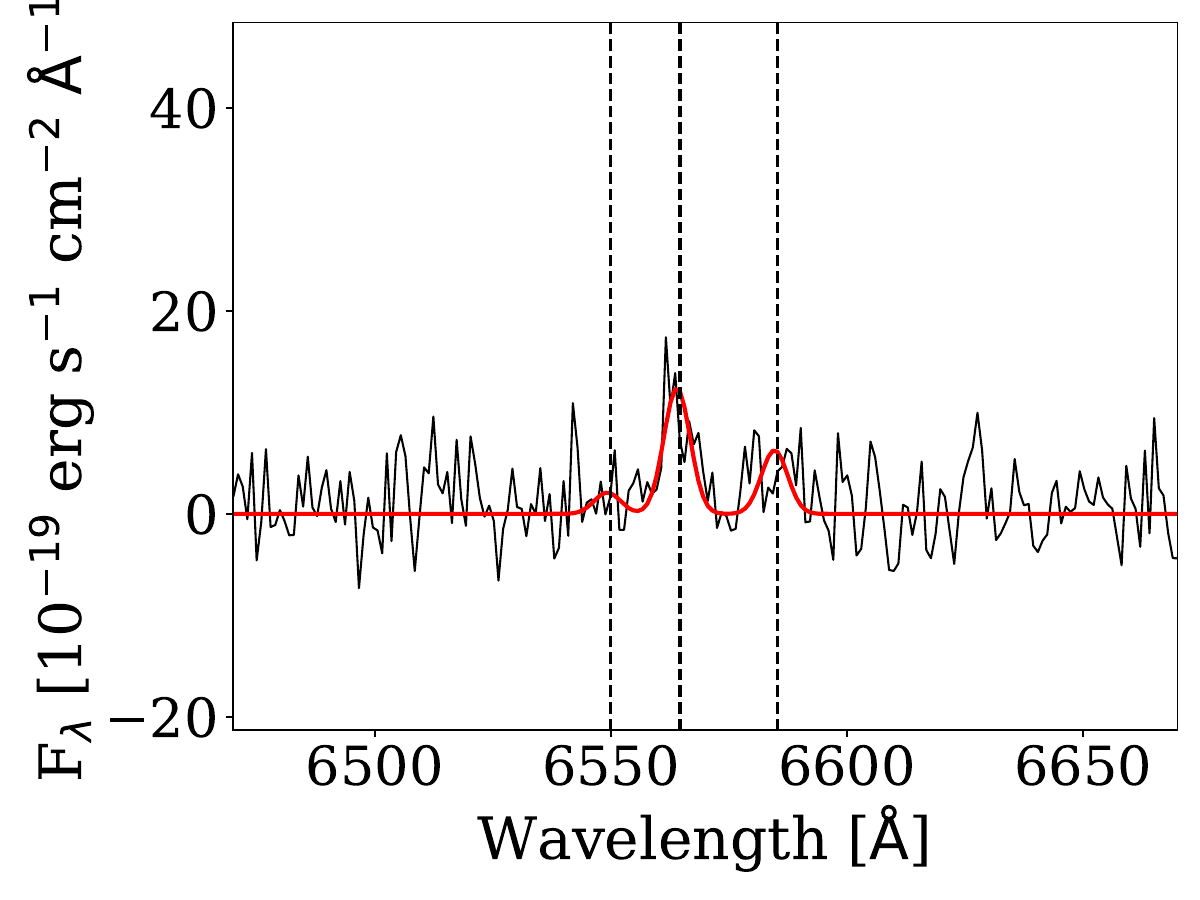}
\end{minipage}
\begin{minipage}[t]{0.33\textwidth}
 \centering
\includegraphics[width=\textwidth]{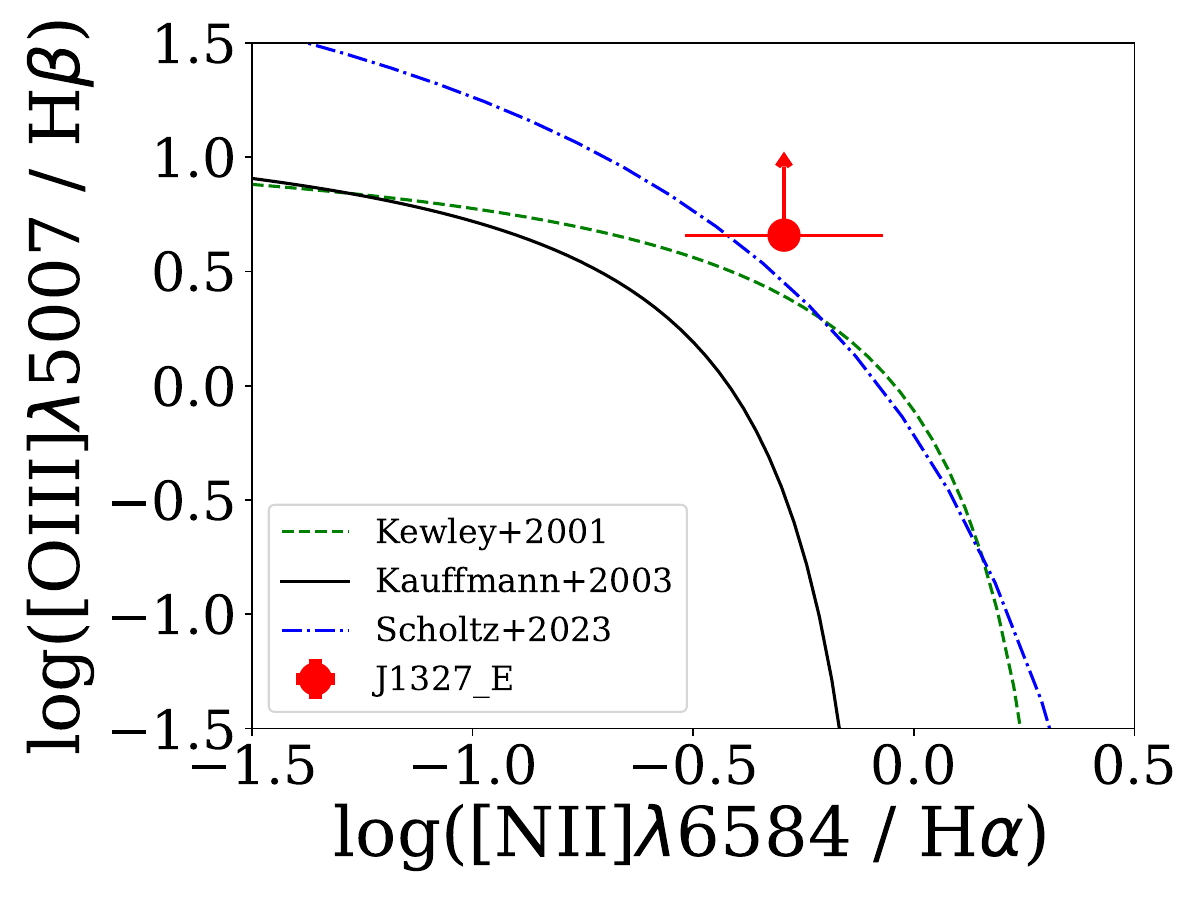}
\end{minipage}
\caption{\textbf{Left and Middle:} Zoom-in spectra and best-fit models for \hb\ (undetected), \oiiiab, \nii\ and \ha\ emission lines for J1327\_E. The location of each emission line at the quasar systemic velocity is indicated by the dashed line. \textbf{Right:} The location of J1327\_E on the BPT diagram. The black and green dashed line indicate the AGN versus star-forming galaxies for low-$z$ objects as defined in \citet{kauf03a} and \citet{kewl01}, respectively. The blue dash-dotted line indicates the new dividing line defined for high-$z$ objects based on JWST results from \citet{Scholtz2023}.}
\label{fig:J1327E}
\end{figure*}

The most extended \oiiitext\ line emission is seen in J1327$+$5732, which extends up to $\sim$5 kpc (the middle row of Fig. \ref{fig:o3map2}). 
A tentative negative-positive velocity gradient is visible in the vicinity of the quasar, which may trace the rotating gas within the quasar host galaxy. 
There is a distinct line-emitting region east of the quasar (marked in Fig. \ref{fig:o3map2}; denoted as J1327\_E hereafter), with a clear luminosity centroid $\sim$2.5 kpc from the quasar and S/N (\oiiitext) $>$ 10 for all spaxels within 0.2\arcsec\ ($\sim$1.2 kpc) from the centroid.
The entire region has modest velocities ($<\pm$200 \kms) and velocity dispersions ($<$200 \kms).
%To examine the properties of this
The integrated spectrum of J1327\_E extracted from an aperture with 0.2\arcsec\, radius from the quasar PSF-subtracted data cube is shown in Fig. \ref{fig:J1327E}, where major emission lines including \oiiiab, \nii\ and \ha\ are detected. We fit the emission line profiles of all major emission lines detected with a single Gaussian profile adopting the same kinematics (i.e., velocity and velocity dispersion). {J1327\_E is blueshifted by $\sim$20 \kms\ with respect to the quasar systemic velocity as determined from the narrow emission lines, and exhibits a line width of $\sim$140 \kms.} We see no evidence of broad \ha\ emission lines tracing AGN activity in our spectrum.
We obtain an observed \oiiitext\ luminosity of $\sim$($5.8\pm{0.2})\times$10$^{42}$ \ergs\ and \ha\ luminosity of $\sim$$(2.1\pm{0.3})\times$10$^{42}$ \ergs, without dust-extinction correction. These are comparable to those found for star-forming galaxies near quasars at $z \sim$ 6 \citep[e.g.,][]{wang_spectroscopic_2023,Champagne2025a,Decarli2024,Marshall2023,Marshall2025a,Marshall2025b}. {Based on all these observational evidence above, J1327\_E is likely a companion galaxy that is merging with the quasar host galaxy. However, we cannot formally rule out the possibility that it traces {a very luminous ionized gas clump} within the quasar host galaxy itself.} 

{If J1327\_E is indeed a companion galaxy, it provides another example of such systems near $z \gtrsim$ 5 quasars. Previous studies have found various numbers of companion galaxies around $z\sim6$ quasars \citep[e.g.,][and references therein]{Decarli2024,Costa2024,wang_spectroscopic_2023,Champagne2025a}. According to a recent compilation \citep{Costa2024}, the number ranges from 0 to more than 30 within a line-of-sight (LoS) velocity difference of $|\Delta v|$ $<$ 4000\kms.
Our data are very shallow and thus only provide a conservative lower limit (1 in 27 quasars) to the number of optical line-emitting companion galaxies in our sample. Therefore, our current results are consistent with the model predictions and previous observations. A more detailed comparison of our data with theoretical predictions is saved for a future study.
}

{In the following discussions, we explore further the nature of J1327\_E assuming that it is indeed a companion galaxy in the process of merging with the quasar host galaxy. The location of J1327\_E on the BPT diagram \citep{bpt} is shown in the right panel of Fig. \ref{fig:J1327E}. The 3-$\sigma$ upper limit of the \hb\ luminosity is adopted to derive the lower limit for the \oiiihb\ ratio. J1327\_E is located in the AGN area (i.e, above the maximum starburst line) defined for galaxies in the low-$z$ universe \citep{Kewley2001,kauf03a}. It is still in the AGN region when considering the modified AGN vs star-forming galaxy dividing line at $z>3$ as defined in \citet{Scholtz2023}, where the shift of the dividing line is mainly caused by the low metallicity of high-$z$ AGN. Therefore, one simple explanation is that J1327\_E hosts an AGN within itself. If confirmed, this system, to our knowledge, will be the first close quasar-AGN pair reported at this redshift, adding another intriguing example of kpc-scale dual/multiple AGN at $z \gtrsim 5$ \citep[e.g.,][]{Maiolino2024,Li2025,Ubler2025}. Future deeper JWST/NIRSpec and/or JWST/MIRI observations could better diagnose the existence of an AGN within J1327\_E through broad emission lines and/or AGN torus emission.}

{Alternatively, the AGN-like line ratios of J1327\_E may also be caused by the photoionization from the quasar. It has been observed in other systems that a central quasar/AGN may photoionize nearby companion galaxies \citep[e.g.,][]{Moran1992, daSilva2011,Merluzzi2018,Keel2019,Moiseev2023,Protusova2024,Marshall2025b}, which are also seen in some simulations \citep[e.g.,][]{Costa2014a,Chen2020}. Similar to the approach presented in Sec. 6.4 of \citet{Marshall2025b}, we estimate whether the quasar is luminous enough to ionize the gas within J1327\_E. The ionizing photon rate required to generate the observed \ha\ luminosity of J1327\_E is
\begin{equation}
Q_{\rm{ionizing}}=L_{\rm{H}\alpha}/(1.4\times10^{-12}\rm{\,erg})\quad,
\end{equation}
assuming that one \ha\ photon is produced for every 2.98 recombinations \citep{Hummer1987}.
This leads to an incident ionizing-photon rate of $Q_{\rm{ionizing}}\simeq1.5\times10^{54}$\,s$^{-1}$.}

{The incident ionizing quasar luminosity onto J1327\_E is
\begin{equation}
L_{\rm{incident}} = L_{\rm{bol}} \gamma \frac{\Omega}{4\pi}\quad,
\end{equation}
where $\gamma$ is the ratio of ionizing to bolometric luminosity, and $\Omega$ is the solid angle the companion subtends as seen from the quasar. For small angles, $\Omega\simeq\pi(\arctan{(r/d)})^2$. For a typical type 1 quasar, $\gamma=0.14$ \citep{Keel2019}.
The size $r$ is estimated as the extent of the \oiiitext\ emission of J1327\_E perpendicular to the direction of the quasar on the sky plane, which gives $r=0\farcs2$.
The distance $d$ is calculated from the luminosity centroid of J1327\_E to the quasar.
The ionizing luminosity is further converted to the incident photon rate adopting a mean ionizing photon energy of 2~rydbergs or $4.36\times10^{-11}$\,erg. Therefore,
\begin{equation}
Q_{\rm{incident}} = L_{\rm{incident}}/(4.36\times10^{-11}\,\rm{erg})\quad.
\end{equation}
This gives $Q_{\rm{incident}}\simeq5.5\times10^{54}$\,s$^{-1}$.
The incident photon rate from the quasar is thus sufficient to photoionize J1327\_E, with $Q_{\rm{incident}}/Q_{\rm{ionizing}} \simeq 3.6$.
We note that there are several caveats associated with our estimation:
i) The distance between J1327\_E and the quasar is not de-projected.
The physical distance could be larger and require higher ionizing luminosity from the quasar. ii) The \ha\ luminosity is not dust extinction corrected, which will also require higher quasar ionizing luminosity.
iii) The quasar could be more luminous in the past and has now faded, as seen for Hanny's Voorwerp and other AGN-ionized clouds on timescales of ${\sim}10^4$--$10^5$\,yr. \citep{Lintott2009,Schawinski2010,Keel2012b,Keel2012,Schirmer2013}. This will give a higher quasar ionizing luminosity than currently observed. Overall, our calculation suggests that the quasar, in principle, has enough ionizing photons to photoionize the gas within J1327\_E, which naturally explains that J1327\_E is in the AGN region of the BPT diagram. This provides tentative evidence that quasar radiative feedback may be capable of shaping the physical properties of the gas within the merging companion galaxies. Similar radiative feedback but on larger spatial scales is also speculated for a luminous $z \sim$ 6 quasar \citep{Zhu2025d}. }

\section{Conclusion}
\label{sec:conclusion}
In this paper, we present the results from our JWST survey program (Q-IFU) of 27 luminous quasars at $z$ $
\sim$ 4.74 - 4.88 and $z$ $\sim$ 5.66 – 6 with NIRSpec/IFU. We derive and analyze rest-frame optical nuclear properties of these quasars and extended line emission within these systems for the first time. Our main findings are summarized below.

\begin{itemize}
\item 
We determine the systemic redshifts of our objects through the peak of the narrow \oiiitext\ component or the overall \hb\ component when no narrow \oiiitext\ emission line is detected. Our mean and median redshifts $\sim$390 and $\sim$40 \kms\ larger than the rest-frame UV emission line-based redshifts (including \civtext, \ciiitext\ and/or \siivtext) in the literature, respectively, despite the large scatter.

\item 
We build a composite spectrum for our whole sample with median combining, which is overall similar to that of {\textit{S16 sample}} at $z \sim$ 1.5--3.5 sharing the same bolometric luminosity range as our sample. The most obvious difference lies in the \oiiitext\ line emission which is more blueshifted and broader yet weaker in our sample when compared to the {\textit{S16 sample}}. This is consistent with the enhanced frequency of extremely fast outflows in our sample as reported in \citet{Liu2026a}. In addition, while the differences are more subtle with respect to the case of \oiiitext\, the Balmer emission lines are broader and the Fe emission is stronger in our sample.

\item 
We obtain bolometric luminosities of log$($L$_{\rm bol}$/erg s$^{-1}$) $\sim$ 46.8--47.7, \hb-based BH masses of $\log(M_{\mathrm{BH}}/M_\odot) \sim 8.6$--$9.7$ and Eddington ratios of $\sim$ 0.1--2.6 for our sample. Our results confirm that these luminous quasars are powered by actively accreting, $\sim$10$^9$ \msun\ BH holes within $\sim$1 Gyr after the Big Bang. Comparing to the {\textit{S16 sample}} at $z \sim$ 1.5--3.5 within the same bolometric luminosity range, our sample shows larger median BH mass (9.10 vs 9.32) and Eddington ratio (0.86 vs 0.44). Nevertheless, such a difference may be insignificant as the 90\% confidence levels of the median BH masses are 8.90--9.31 vs 9.22--9.51, and those of median Eddington ratios are 0.48–-1.25 vs 0.28–-0.54. In addition, for the two objects in our sample with \mgiitext-based BH masses in the literature, our \hb-based BH masses agree with the \mgiitext-based values.

\item 
The \ha\ emission lines are covered for 12 objects. The FWHM of \hb\ and \ha\ lines agree with each other well and fall within the $\sim$0.1 dex rms scatter of a low-$z$ study ($z\leq0.35$). The mean \ha/\hb\ ratio is $\sim$3.8, close to the value ($\sim$3.5) obtained from the aforementioned low-$z$ study of luminous quasars and the theoretical limit of 3.1. The \hb-based and \ha-based BH masses are broadly consistent with each other, despite a potential small mean offset of $\sim$$+$0.2 dex and a maximum difference of $\sim$0.4 dex. {Nevertheless, this difference can be primarily attributed to the different ``f-factors'' adopted in the two BH mass recipes, and the discrepancy will decrease by 0.125 dex if an ``f-factor'' of 1 is used for both recipes}.

\item 
The dependence of \ewo\ on the quasar luminosities and Eddington ratios in general follow the trends formed by low-$z$ quasars, except that several objects show boosted \ewo, especially in the \ewo\ vs quasar luminosity plane. 
Our objects also generally fall within the same region as occupied by low-$z$ quasars in the Eigenvector 1 planes. In the \ewo\ vs \RFe\ plane, there are three objects with higher \ewo\ than the general trend. In both cases, the \ewo\ excesses seen in certain objects of our sample are primarily related to the emission from the fast outflows.

\item 
We see no evidence of dual/binary quasars and lensed quasars in our shallow survey. We have uncovered extended line emission in 6 out of the 27 objects after PSF subtraction, which should be deemed as a lower limit given the shallowness of our IFU data. The extended emission shows asymmetric morphologies in all 6 objects to varying degrees, with absolute velocities $\lesssim$300 \kms\ and velocity dispersions of $\sim$200--500 \kms. These characteristics may trace late-stage merger activities, turbulent and clumpy ISM, and/or companion galaxies.

\item
Among them, J1327$+$5732 exhibits the most extended line-emitting region (up to $\sim$5 kpc radially). The gas in the vicinity of the quasar shows a velocity gradient potentially tracing gas rotation. We discover a distinct line emitting region, J1327\_E, {which is $\sim$2.5 kpc away from the quasar in projection and blueshifted by $\sim$20 \kms\ with respect to the quasar redshift. This source likely traces a merging companion galaxy. It may host an AGN, thereby providing an example of quasar-AGN pair candidates at $z>5$. Alternatively, it could also be photoionized by the quasar radiation, providing tentative evidence of on-going quasar radiative feedback shaping the ISM properties of a merging companion galaxy. }

\end{itemize}

\begin{acknowledgments}
W.L. acknowledges funding from the JWST Arizona/Steward Postdoc in Early galaxies and Reionization (JASPER) Scholar contract at the University of Arizona. 
W.L. acknowledges support from NASA through STScI grant JWST-Survey-3428.
This work is based on observations made with the NASA/ESA/CSA James Webb Space Telescope. The data were obtained from the Mikulski Archive for Space Telescopes at the Space Telescope Science Institute, which is operated by the Association of Universities for Research in Astronomy, Inc., under NASA contract NAS 5-03127 for JWST. These observations are associated with programs \#3428. Support for these programs was given through a grant from the Space Telescope Science Institute, which is operated by the Association of Universities for Research in Astronomy, Inc., under NASA contract NAS 5-03127.
\end{acknowledgments}

\begin{contribution}
W.L. conceived the project, carried out the data reduction and analysis, and wrote the manuscript. He is also the PI of the JWST survey program \#3428 where the data come from. X.F. helped revise the manuscript and provided important suggestions for data analysis and interpretation. All authors reviewed and/or commented on the manuscript.
\end{contribution}

\appendix

\section{Spectra and Best-fits of Individual Objects}
\label{sec:appendix}
The spectra and their best-fits covering the \hg, \hg, and \oiiitext\ regions of our sample are shown in Fig. \ref{fig:specfit}, except for the 6 objects with the most extreme outflows {(J0759$+$1800, J0829$+$0303, J0840$+$5624, J0859$+$2520, J1612$+$5202)} that are presented in \citet{Liu2026a}. Those of the \ha\ regions for the 12 objects with corresponding spectral coverages are shown in Fig. \ref{fig:haexample}.

\begin{figure*}[!ht]
\begin{minipage}[t]{0.5\textwidth}

\centering
\includegraphics[width=\textwidth]{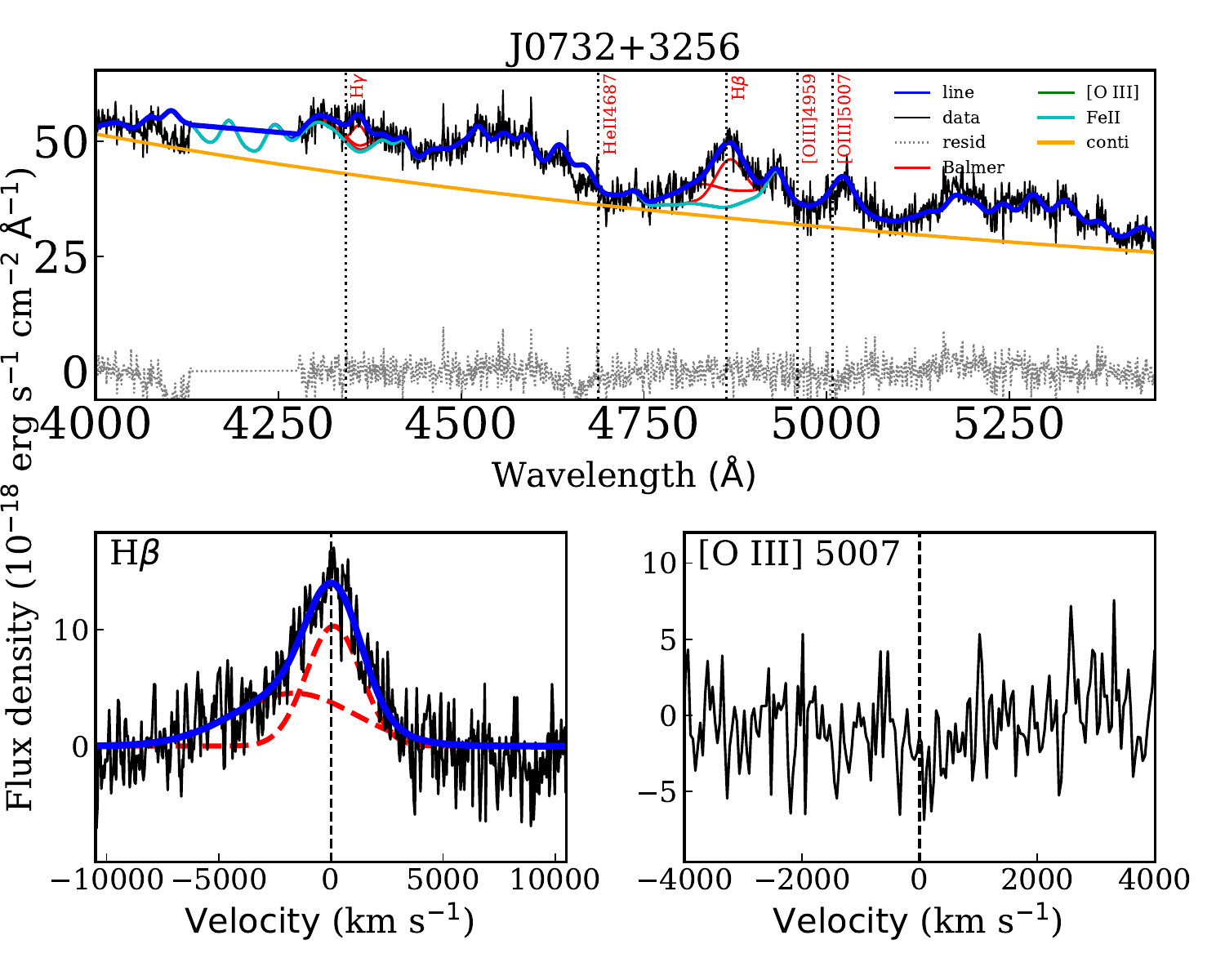}
\end{minipage}
\begin{minipage}[t]{0.5\textwidth}
\centering
\includegraphics[width=\textwidth]{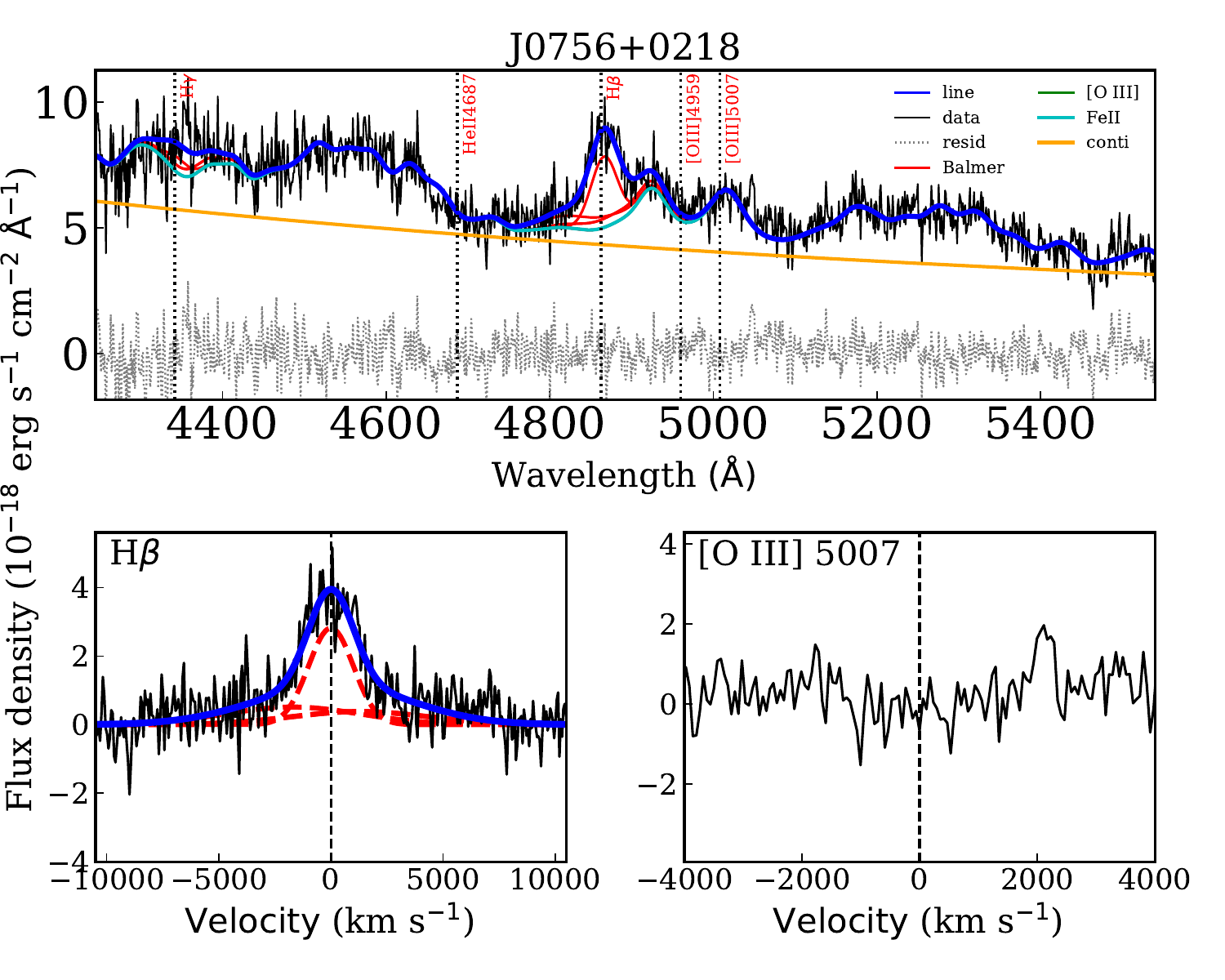}
\end{minipage}

\begin{minipage}[t]{0.5\textwidth}
\centering
\includegraphics[width=\textwidth]{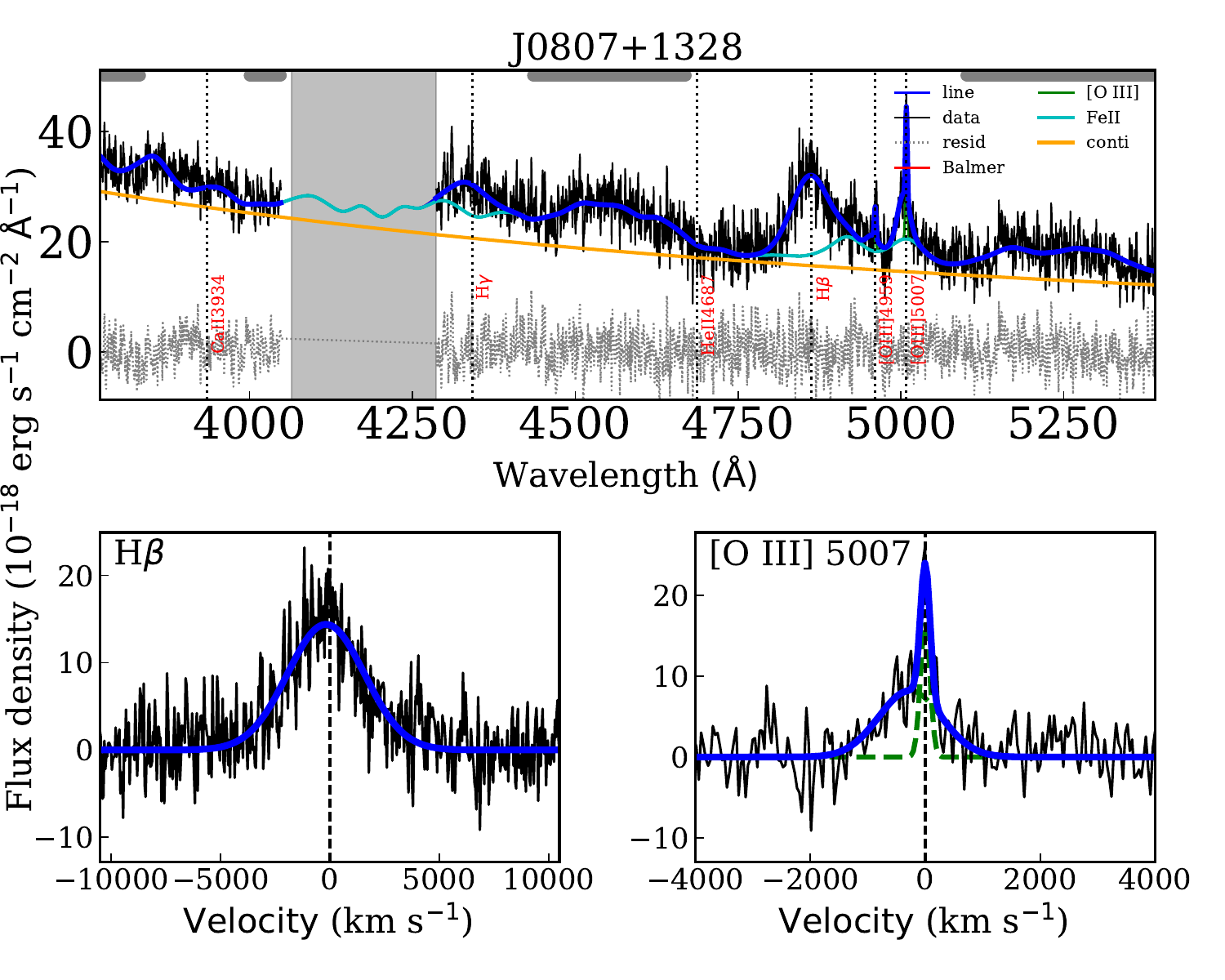}
\end{minipage}
\begin{minipage}[t]{0.5\textwidth}
\centering
\includegraphics[width=\textwidth]{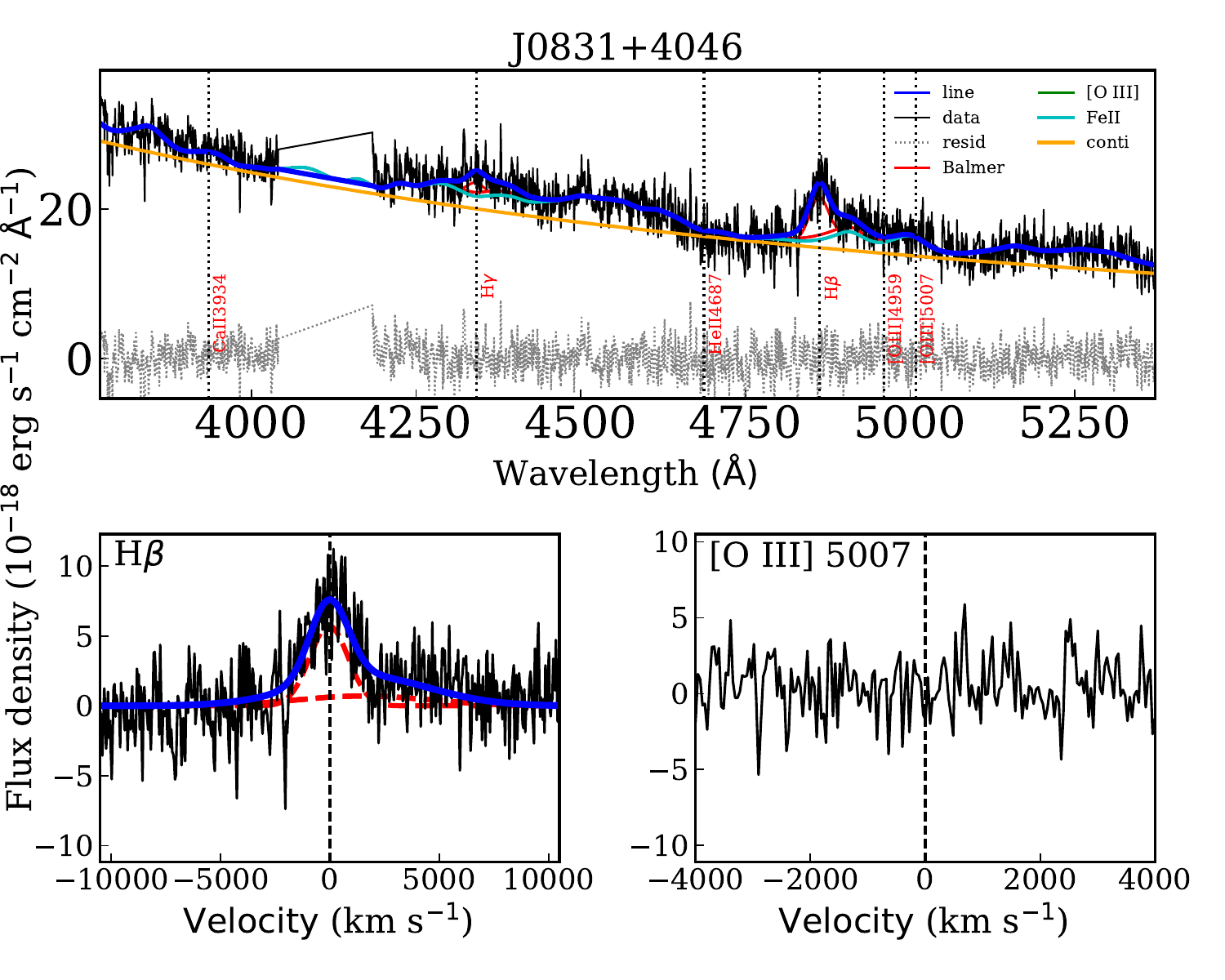}
\end{minipage}

\begin{minipage}[t]{0.5\textwidth}
\centering
\includegraphics[width=\textwidth]{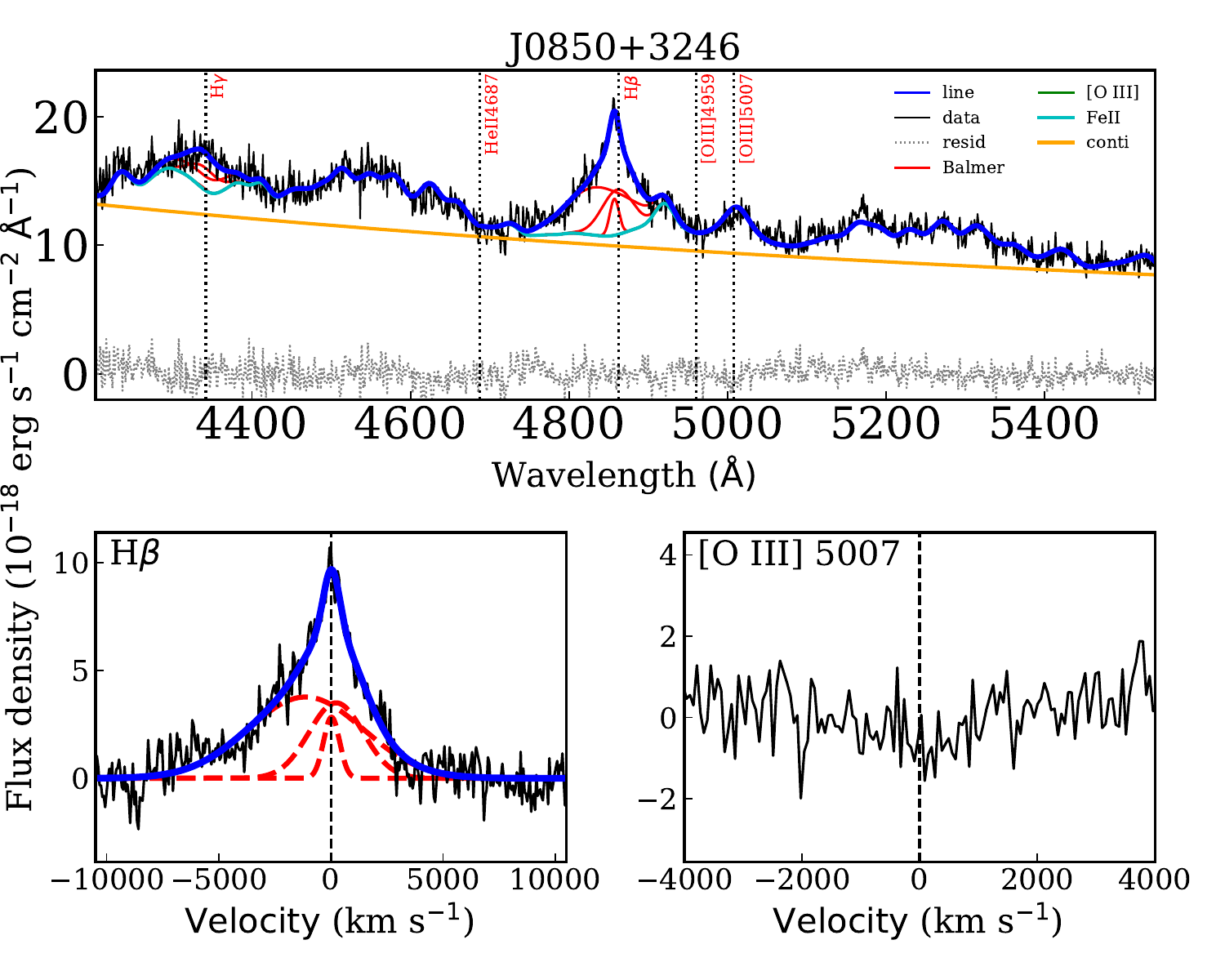}
\end{minipage}
\begin{minipage}[t]{0.5\textwidth}
\centering
\includegraphics[width=\textwidth]{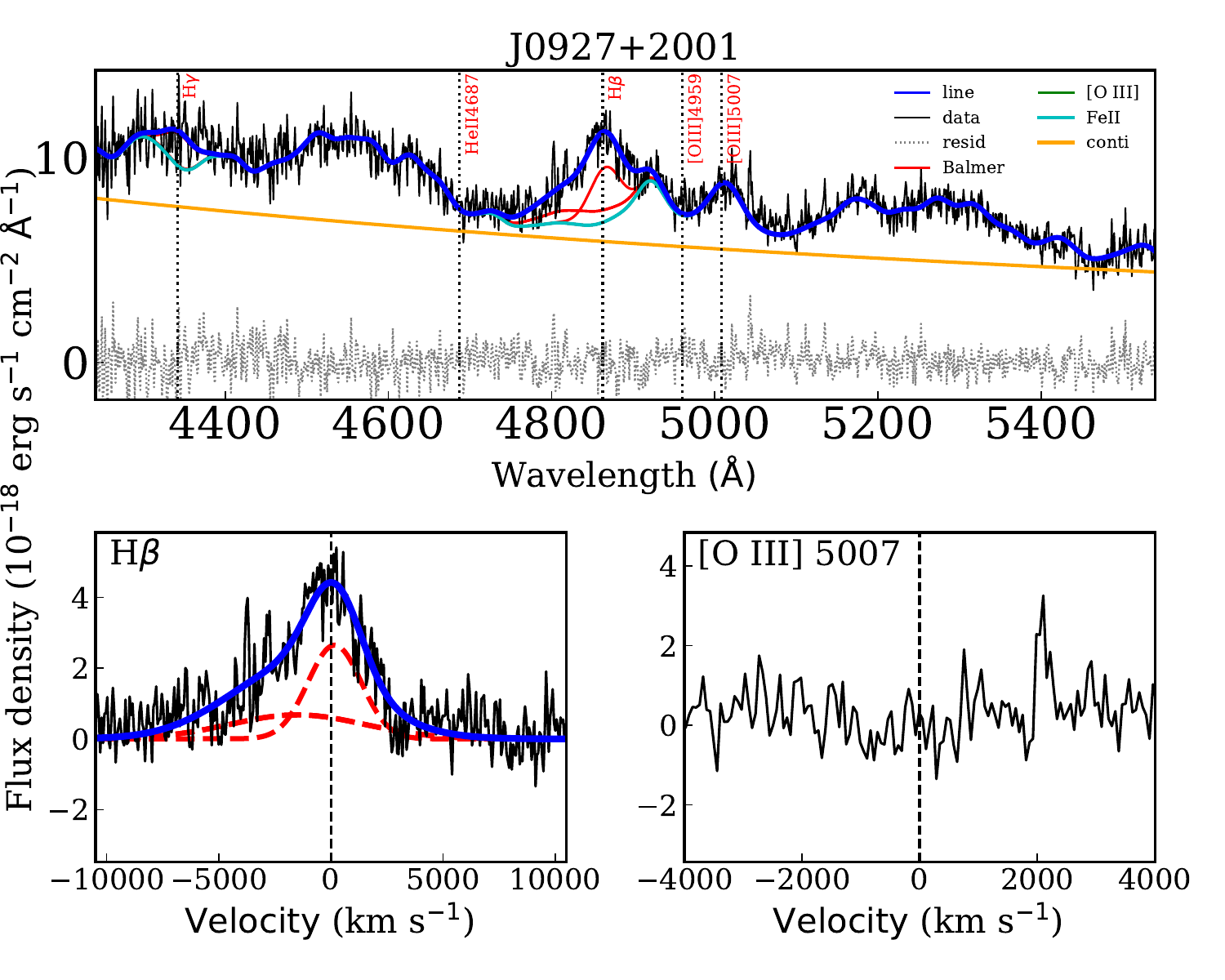}
\end{minipage}
\caption{For each frame of an individual object: \textbf{Top panel:} JWST spectrum (black), best-fit emission line profiles (blue), Fe emission (cyan), continuum (orange), and residual (gray dotted line). The best-fit individual Gaussian components for \hb\ and \hg\ are shown in red and those for \oiiiab\ are shown in green. Systemic velocities of individual emission lines are shown in vertical black dotted lines. The spectral windows adopted for fitting the quasar pseudo continuum are marked by the gray thick bars. The detector gap and adjacent noisy regions not used in the fitting are masked by the vertical gray shaded region. \textbf{Bottom panels:} \hb\ (left) and \oiii\ (right) line profiles with their best-fit models (blue solid lines) and individual components (dashed lines). For \oiiitext, only the residuals are shown when the line is not detected.}
\label{fig:specfit}
\end{figure*}

\begin{figure*}
\figurenum{11}
\begin{minipage}[t]{0.5\textwidth}
\centering
\includegraphics[width=\textwidth]{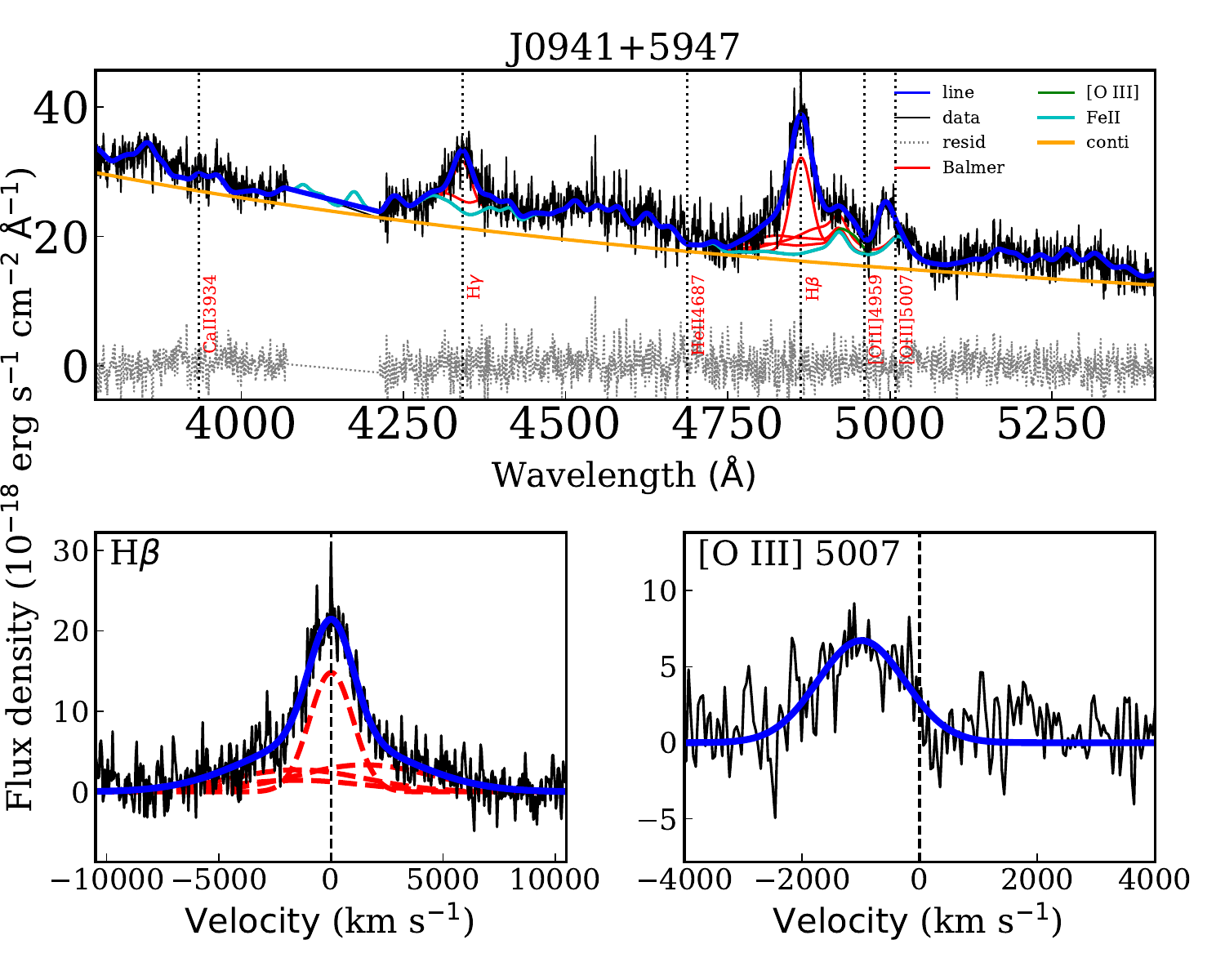}
\end{minipage}
\begin{minipage}[t]{0.5\textwidth}
\centering
\includegraphics[width=\textwidth]{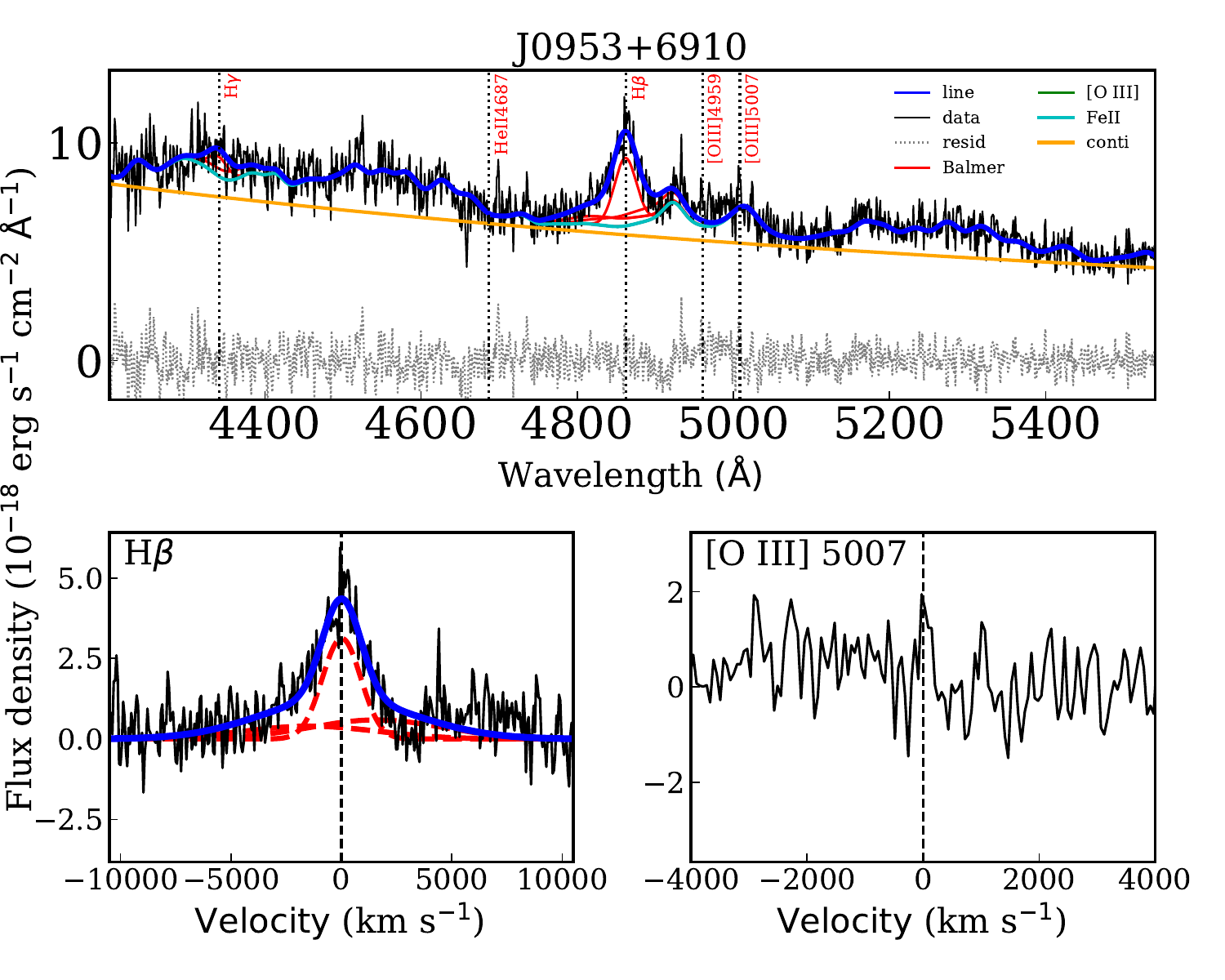}
\end{minipage}

\begin{minipage}[t]{0.5\textwidth}
\centering
\includegraphics[width=\textwidth]{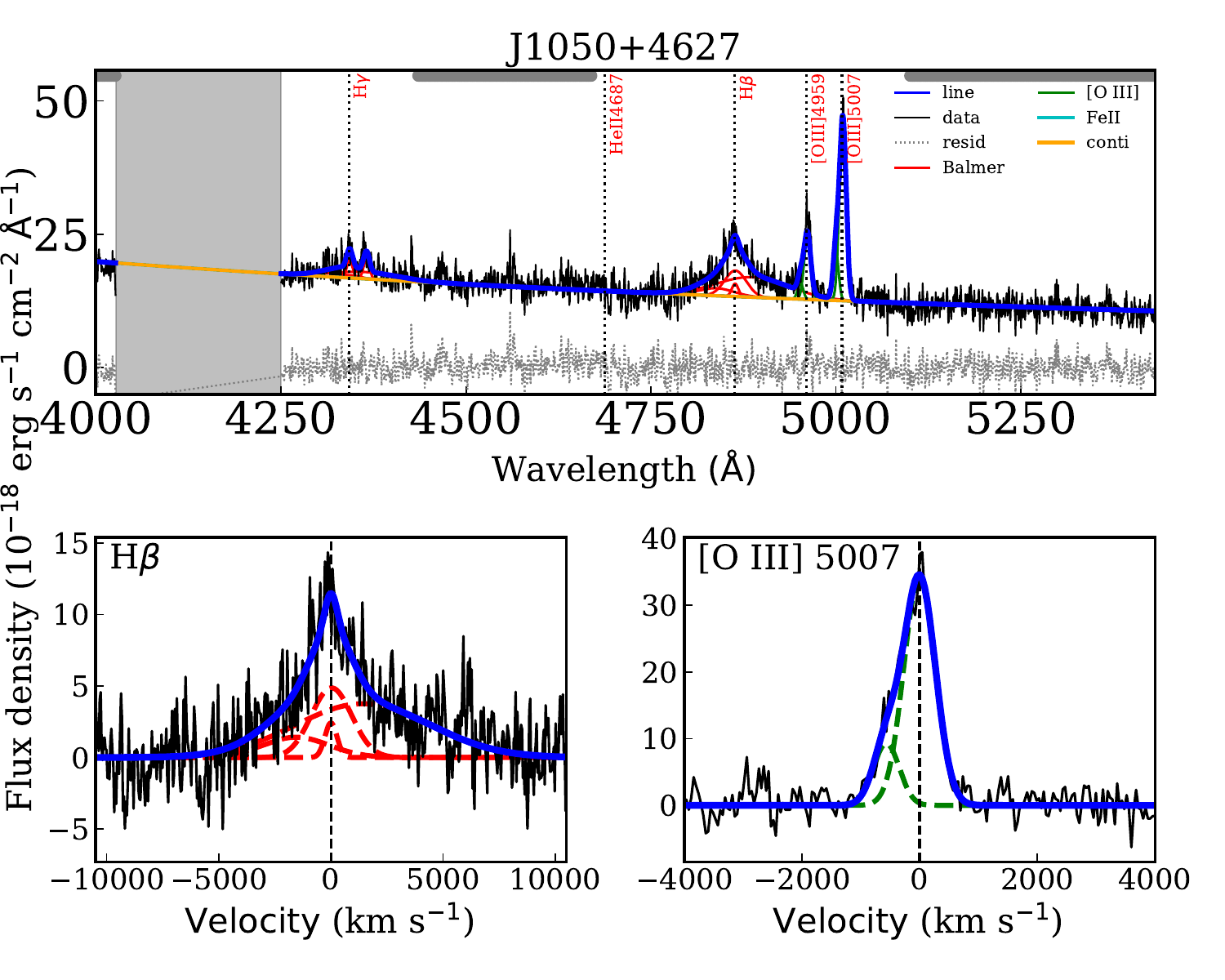}
\end{minipage}
\begin{minipage}[t]{0.5\textwidth}
\centering
\includegraphics[width=\textwidth]{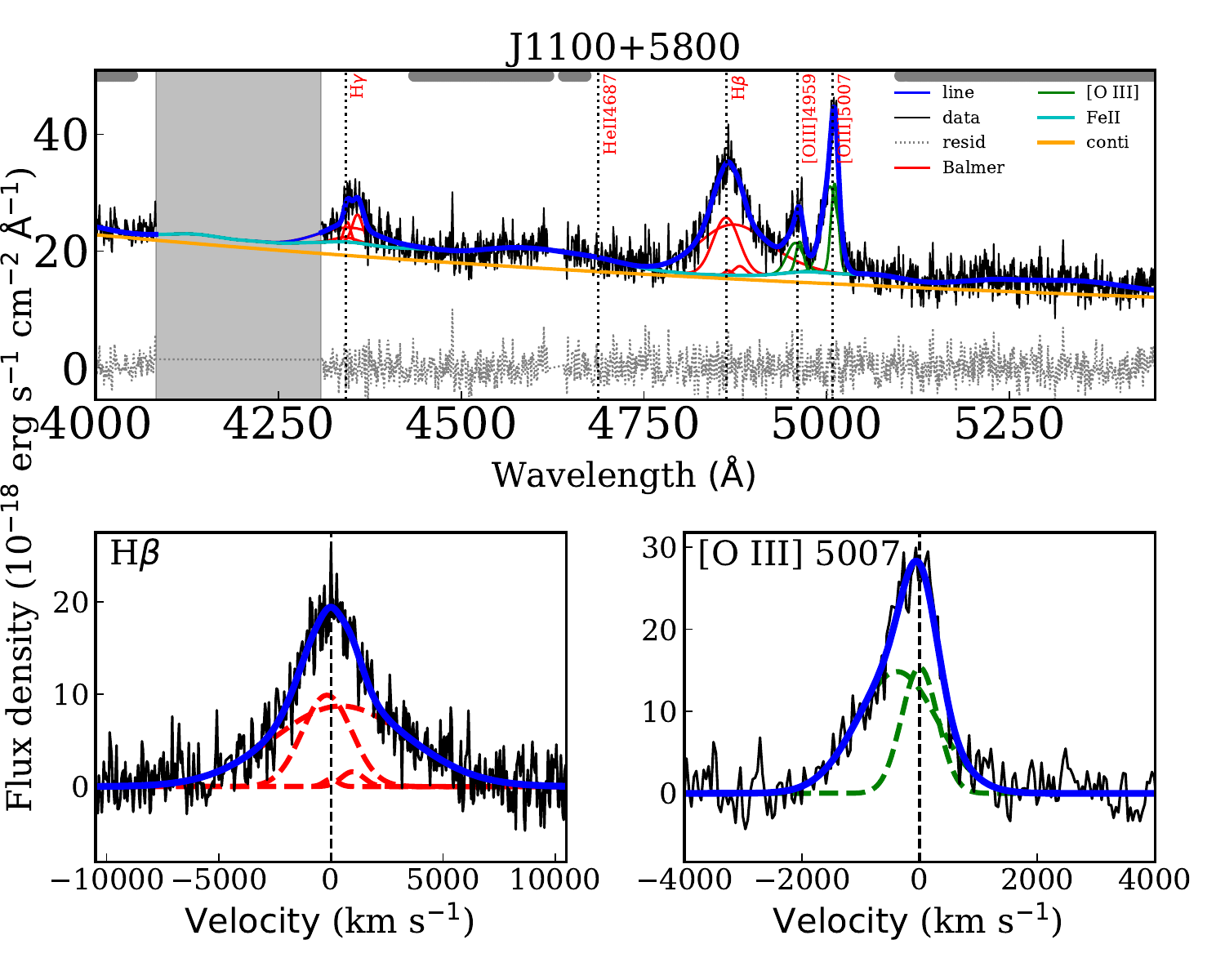}
\end{minipage}

\begin{minipage}[t]{0.5\textwidth}
\figurenum{11}
\centering
\includegraphics[width=\textwidth]{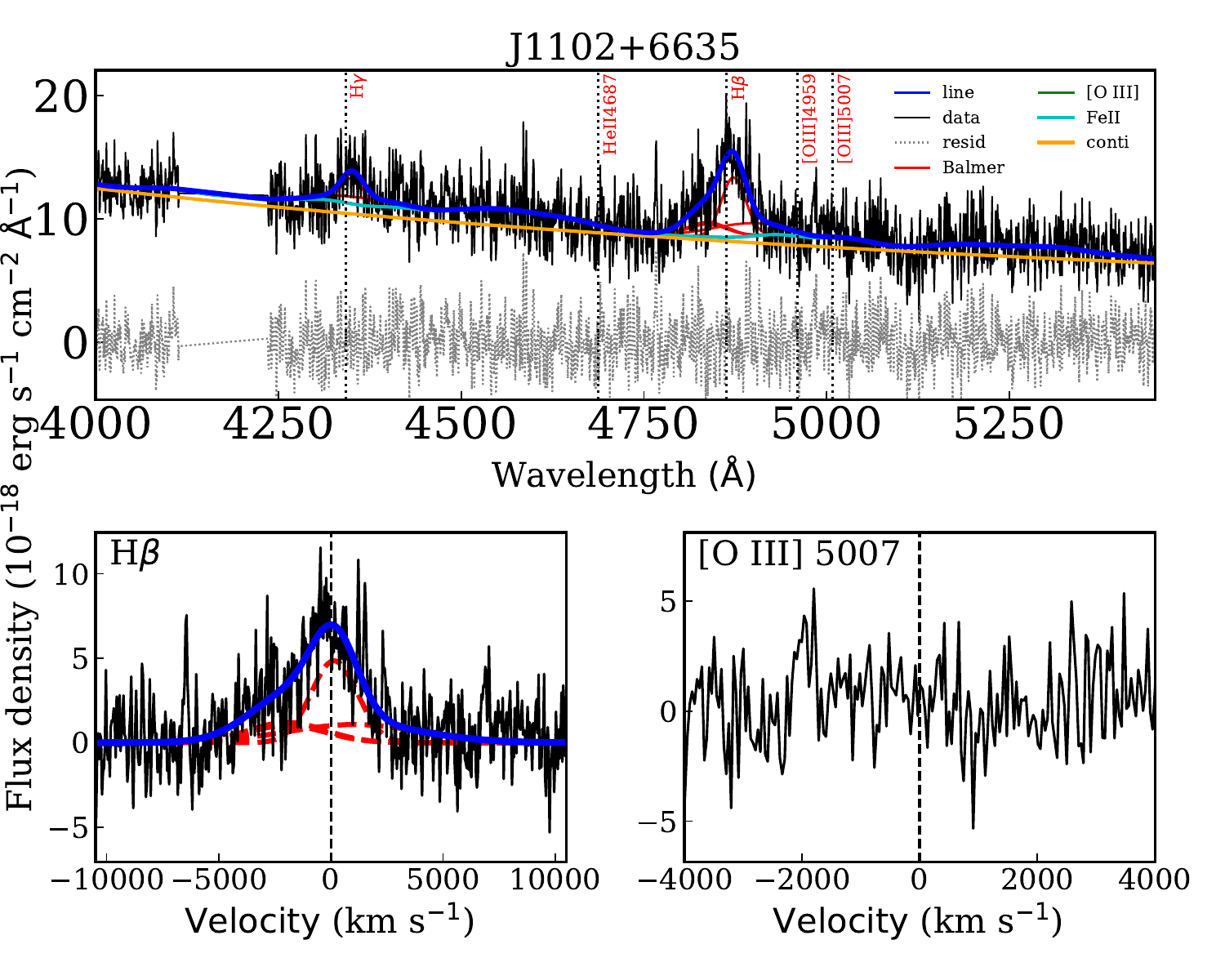}
\end{minipage}
\begin{minipage}[t]{0.5\textwidth}
\centering
\includegraphics[width=\textwidth]{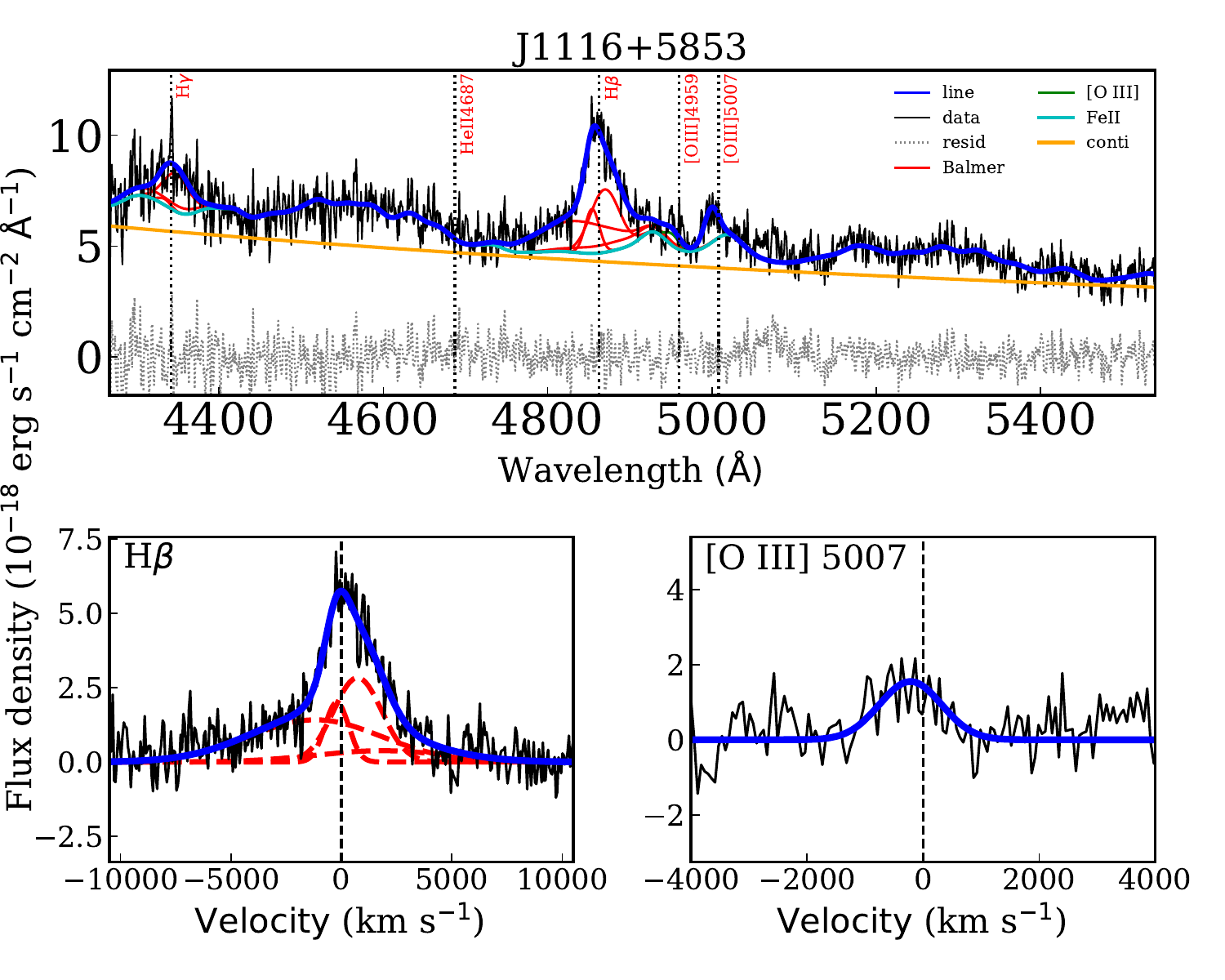}
\end{minipage}
\caption{Continued}
% \caption{Same as Fig. \ref{fig:specfit} but for objects: }
%\label{fig:specfit2}
\end{figure*}

\begin{figure*}[!ht]
\figurenum{11}
\begin{minipage}[t]{0.5\textwidth}
\centering
\includegraphics[width=\textwidth]{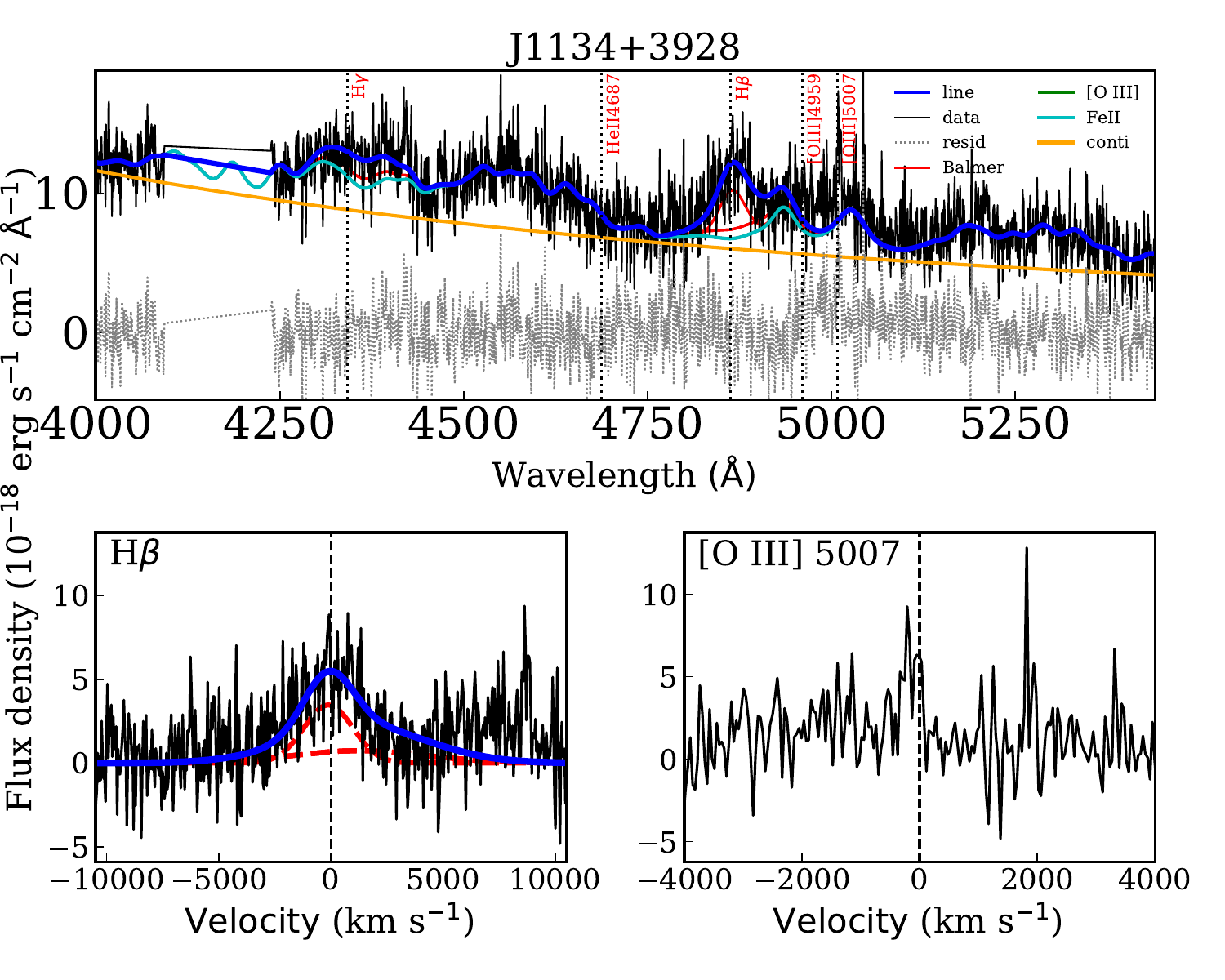}
\end{minipage}
\begin{minipage}[t]{0.5\textwidth}
\centering
\includegraphics[width=\textwidth]{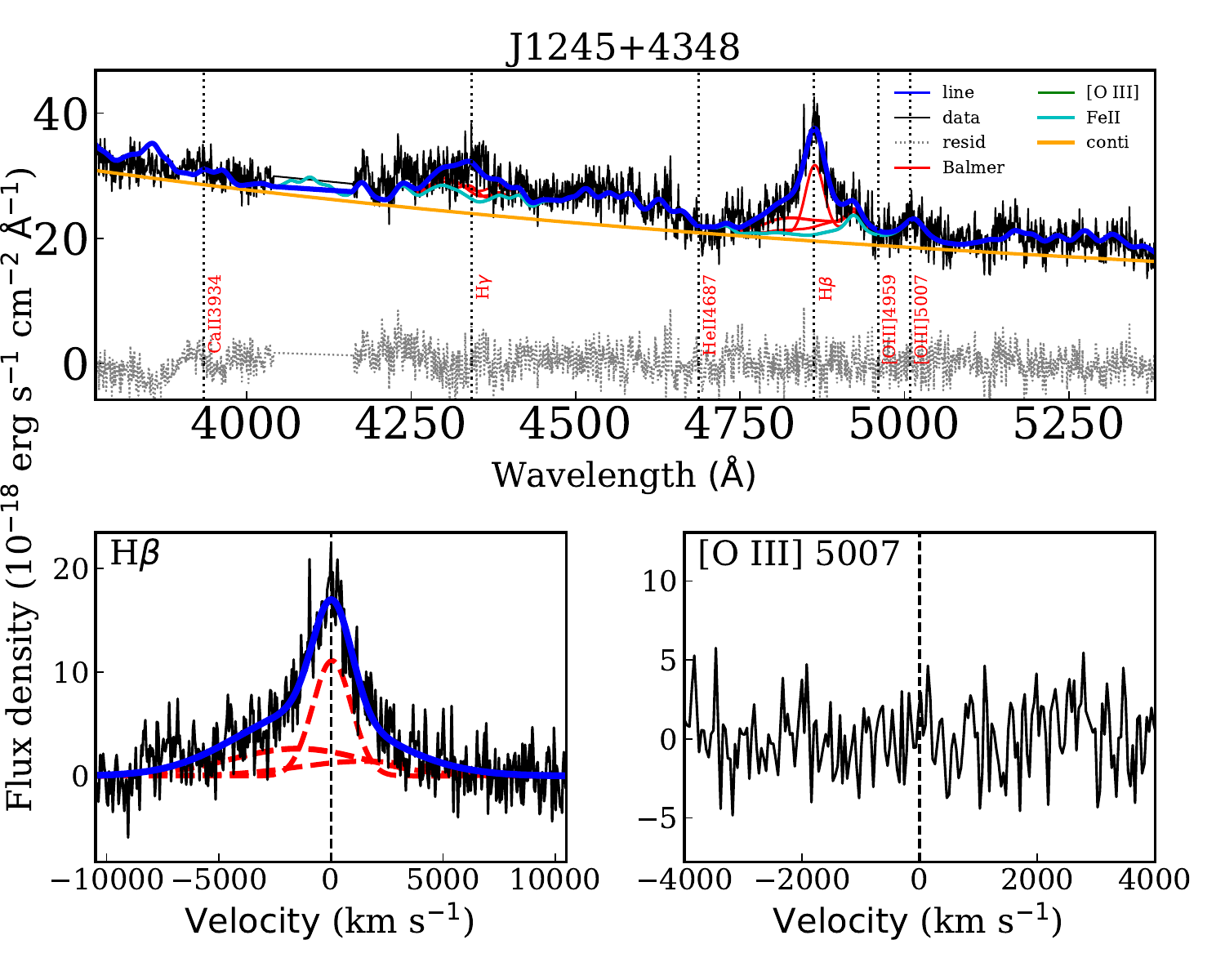}
\end{minipage}

\begin{minipage}[t]{0.5\textwidth}
\centering
\includegraphics[width=\textwidth]{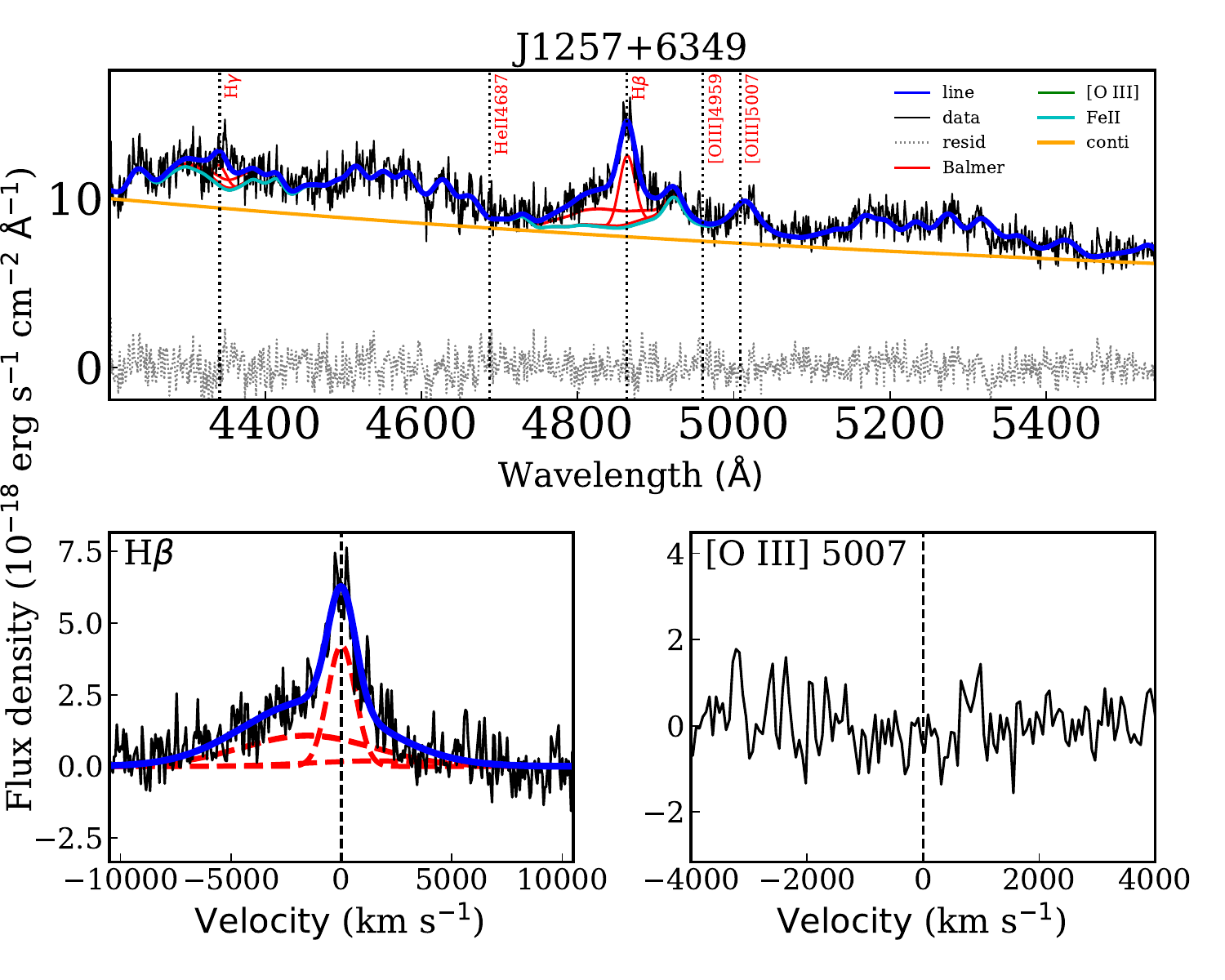}
\end{minipage}
\begin{minipage}[t]{0.5\textwidth}
\centering
\includegraphics[width=\textwidth]{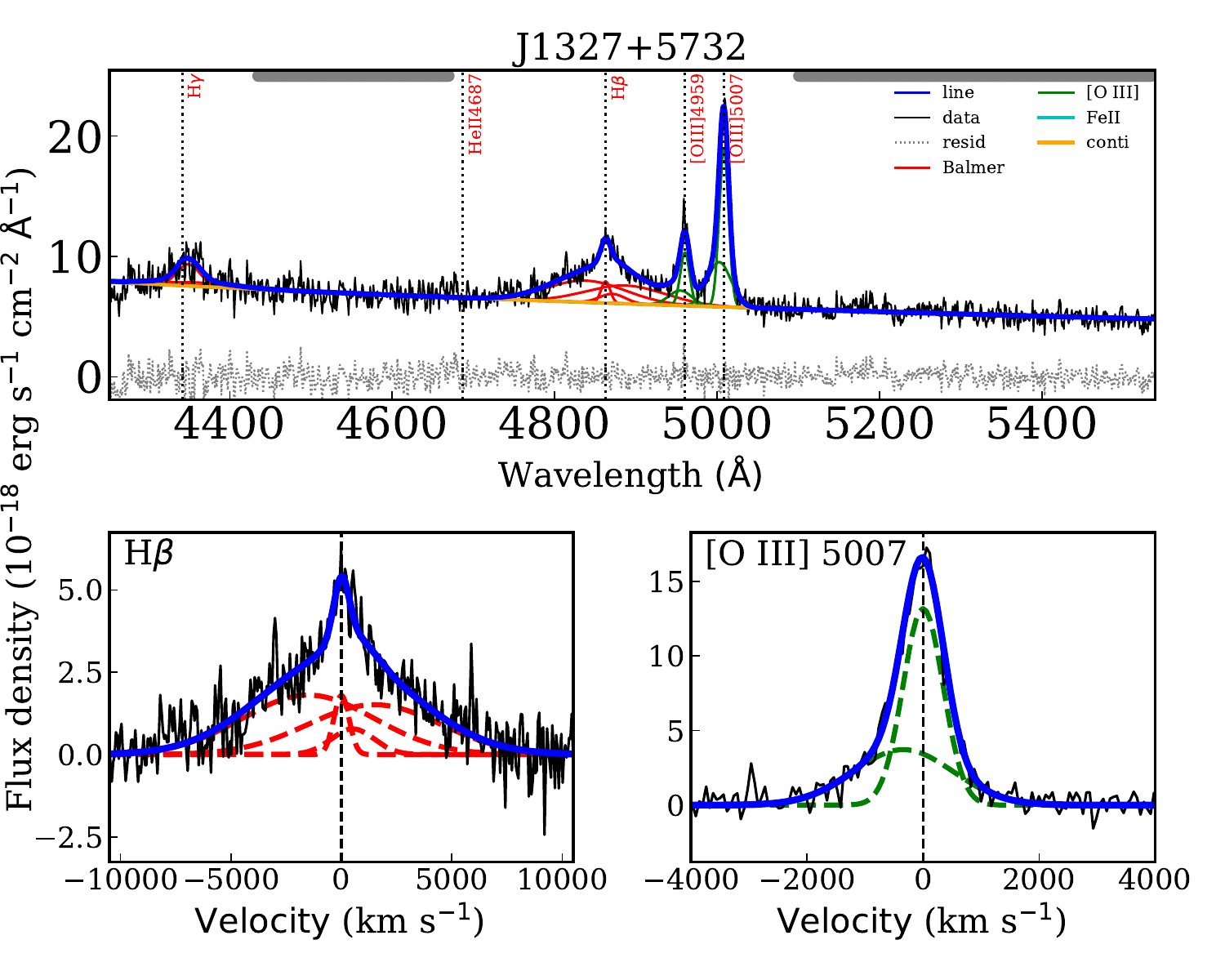}
\end{minipage}

\begin{minipage}[t]{0.5\textwidth}
\centering
\includegraphics[width=\textwidth]{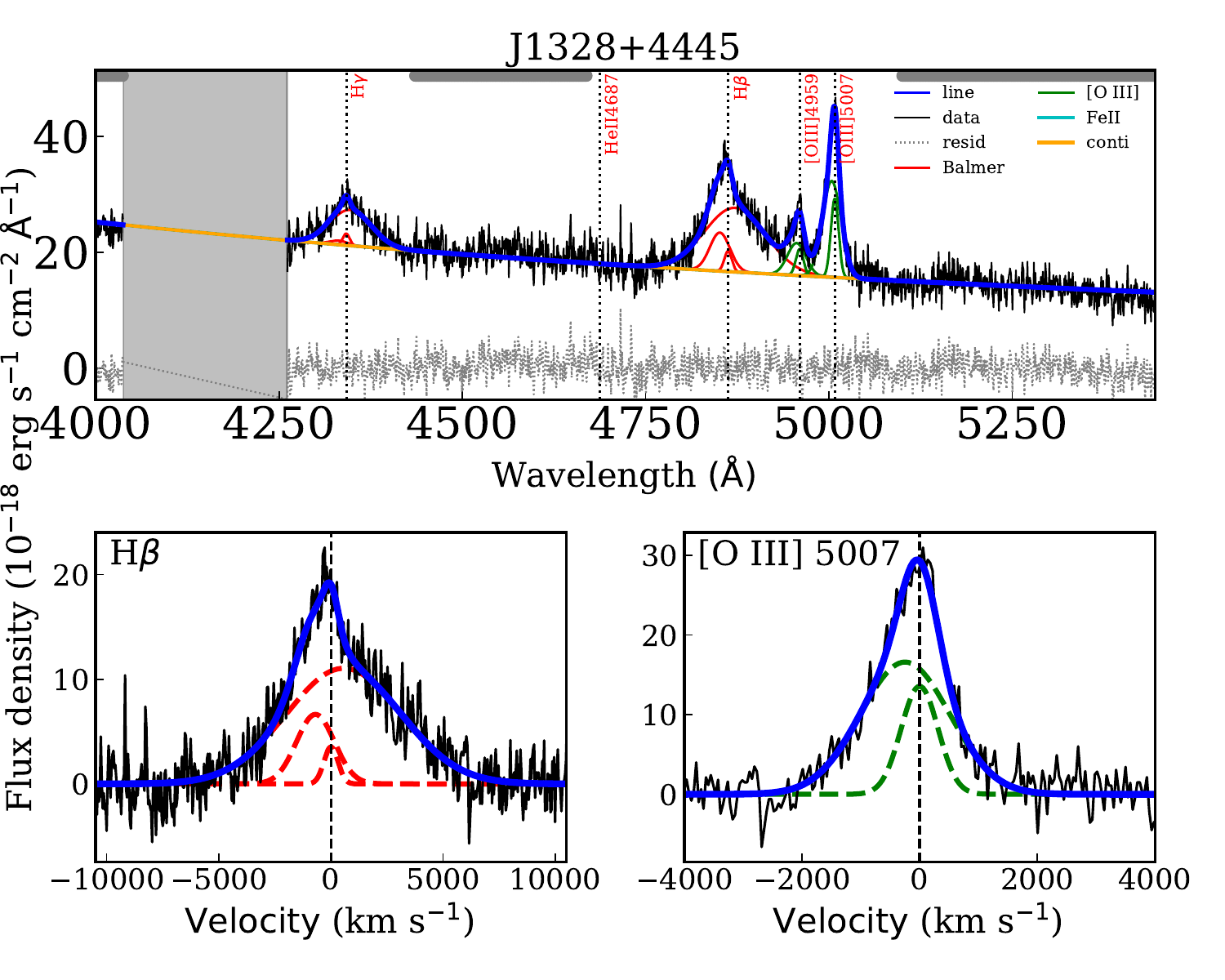}
\end{minipage}
\begin{minipage}[t]{0.5\textwidth}
\centering
\includegraphics[width=\textwidth]{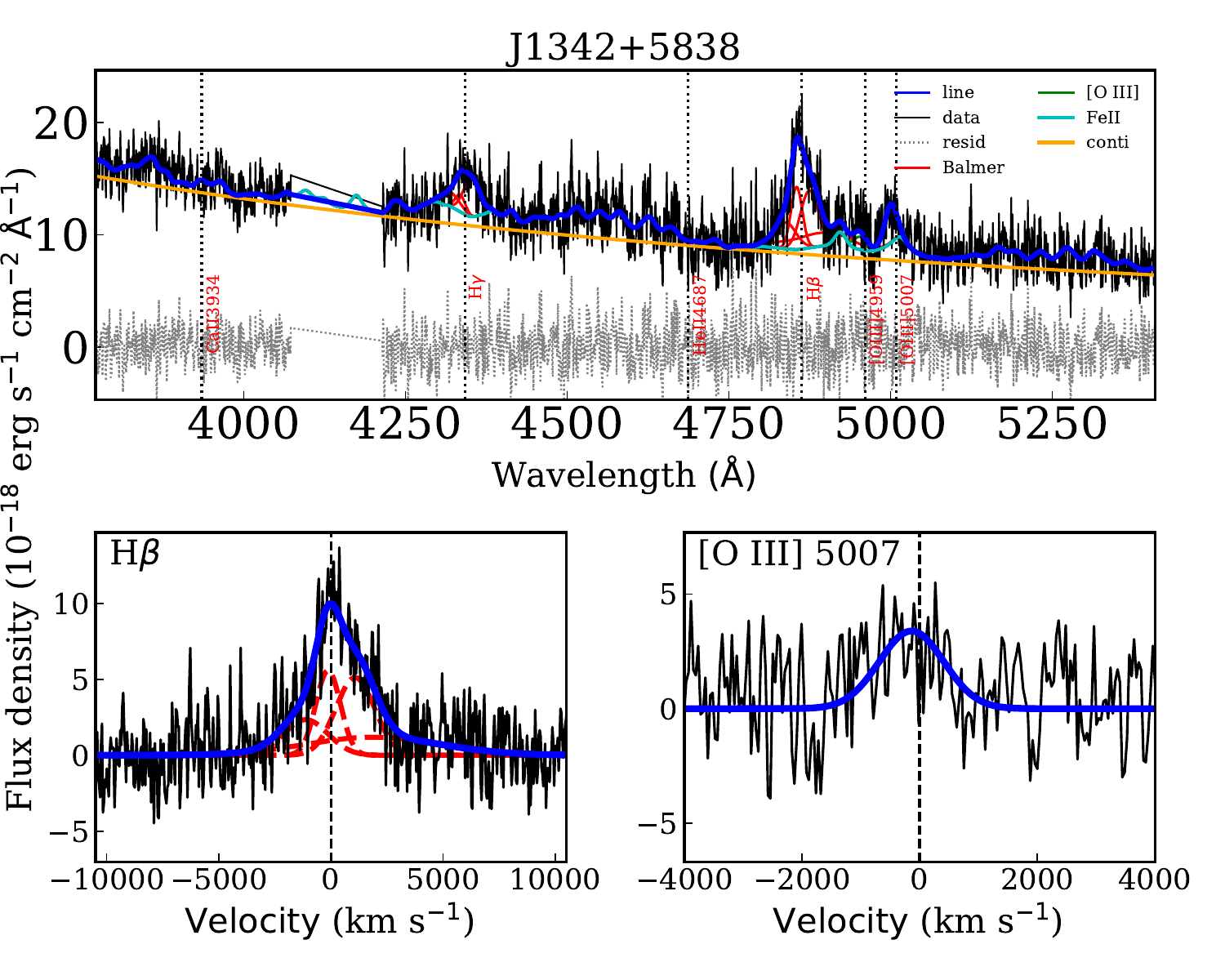}
\end{minipage}
\caption{Continued}
%\label{fig:specfit3}
\end{figure*}

\begin{figure*}
\figurenum{11}
\begin{minipage}[t]{0.5\textwidth}
\centering
\includegraphics[width=\textwidth]{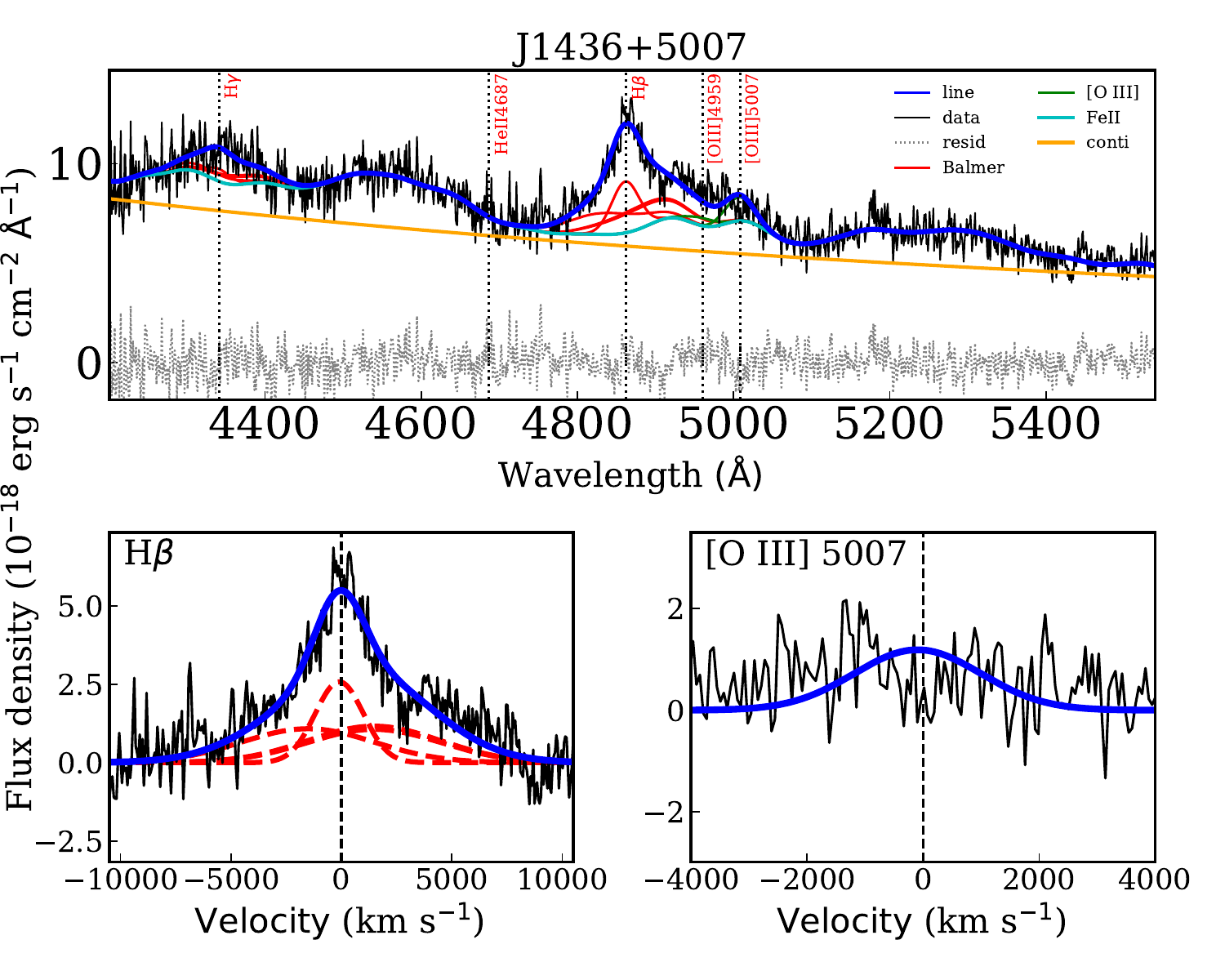}
\end{minipage}
\begin{minipage}[t]{0.5\textwidth}
\centering
\includegraphics[width=\textwidth]{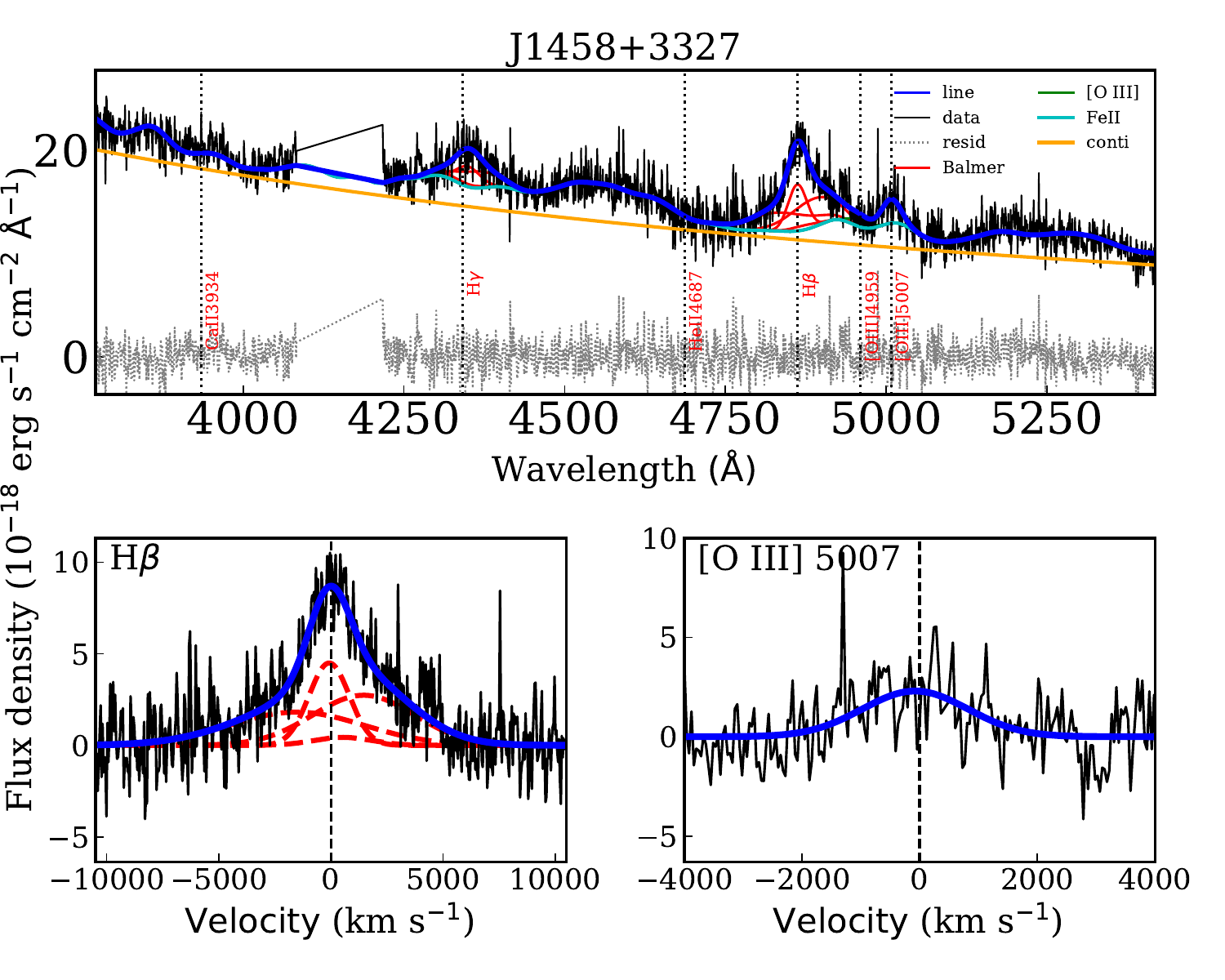}
\end{minipage}

\begin{minipage}[t]{0.5\textwidth}
\centering
\includegraphics[width=\textwidth]{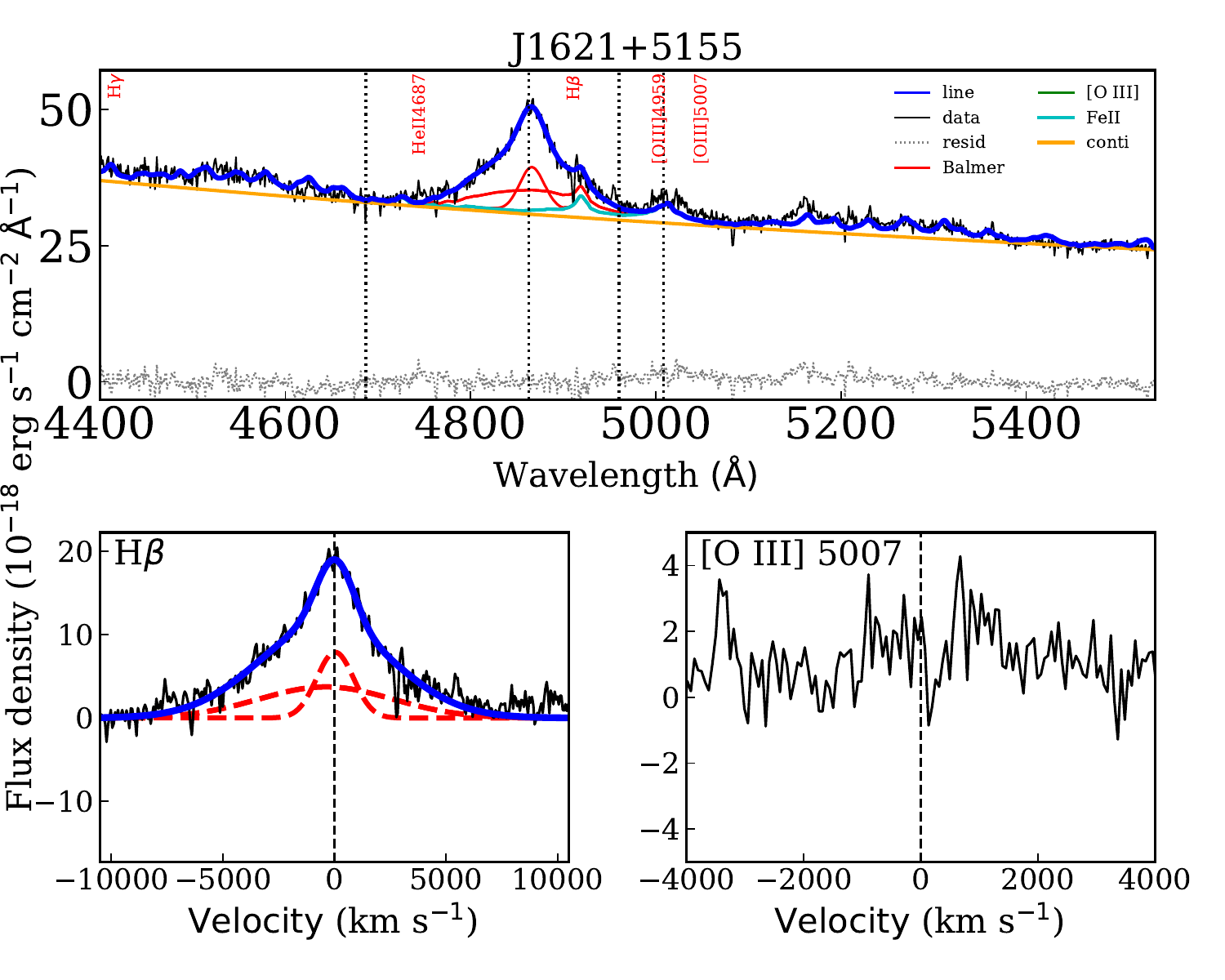}
\end{minipage}
\caption{Continued}
\end{figure*}

\begin{figure*}[!ht]
\figurenum{12}

\begin{minipage}[t]{0.5\textwidth}
\centering
\includegraphics[width=\textwidth]{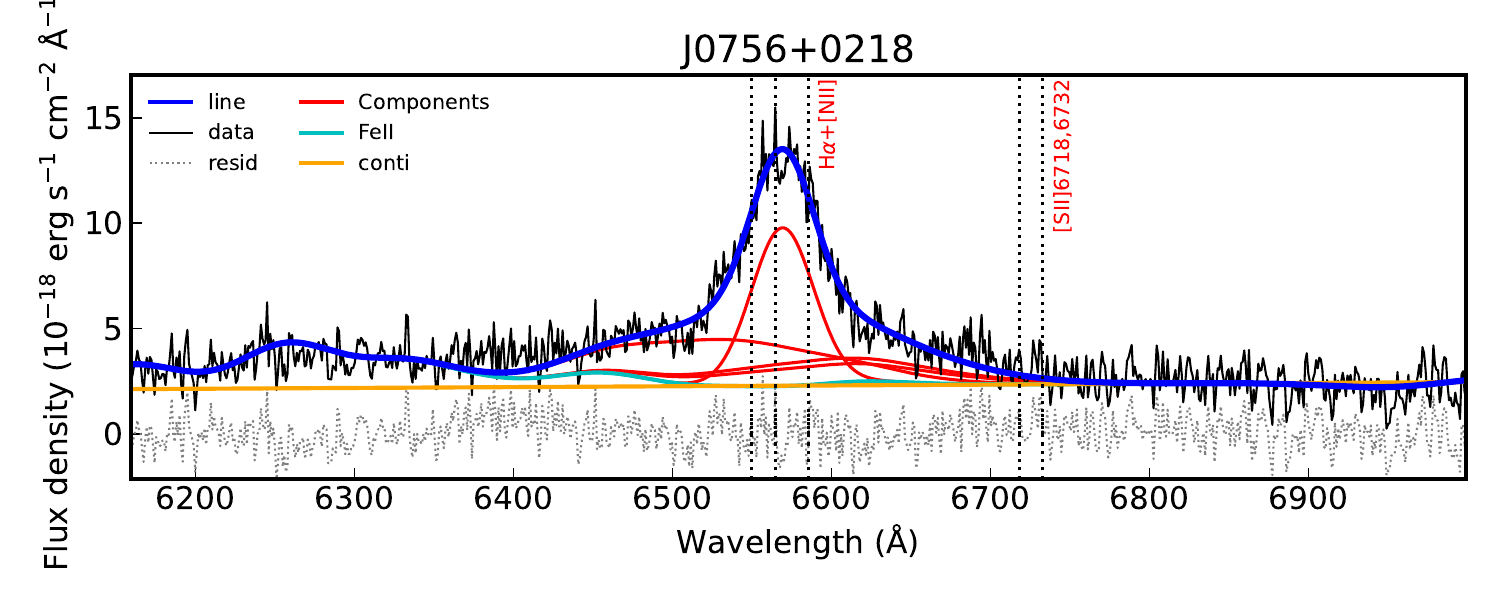}
\end{minipage}
\begin{minipage}[t]{0.5\textwidth}
\centering
\includegraphics[width=\textwidth]{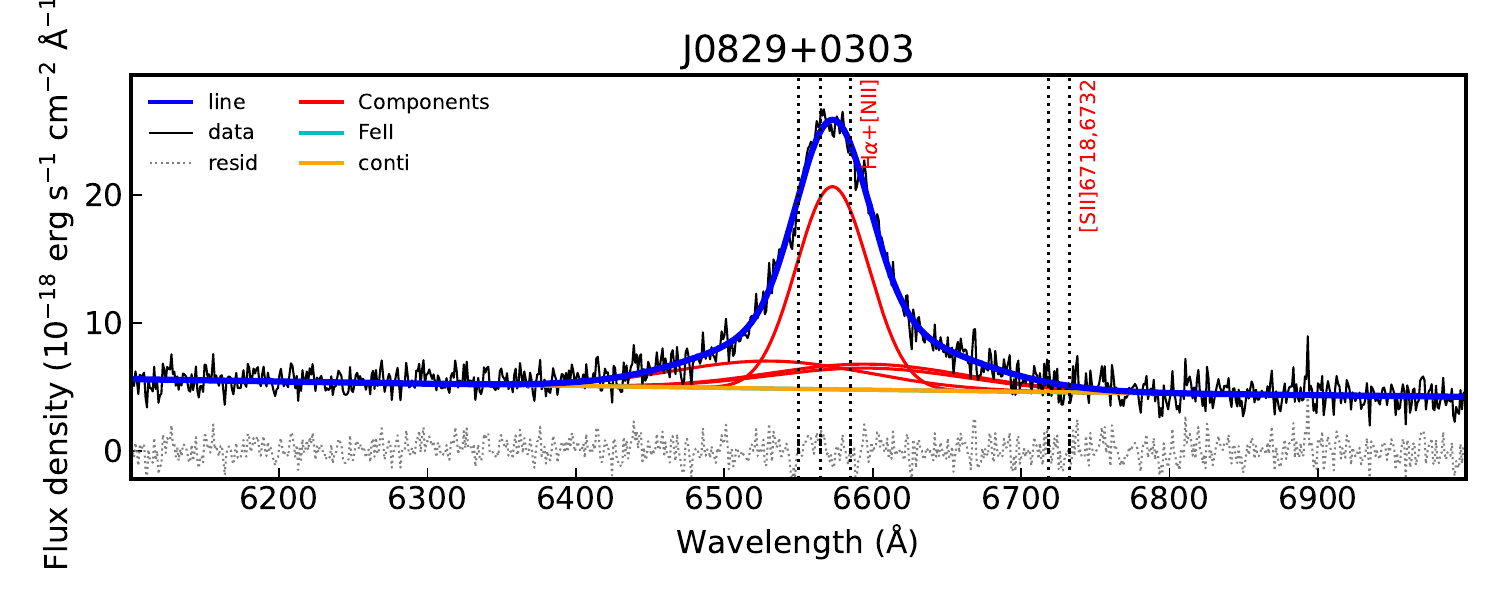}
\end{minipage}

\begin{minipage}[t]{0.5\textwidth}
\centering
\includegraphics[width=\textwidth]{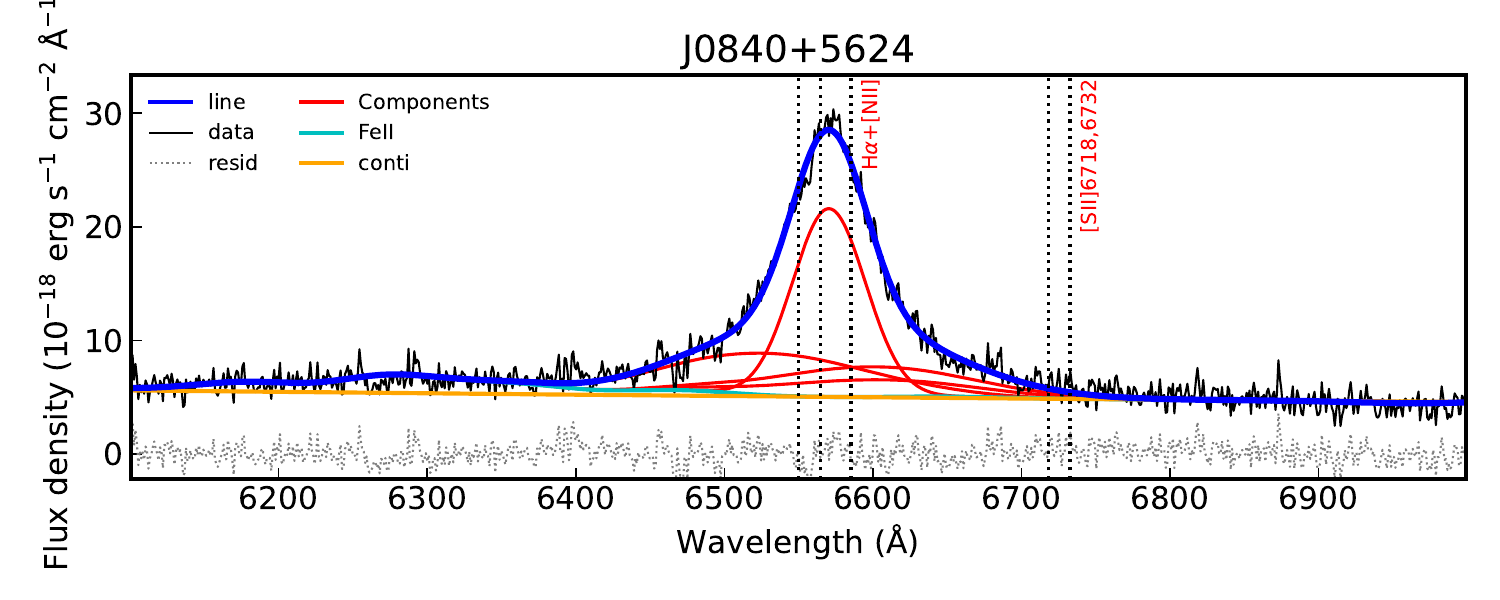}
\end{minipage}
\begin{minipage}[t]{0.5\textwidth}
\centering
\includegraphics[width=\textwidth]{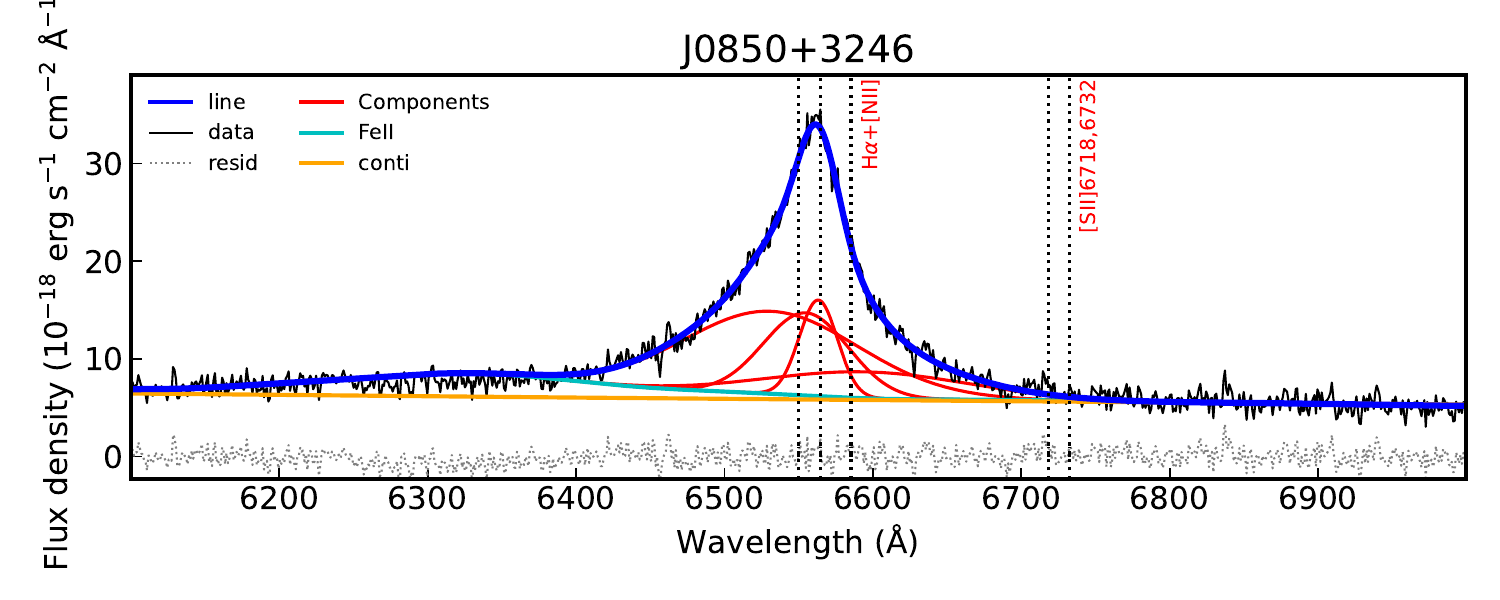}
\end{minipage}

\begin{minipage}[t]{0.5\textwidth}
\centering
\includegraphics[width=\textwidth]{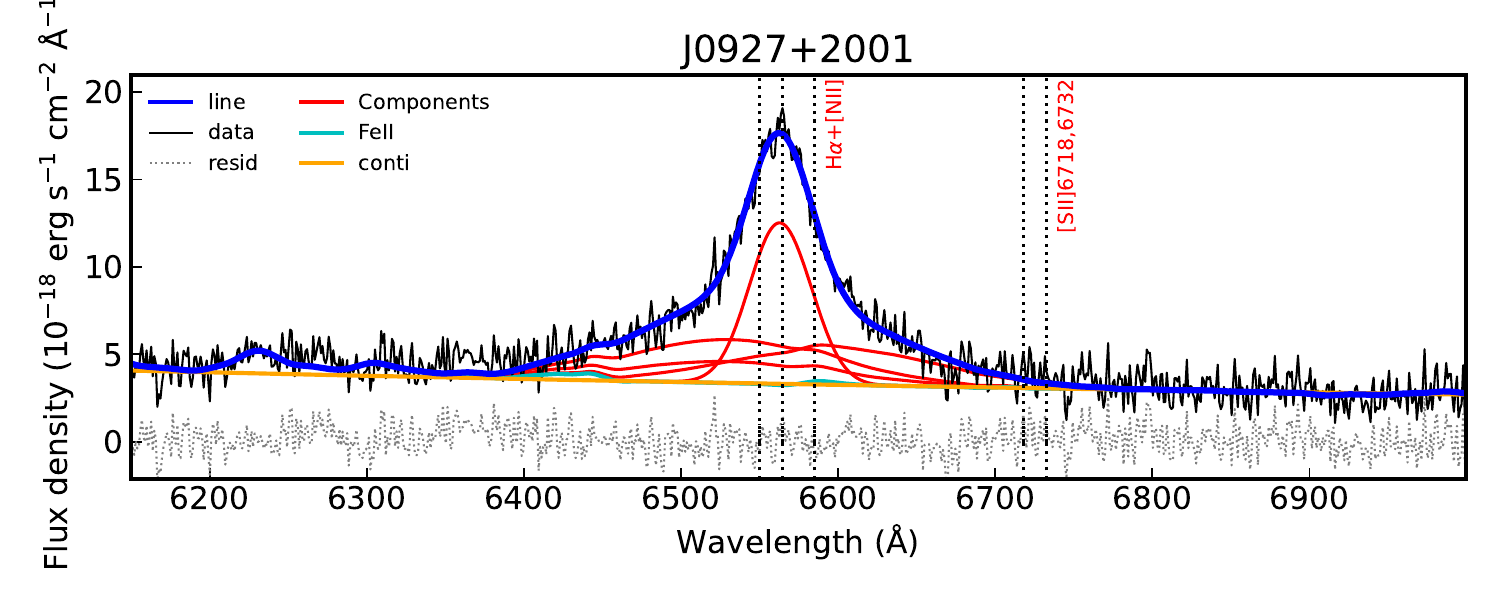}
\end{minipage}
\begin{minipage}[t]{0.5\textwidth}
\centering
\includegraphics[width=\textwidth]{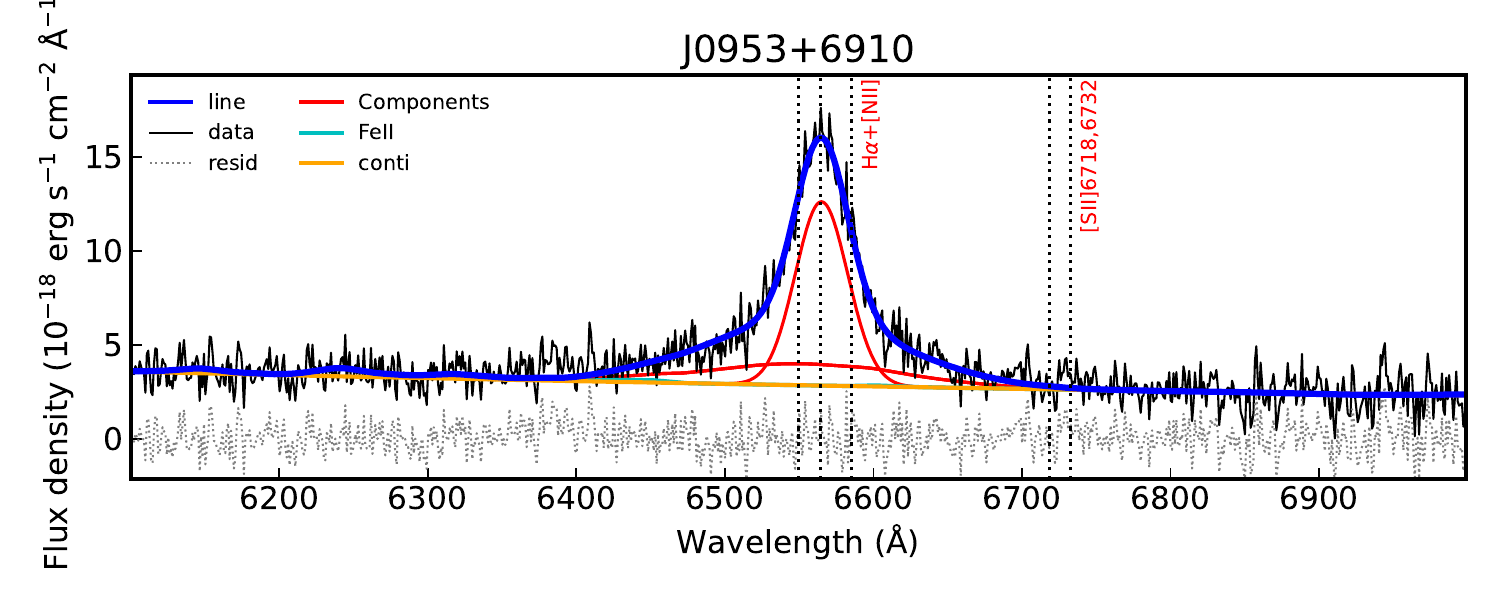}
\end{minipage}

\begin{minipage}[t]{0.5\textwidth}
\centering
\includegraphics[width=\textwidth]{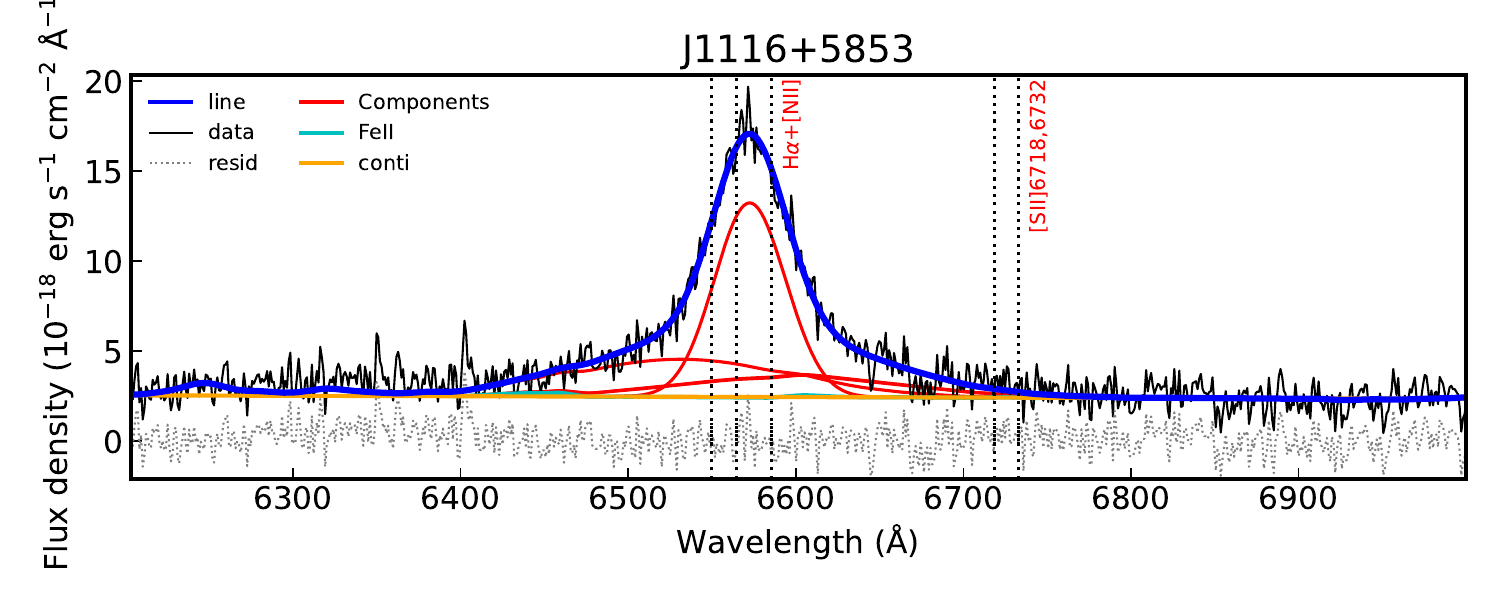}
\end{minipage}
\begin{minipage}[t]{0.5\textwidth}
\centering
\includegraphics[width=\textwidth]{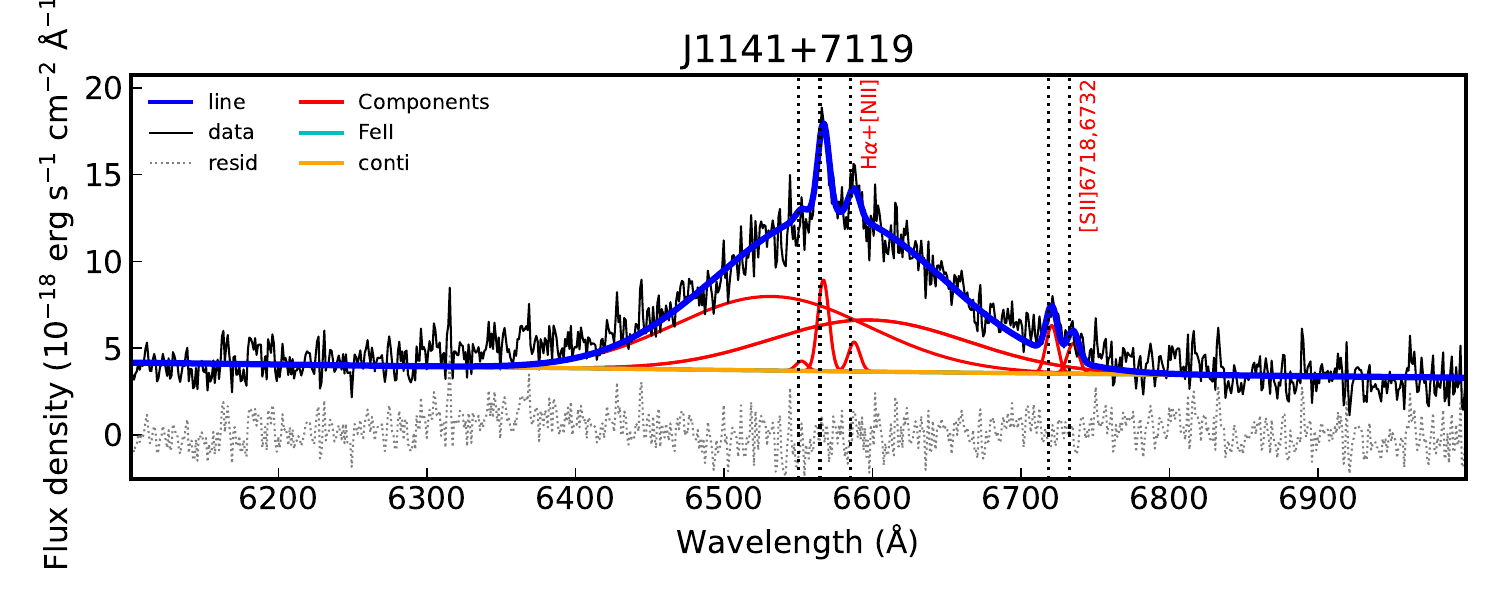}
\end{minipage}

\begin{minipage}[t]{0.5\textwidth}
\centering
\includegraphics[width=\textwidth]{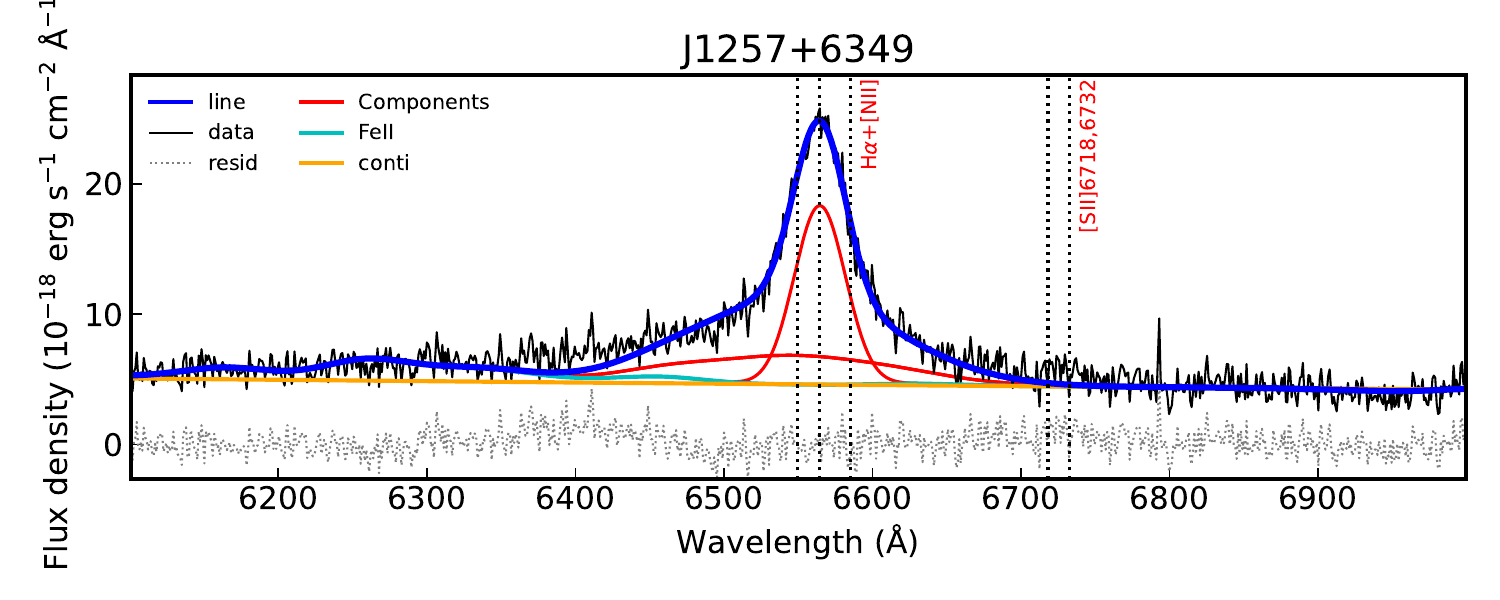}
\end{minipage}
\begin{minipage}[t]{0.5\textwidth}
\centering
\includegraphics[width=\textwidth]{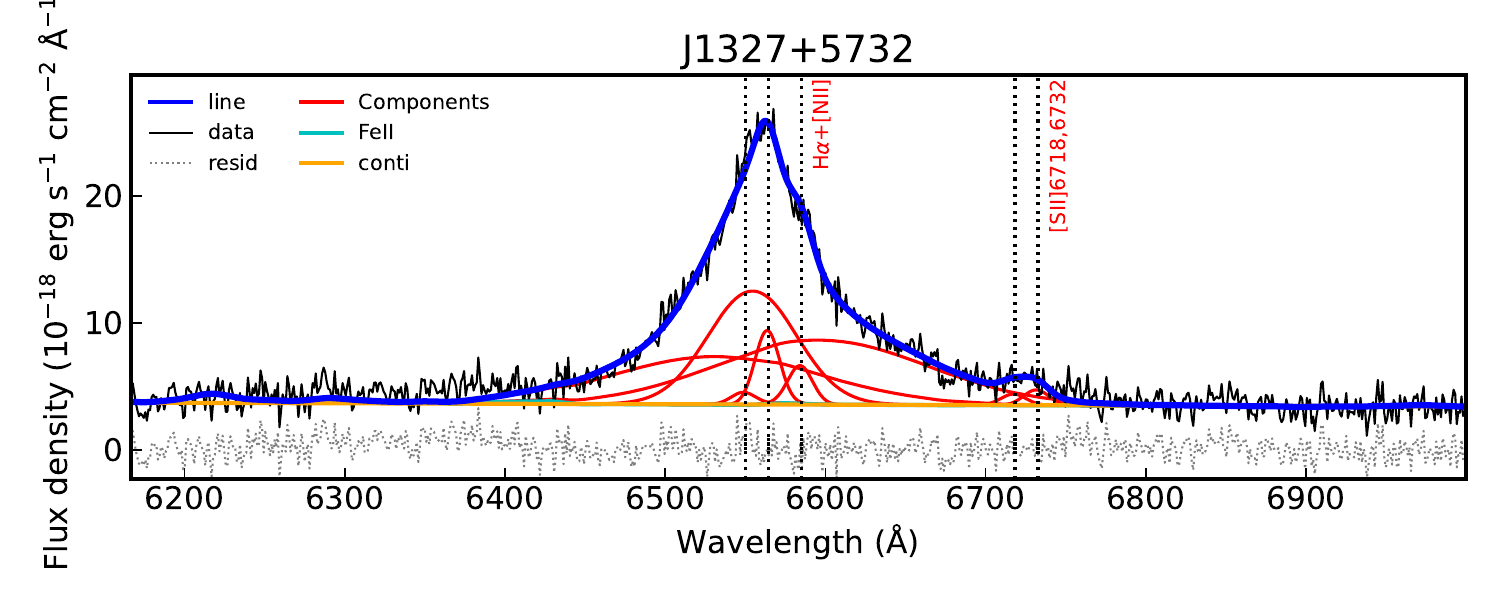}
\end{minipage}

\begin{minipage}[t]{0.5\textwidth}
\centering
\includegraphics[width=\textwidth]{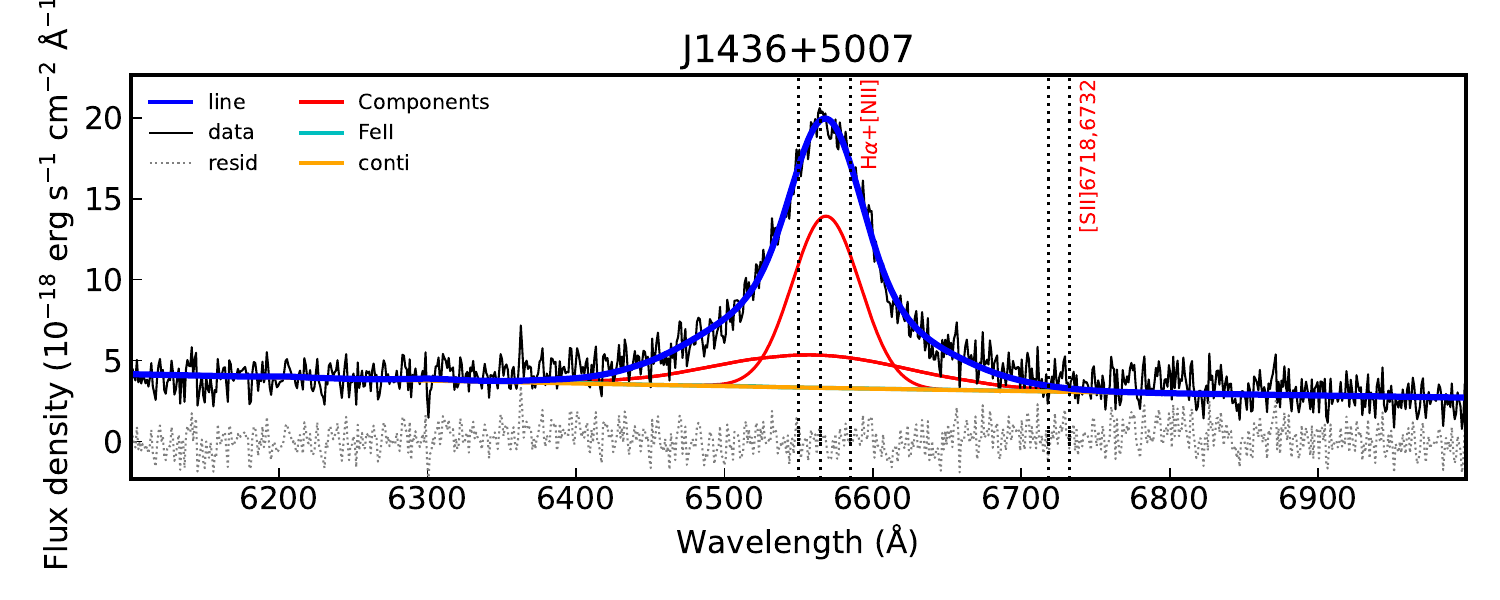}
\end{minipage}
\begin{minipage}[t]{0.5\textwidth}
\centering
\includegraphics[width=\textwidth]{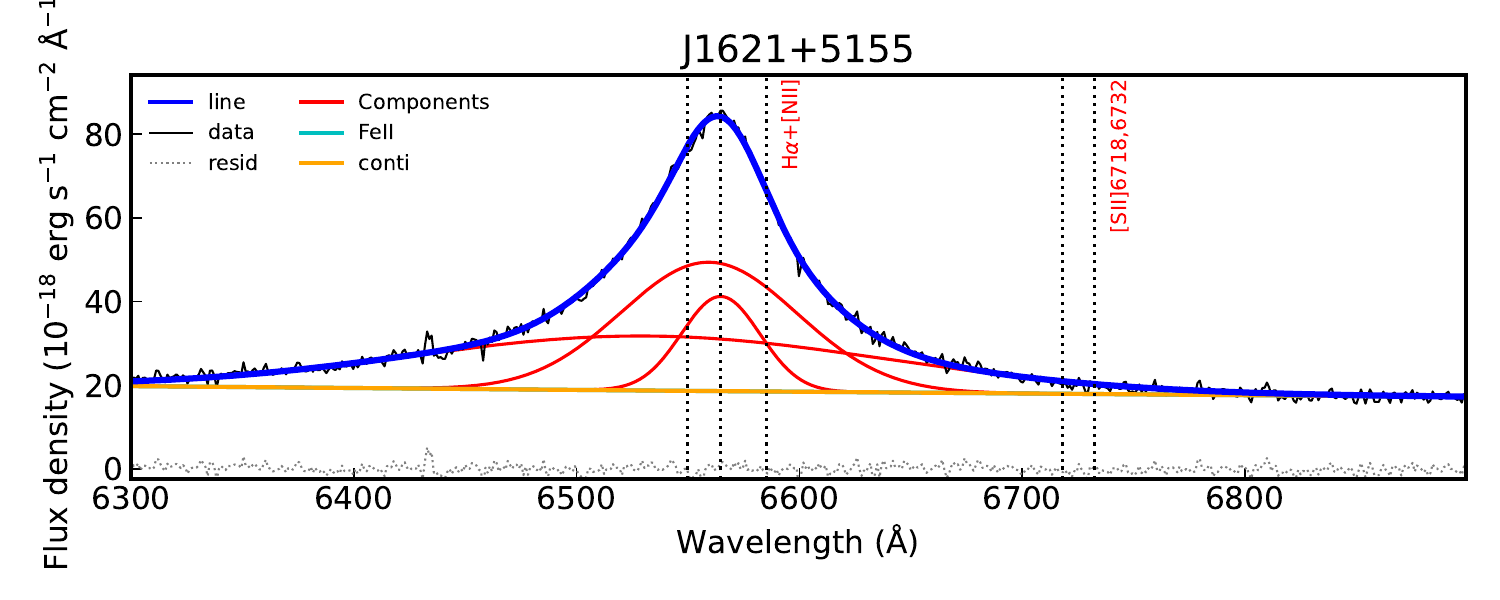}
\end{minipage}

\caption{Spectra and best-fit models for the \ha\ region of the $z \sim 6$ objects. The plotting styles are the same as in Fig. \ref{fig:specfit}.}
\label{fig:haexample}
\end{figure*}

\vspace{5mm}
\facilities{JWST(NIRSpec)}


\software{astropy \citep{Astropy2018},   
          PyQSOFit \citep{pyqsofit,Shen2019}
          }



\bibliography{BH_5_6}{}
\bibliographystyle{aasjournal}



\end{document}